%% file: WOLA_subband_processing.tex
\documentclass[letterpaper,journal]{IEEEtran}

\newcommand{\normsq}[1]{\left\lVert #1 \right\rVert^{2}}

\newcommand{\trans}{\intercal}

\newcommand{\pinv}{\ddag}

\newcommand{\idft}[1]{\mathbb{F}_{#1}^{*}}
\newcommand{\dft}[1]{\mathbb{F}_{#1}}
\newcommand{\rectwin}[1]{$\sqcap_{#1}$}
\newcommand{\cosinewin}[1]{$\cap_{#1}$}
\newcommand{\roothannwin}[1]{$\Cap_{#1}$}
\newcommand{\erl}{\text{ERLE}}
\newcommand{\E}[1]{\mathbb{E}\left\{ #1 \right\}}

\usepackage{microtype}
\usepackage{anyfontsize}
\usepackage{cite}
\usepackage{amsmath,amssymb,amsfonts,mathrsfs}
\usepackage{algorithm,algpseudocode}
\usepackage{optidef}
\usepackage{graphicx}
\usepackage[caption=false,font=normalsize,labelfont=sf,textfont=sf]{subfig}
\usepackage{textcomp}
\usepackage{multirow}
\usepackage{tikz}
\usepackage{url}		
\usepackage{float}
\usepackage{xcolor}
\usepackage{hyperref}
\hypersetup{hidelinks}
\def\BibTeX{{\rm B\kern-.05em{\sc i\kern-.025em b}\kern-.08em
    T\kern-.1667em\lower.7ex\hbox{E}\kern-.125emX}}
\usepackage{balance}
\usepackage{arydshln}   
\usepackage{dsfont}
\usepackage{booktabs}

\usepackage{pgfplots}
\usepgfplotslibrary{polar}
\pgfplotsset{compat=newest}
\usetikzlibrary{plotmarks}
\usetikzlibrary{calc}
\usetikzlibrary{arrows.meta}
\usepgfplotslibrary{patchplots}
\usepackage{grffile}
\newlength\fheight
\newlength\fwidth
\newcommand{\plotlinewidth}{0.8pt}
\usepackage{enumitem}
\usepackage[acronym]{glossaries}
\makeatletter
\def\endthebibliography{%
	\def\@noitemerr{\@latex@warning{Empty `thebibliography' environment}}%
	\endlist
}
\makeatother

\begin{document}
\title{A Generalized Weighted Overlap-Add (WOLA) Filter Bank for Improved Subband System Identification}
\author{Mohit Sharma\IEEEauthorrefmark{1},  Robbe Van Rompaey\IEEEauthorrefmark{2}, Wouter Lanneer\IEEEauthorrefmark{2} and Marc Moonen\IEEEauthorrefmark{1} ~\IEEEmembership{Fellow,~IEEE}
\thanks{This research work was carried out at the ESAT Laboratory of KU Leuven, in the frame of  FWO Research Project nr. G0A9921N ``Per-Tone Equalization and Per-Tone Precoding for Speech and Audio Signal Processing", Research Council KU Leuven Project C14-21-0075 ``A holistic approach to the design of integrated and distributed digital signal processing algorithms for audio and speech communication devices", and VLAIO Project nr. HBC.2022.0428 ``SPARTA- Signal Processing for next-generation opticAl netwoRks and remoTe Audio experiences".  The scientific responsibility is assumed by its authors.\\
\noindent Mohit Sharma and Marc Moonen are with KU Leuven, Department of Electrical Engineering (ESAT), KU Leuven, 3000 Leuven, Belgium.\\
\noindent Robbe Van Rompaey and Wouter Lanneer are with Nokia Bell Labs, Copernicuslaan 50, 2018 Antwerp, Belgium.}}

\markboth{IEEE TRANSACTIONS ON SIGNAL PROCESSING,~Vol.~xx, No.~xx, August~xx}{}%

\maketitle
{\tiny }
\input{WOLA_win_design_Acros.tex}%
\begin{abstract}
    This paper addresses the challenges in \gls{STFT} domain subband adaptive filtering, in particular, subband system identification. 
    Previous studies in this area have primarily focused on setups with subband filtering at a downsampled rate, implemented using the \gls{WOLA} filter bank, popular in audio and speech-processing for its reduced complexity. However, this traditional approach imposes constraints on the subband filters when transformed to their full-rate representation.
    This paper makes three key contributions.  First, it introduces a generalized \gls{WOLA} filter bank that repositions subband filters before the downsampling operation, eliminating the constraints on subband filters inherent in the conventional \gls{WOLA} filter bank. 
    Second, it investigates the \gls{MSE} performance of the generalized \gls{WOLA} filter bank for full-band system identification, establishing analytical ties between the order of subband filters, the full-band system impulse response length, the decimation factor, and the prototype filters.
    Third, to address the increased computational complexity of the generalized \gls{WOLA}, the paper proposes a low-complexity implementation termed \gls{PTWOLA}, which maintains computational complexity on par with conventional \gls{WOLA}. 
    Analytical and empirical evidence demonstrates that the proposed generalized \gls{WOLA} filter bank significantly enhances the performance of subband system identification.
\end{abstract}


\glsresetall 
\input{WOLA_subband_introduction.tex}
\input{WOLA_subband_ideal_filter.tex}

\input{WOLA_subband_low_order_filter.tex}
\input{WOLA_subband_domain_steady_state_solution.tex}
\input{WOLA_subband_domain_system_identification.tex}

\input{WOLA_subband_performance_analysis.tex}
\input{WOLA_subband_PTEQ_WOLA.tex}%
\input{WOLA_subband_results.tex}
\input{WOLA_subband_conclusion.tex}
\bibliography{WOLA_win_design_bib.bib}%
\bibliographystyle{ieeetr}
\vfill\pagebreak

\end{document}

%% file: WOLA_win_design_Acros.tex
\newacronym{WOLA}{WOLA}{weighted overlap-add}
\newacronym{PTEQ}{PTEQ}{per-tone equalizer}
\newacronym{ADSL}{ADSL}{asymmetric digital subscriber line }
\newacronym{PTWOLA}{PT-WOLA}{per-tone weighted overlap-add}
\newacronym{FBLMS}{FBLMS}{frequency-domain block least-mean-square}
\newacronym{OLS}{OLS}{overlap-save}
\newacronym{AEC}{AEC}{acoustic echo cancellation}
\newacronym{AFC}{AFC}{acoustic feedback cancellation}
\newacronym{NR}{NR}{noise reduction}
\newacronym{DFT}{DFT}{discrete Fourier transform}
\newacronym{STFT}{STFT}{short-time Fourier transform}
\newacronym{PR}{PR}{perfect reconstruction}
\newacronym{FIR}{FIR}{finite impulse response}
\newacronym{ERLE}{ERLE}{echo return loss enhancement}
\newacronym{RIR}{RIR}{room impulse response}
\newacronym{LEM}{LEM}{loudspeaker-enclosure-microphone}
\newacronym{RLS}{RLS}{recursive least squares}
\newacronym{LTI}{LTI}{linear time-invariant}
\newacronym{IDFT}{IDFT}{inverse discrete Fourier transform}
\newacronym{EBR}{EBR}{echo-to-background ratio}
\newacronym{FDAF}{FDAF}{frequency-domain adaptive filtering}
\newacronym{MMSE}{MMSE}{minimum mean square error}
\newacronym{MSE}{MSE}{mean square error}
\newacronym{LS}{LS}{least-squares}

%% file: WOLA_subband_introduction.tex
\section{INTRODUCTION}
\IEEEPARstart{A}{daptive} filtering problems appear in various signal processing fields. Particularly in system identification problems when dealing with systems with long impulse responses, such as in audio and speech signal processing applications, adaptive filtering is performed in the \gls{STFT}-domain to reduce computational complexity \cite{1163353,109205}. \gls{STFT}-domain subband adaptive filtering typically employs a uniform \gls{DFT} modulated filter bank for efficient implementation.

Subband adaptive filtering utilizing a uniform \gls{DFT} modulated filter bank with specific analysis and synthesis filters along with constrained subband processing, can be represented equivalently as a full-band time-domain adaptive filtering operation. Such subband adaptive filtering is typically labeled as \gls{FDAF} \cite{109205, 1164216, Diniz2013}.

\begingroup%
  \makeatletter%
  \providecommand\color[2][]{%
    \errmessage{(Inkscape) Color is used for the text in Inkscape, but the package 'color.sty' is not loaded}%
    \renewcommand\color[2][]{}%
  }%
  \providecommand\transparent[1]{%
    \errmessage{(Inkscape) Transparency is used (non-zero) for the text in Inkscape, but the package 'transparent.sty' is not loaded}%
    \renewcommand\transparent[1]{}%
  }%
  \providecommand\rotatebox[2]{#2}%
  \newcommand*\fsize{\dimexpr\f@size pt\relax}%
  \newcommand*\lineheight[1]{\fontsize{\fsize}{#1\fsize}\selectfont}%
  \ifx\svgwidth\undefined%
    \setlength{\unitlength}{485.08734756bp}%
    \ifx\svgscale\undefined%
      \relax%
    \else%
      \setlength{\unitlength}{\unitlength * \real{\svgscale}}%
    \fi%
  \else%
    \setlength{\unitlength}{\svgwidth}%
  \fi%
  \global\let\svgwidth\undefined%
  \global\let\svgscale\undefined%
  \makeatother%
  \begin{figure*}
    \centering
    \begin{picture}(1,0.25346742)%
      \lineheight{1}%
      \setlength\tabcolsep{0pt}%
    \put(0,0){\includegraphics[width=\unitlength,page=1]{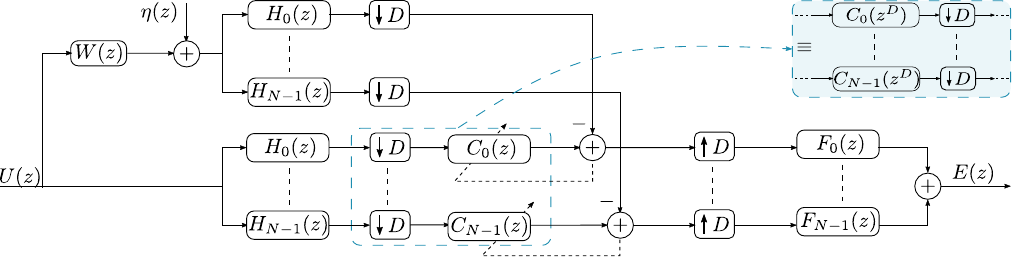}}%
    \put(0.61,0.188){\color[rgb]{0,0,0}\makebox(0,0)[lt]{\lineheight{1.25}\smash{\begin{tabular}[t]{l}{\footnotesize (Multirate noble identity)}\end{tabular}}}}%
  \end{picture}%
  \caption{\Acrfull{WOLA} filter bank based subband system identification.}
  \label{fig:Conventional_WOLA_subband_system_identification_block_diagram}
  \end{figure*}
\endgroup%

General oversampled uniform \gls{DFT} modulated filter bank-based subband adaptive filtering, with zero-order analysis and synthesis filter polyphase components, is commonly referred to as \gls{WOLA} based subband adaptive filtering. The term `zero-order' implies that the prototype filter length does not exceed the number of subbands $N$, resulting in scalar polyphase components \cite{vaidyanathan2006multirate}. Fig. \ref{fig:Conventional_WOLA_subband_system_identification_block_diagram} illustrates the state-of-the-art setup for \gls{WOLA} based subband system identification, where $W(z)$  is the system to be identified, with input signal $U(z)$ and noise $\eta(z)$. Moreover, it incorporates analysis filter $H_k(z)$, synthesis filter $F_k(z)$ and subband filter $C_{k}(z)$ associated with $k^{th}$ subband. When the subband filters are one-tap filters, the setup represents conventional \gls{WOLA} based subband processing. With higher-order subband filters, the setup represents \gls{WOLA} based subband processing with inter-frame filtering.

\gls{WOLA} based subband adaptive filtering offers reduced computational complexity by closely approximating the long time-domain impulse response of the system with a lower-order subband filter in each subband, and by performing the subband adaptive filtering at a downsampled rate. Hence, it is widely popular in audio and speech signal processing applications \cite{196909}. However, this reduction in computational complexity comes at the cost of undesirable aliasing and distortion effects. Consequently, \gls{WOLA} based subband adaptive filtering does not correspond to a full-band time-domain adaptive filtering operation, which limits its overall modeling accuracy.

Various techniques have been proposed to improve the modeling accuracy of \gls{WOLA} based subband adaptive filtering. These techniques can be broadly classified into two categories: those aimed at reducing aliasing effects originating from the downsampling and non-ideal prototype filters in the analysis filter bank, and those aimed at reducing cyclic distortion effects originating from \gls{STFT}-domain processing of full-band filtering.

To reduce the impact of aliasing effects, adaptive cross-band filters have been proposed in \cite{149989} and analyzed in \cite{avargel2006performance, 4156182}. In \cite{6287813}, a post-processing stage has been proposed that generates additional subband filter coefficients for time-frequency interpolated reference signals. Additionally, multiple prototype filter design techniques have been proposed in \cite{1168464, 9413836, 8683473, 8521343, 1658256, ENEMAN20011947, 4202622, 6834830, 4637900} to minimize the inter-subband aliasing, thereby reducing the aliasing effects.

To mitigate cyclic distortion effects, constrained subband processing, similar to \gls{FDAF}, has been proposed in \cite{109205}. However, the constraints introduced by this method add complexity and limit ``the degrees of freedom", restricting the means of improvement in filter bank design. In \cite{ENEMAN2001117, 863066}, the relation between full-band and \gls{STFT}-domain subband adaptive filtering has been analyzed, motivating the use of full-band errors for adaptation, similar to a closed-loop structure. However, this method introduces additional delay to the adaptation path, leading to slow and unreliable convergence, as noted in \cite{Diniz2013}. In \cite{6853808}, a novel scheme for \gls{WOLA} based subband adaptive filtering has been proposed. The proposed adaptive filter structure in \cite{6853808} employs an additional set of adaptive subband filters and leverages the relation between linear convolution and the root-Hann prototype analysis filter. However, this approach introduces additional computational complexity due to the additional set of subband filters and limits the filter bank design to the root-Hann prototype analysis filter. 

To the best of our knowledge, no prior work has specifically addressed the impact of the prototype synthesis filter design on cyclic distortions. 
Additionally, there is a notable gap in research concerning the theoretical performance analysis of \gls{WOLA} filter banks from a time-domain perspective. Previous studies on \gls{WOLA} filter bank-based adaptive filtering have predominantly focused on subband filter adaptation at a downsampled rate, as illustrated in Fig. \ref{fig:Conventional_WOLA_subband_system_identification_block_diagram}. By applying the so-called multirate noble identities \cite{vaidyanathan2006multirate} to move the subband filters before the downsampling operation, it is seen that, at full rate, the subband filters are constrained to the $C_{k}(z^{D})$ structure. This structure imposes  $D - 1$  zeroes between subsequent filter coefficients, thus restricting the conventional \gls{WOLA} filter bank to inter-frame filtering.
The predominant focus on subband adaptive filtering at a downsampled rate in existing research can be attributed to the practical challenges associated with the computational complexity of performing subband adaptive filtering at full rate. Although it is theoretically feasible to position the subband filters before the downsampling operation, as shown in Fig. \ref{fig:generalized_WOLA_subband_system_identification_block_diagram}, the high computational demands have rendered such a system impractical for real-world applications.

Nevertheless, this paper aims to analyze a generalized \gls{WOLA} filter bank with full rate subband adaptive filtering, as shown in Fig. \ref{fig:generalized_WOLA_subband_system_identification_block_diagram}, and analyze the impact of various factors on the overall system performance from a time-domain perspective. The paper also aims at seeking a low-complexity implementation of such a system.
The paper makes three key contributions: \\
\noindent \textit{First}, the paper introduces a generalized \gls{WOLA} filter bank for subband adaptive filtering, which is motivated by the \gls{PTEQ} structure in \gls{ADSL} systems \cite{Per_tone_equalizer_katleen,9659802} and  fundamentally removes the primary limitation of conventional \gls{WOLA}. As shown in Fig. \ref{fig:generalized_WOLA_subband_system_identification_block_diagram}, by repositioning the subband filters to operate at the full rate before the downsampling operation, the filters are no longer constrained to the sparse $C_{k}(z^{D})$ form, but become a general, unconstrained $C_{k}(z)$ filter, providing the necessary degrees of freedom for the accurate system modeling that was previously not possible\footnote{The conventional \gls{WOLA} with inter-frame filtering in Fig. \ref{fig:Conventional_WOLA_subband_system_identification_block_diagram} can be represented by the proposed generalized \gls{WOLA}, by constraining its subband filters $C_{k}(z)$ in Fig. \ref{fig:generalized_WOLA_subband_system_identification_block_diagram} to follow $C_{k}(z^{D})$.}.
The paper further investigates the application of the generalized \gls{WOLA} filter bank in subband system identification, with a particular focus on its time-domain representation, addressing the limitations of conventional WOLA filter banks in subband system identification. 

\noindent \textit{Second}, the paper examines the \gls{MSE} performance of the generalized \gls{WOLA} filter bank in full-band system identification and establishes analytical ties between various factors, such as the order of subband filters, the length of the full-band system impulse response, the decimation factor, and the design of prototype analysis and synthesis filters.

\noindent \textit{Third}, since the proposed approach leads to a substantial increase in computational complexity, to address this, the paper presents an alternative low-complexity implementation of the generalized \gls{WOLA} filter bank, referred to as \gls{PTWOLA}, inspired by the term \gls{PTEQ} from \cite{Per_tone_equalizer_katleen}. The \gls{PTWOLA} maintains a computational complexity comparable to that of conventional \gls{WOLA}, while demonstrating performance that is comparable to, or slightly exceeds,  that of the generalized \gls{WOLA} filter bank for the same subband filter order.

The structure of the paper is as follows: 
Section \ref{WOLA_subband_filtering_Sec:Preliminaries_TD_rep_low_order_fitlering} presents the system model for generalized \gls{WOLA} filter bank-based subband adaptive filtering, also establishing a time-domain representation of the subband filtering. 
Section \ref{WOLA_subband_filtering_Sec:Steady_state_solution} presents the steady-state solution for the generalized \gls{WOLA} filter bank based subband system identification, while Section \ref{WOLA_subband_filtering_Sec:MSE_performance_analysis} analyzes its the steady-state \gls{MSE} performance. 
Section \ref{WOLA_subband_filtering_Sec:Per-Tone_WOLA_comp_eff_multi_tap_WOLA} discusses the computational complexity of the generalized \gls{WOLA} filter bank and proposes a computationally efficient implementation, termed \gls{PTWOLA}. 
Section \ref{WOLA_subband_filtering_Sec:Simulation_results_AEC} presents simulation results for \gls{AEC} as a use-case scenario. 
Finally, Section \ref{WOLA_subband_filtering_Sec:conclusions} concludes the paper.

%% file: WOLA_subband_ideal_filter.tex
\section{Generalized  \gls{WOLA} and Time-Domain Representation of Subband Filtering}\label{WOLA_subband_filtering_Sec:Preliminaries_TD_rep_low_order_fitlering}
\noindent  The subband system identification problem comprises two primary components: subband filtering and the adaptation of subband filter coefficients that minimize the subband errors\footnote{In literature, the so-called open-loop subband system identification is widely used due to its faster convergence, which minimizes the subband errors, in contrast to the closed-loop subband system identification, wherein the full-band error after the synthesis filter bank is minimized \cite{862015,Diniz2013}.}. This section focuses on the subband filtering aspect and derives a relation between the subband filtering and the overall time-domain transfer function of the generalized \gls{WOLA} filter bank, shown in Fig. \ref{fig:generalized_WOLA_subband_system_identification_block_diagram}, which will be utilized in subsequent sections.

\begingroup%
  \makeatletter%
  \providecommand\color[2][]{%
    \errmessage{(Inkscape) Color is used for the text in Inkscape, but the package 'color.sty' is not loaded}%
    \renewcommand\color[2][]{}%
  }%
  \providecommand\transparent[1]{%
    \errmessage{(Inkscape) Transparency is used (non-zero) for the text in Inkscape, but the package 'transparent.sty' is not loaded}%
    \renewcommand\transparent[1]{}%
  }%
  \providecommand\rotatebox[2]{#2}%
  \newcommand*\fsize{\dimexpr\f@size pt\relax}%
  \newcommand*\lineheight[1]{\fontsize{\fsize}{#1\fsize}\selectfont}%
  \ifx\svgwidth\undefined%
    \setlength{\unitlength}{488.05149349bp}%
\ifx\svgscale\undefined%
\relax%
\else%
\setlength{\unitlength}{\unitlength * \real{\svgscale}}%
\fi%
\else%
\setlength{\unitlength}{\svgwidth}%
\fi%
\global\let\svgwidth\undefined%
\global\let\svgscale\undefined%
\makeatother%
  \begin{figure*}
    \centering
\begin{picture}(1,0.38259773)%
      \lineheight{1}%
      \setlength\tabcolsep{0pt}%
    \put(0,0){\includegraphics[width=\unitlength,page=1]{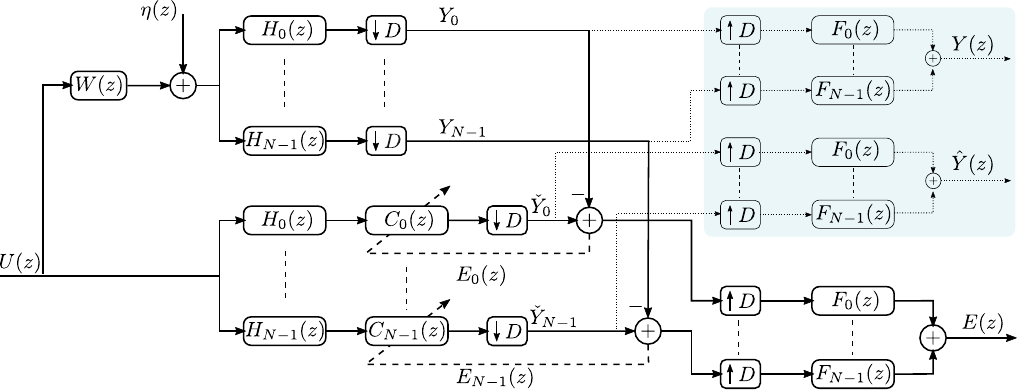}}%
  \end{picture}%
  \caption{Generalized \acrfull{WOLA} filter bank based subband system identification.}
\label{fig:generalized_WOLA_subband_system_identification_block_diagram}
\end{figure*}
\endgroup%

The relation between the input signal $u(n)$ and output signal $y(n)$ for an unknown  \gls{LTI} system $w(n)$, ignoring noise, is first given as
\begin{align} \label{eq:general_LTI_input_output_equation}
	y(n) = u(n) \star w(n)
\end{align}
where $\star$ represents the convolution operation, which can equivalently be expressed in the z-domain as
\begin{align}\label{eq:LTI_system_z_domain}
	\underbrace{\sum_{m = 0}^{\infty}y(m)z^{-m}}_{Y(z)}  = \underbrace{\sum_{p = 0}^{\infty}u(p)z^{-p}}_{U(z)} \underbrace{\sum_{q = 0}^{L-1}w(q)z^{-q}}_{W(z)}
\end{align}
where, $Y(z)$, $U(z)$ and $W(z)$  represent the z-transform of $y(n)$, $u(n)$ and $w(n)$, respectively, and $L$ denotes the \gls{FIR} length of the unknown \gls{LTI} system. 
The unknown \gls{LTI} system, in this case, acts on the full-band signal $U(z)$ as a whole. Therefore, if $L$ is large, accurately approximating the full-band \gls{LTI} system may necessitate a high filter order.

In the generalized \gls{WOLA} filter bank in Fig. \ref{fig:generalized_WOLA_subband_system_identification_block_diagram}, however, $Y(z)$ represents the input signal $U(z)$ convolved with the unknown \gls{LTI} system transfer function $W(z)$ and additionally passed through the \gls{WOLA} filter bank, which is assumed to satisfy the perfect reconstruction property and hence only introduces additional delay. $\hat{Y}(z)$ represents the estimate of $Y(z)$ using a subband filter $C_{k}(z)$ in the $k^{th}$ subband $\left(k \, = \, 0 \, \cdots \, N-1\right)$ and $E_{k}(z)$ represents the $k^{th}$ subband error. Furthermore, the generalized \gls{WOLA} filter bank in Fig. \ref{fig:generalized_WOLA_subband_system_identification_block_diagram} consists of analysis filters $H_{k}(z)$, synthesis filters $F_{k}(z)$, and a $D$-fold decimation and interpolation $\left(D \leq N\right)$. The relationship between the input and output of the filter bank is characterized by the so-called overall transfer function of the filter bank, which includes a distortion function $T(z)$ and alias transfer functions $A_{m}(z)$, i.e.
\begin{align} \label{eq:overall_transfer_function_ideal_FB}
	&Y(z) =  \underbrace{\left\{ \frac{1}{D} \sum_{k = 0}^{N-1} H_{k}(z) F_{k}(z)\right\}}_{T(z)} U(z)W(z)  \nonumber \\
&+  \sum_{m = 1}^{D-1} \left\{\underbrace{\frac{1}{D}\left( \sum_{k = 0}^{N-1}  H_{k}(z \alpha_{D}^{-m}) F_{k}(z) \right)}_{A_{m}(z)} U(z \alpha_{D}^{-m}) W(z \alpha_{D}^{-m})\right\}
\end{align}
where $\alpha_{D} \triangleq \exp^{j2\pi/D}$.

\noindent The analysis and synthesis filters are designed to ensure perfect reconstruction, i.e.
\begin{subequations}\label{eq:PR_conditions}
	\begin{align}
		\begin{split}
			T(z) = z^{-\tau} \ ,
		\end{split}\\
		\begin{split}
			A_{m}(z) = 0, \: \forall m = 1 \cdots D-1
		\end{split}.
	\end{align} 
\end{subequations}
Substituting \eqref{eq:PR_conditions} in \eqref{eq:overall_transfer_function_ideal_FB}, the $Y(z)$ in Fig. \ref{fig:generalized_WOLA_subband_system_identification_block_diagram} is indeed given as
\begin{align} \label{eq:output_no_subband_processing}
	Y(z) = U(z) W(z)z^{-\tau}.
\end{align}
Hence, it can be observed that \eqref{eq:output_no_subband_processing} is equivalent to \eqref{eq:LTI_system_z_domain}, corrected with the (unavoidable) delay represented by $\tau$, i.e.
\begin{equation}\label{eq:general_LTI_input_output_equation_with_delay}
	y(n) = u(n- \tau) \star w(n) \ .
\end{equation}

%% file: WOLA_subband_low_order_filter.tex
\noindent Similarly, for $\hat{Y}(z)$ in Fig. \ref{fig:generalized_WOLA_subband_system_identification_block_diagram}, the input-output relation is given as 
\begin{align}\label{eq:overall_transfer_fun_low_order_subband}
    &\hat{Y}(z) = \underbrace{\left\{ \frac{1}{D} \sum_{k = 0}^{N-1} H_{k}(z) F_{k}(z) C_{k}(z) \right\}}_{\hat{T}(z) } U(z) \nonumber \\
    & +  \sum_{m = 1}^{D-1} \underbrace{\left\{\frac{1}{D}\sum_{k = 0}^{N-1}  H_{k}(z \alpha_{D}^{-m}) F_{k}(z) C_{k}(z \alpha_{D}^{-m})\right\}}_{\hat{A}_{m}(z)} U(z \alpha_{D}^{-m})
\end{align}
where, $\hat{T}(z)$ and $\hat{A}_{m}(z)$ are the so-called distortion function and alias transfer functions of the filter bank with subband processing, respectively. Moreover, $C_{k}(z)$ represents the $k^{th}$ subband \gls{FIR} filter with length $T$, i.e.
\begin{align}
    C_{k}(z) = \sum_{q = 0}^{T-1}c_{k}(q)z^{-q}
\end{align} 
where $c_{k}(q)$ is the $q^{th}$ filter coefficient of the $k^{th}$ subband filter and $\mathbf{c}_{k} \triangleq \begin{bmatrix} c_{k}(0), \cdots, c_{k}(T-1) \end{bmatrix}$.
The time-domain equivalent of \eqref{eq:overall_transfer_fun_low_order_subband} is then given as
\begin{align}\label{eq:PTWOLA_estimated_TD_echo_signal}
	\hat{y}(n) = u(n) \star \hat{t}(n)  + \sum_{m=1}^{D-1}\left(  u(n) \star \hat{a}_{m}(n)  \alpha_{D}^{nm} \right)
\end{align}
The following subsections provide a separate analysis for the distortion function and the aliasing transfer functions.

\subsection{Distortion Function}\label{WOLA_subband_filtering_Sec:Low_order_filter_Subsec:Distortion_transfer_fcn}
In a uniform \gls{DFT} modulated filter bank, the $k^{th}$ analysis and synthesis filter are derived from a prototype analysis filter $\mathbf{h}_{0} \triangleq \begin{bmatrix} h_{0}(0), \cdots, h_{0}(N-1) \end{bmatrix}$ and a prototype synthesis filter $\mathbf{f}_{0} \triangleq \begin{bmatrix} f_{0}(0), \cdots, f_{0}(N-1) \end{bmatrix}$, respectively, i.e.\footnote{The scaling factor $\alpha_{N}^{k}$ in \eqref{eq:unifromDFT_filter_bank_proporty_prototype} for the synthesis filters is introduced primarily for the sake of mathematical convenience \cite{vaidyanathan2006multirate}.}
\begin{align}\label{eq:unifromDFT_filter_bank_proporty_prototype}
	h_{k}(n) = h_{0}(n) \alpha_{N}^{nk}   &\quad  \text{and} \quad f_{k}(n) = f_{0}(n)\alpha_{N}^{nk}\alpha_{N}^{k}\nonumber \\
	&\Leftrightarrow  \nonumber \\
	H_{k}(z) = H_{0}(z \alpha_{N}^{-k})   &\quad \text{and} \quad F_{k}(z) = F_{0}(z \alpha_{N}^{-k})\alpha_{N}^{k} .
\end{align}
Therefore the distortion function in \eqref{eq:overall_transfer_fun_low_order_subband} can be expressed as
\begin{align}
    \hat{T}(z) = \frac{1}{D} \sum_{k = 0}^{N-1} H_{0}(z \alpha_{N}^{-k}) F_{0}(z \alpha_{N}^{-k}) \alpha_{N}^{k} C_{k}(z)  .
\end{align}
The corresponding time-domain filter coefficients $\mathbf{\hat{t}} \triangleq \begin{bmatrix} \hat{t}(0),  \ \cdots,  \ \hat{t}(2N+T-3) \end{bmatrix}$ are then given as
\begin{align}\label{eq:hat_t_g0_n_sums}
	\hat{t}&(n) = \frac{1}{D} \sum_{k = 0}^{N-1} \left[ \left(h_{0}(n) \alpha_{N}^{nk}\right) \star \left(f_{0}(n) \alpha_{N}^{nk} \alpha_{N}^{k}\right)  \star c_{k}(n) \right]  \\[0.5ex]
	&=\! \frac{1}{D} \!\sum_{k=0}^{N-1}\!\sum_{l=0}^{T-1} \!\sum_{m=0}^{N-1} \! \!h_{0}(n\!-\!m\!-\!l) \alpha_{N}^{k \left(n-m-l \right)}\! f_{0}(m) \alpha_{N}^{k m}\alpha_{N}^{k} c_{k}(l) \nonumber \\[1ex] \label{eq:LTI_filter_for_distortion_function_DFT_FB_subband}
	&= \sum_{k=0}^{N-1} \sum_{l=0}^{T-1}  \underbrace{\left\{\frac{1}{D}\sum_{m=0}^{N-1} h_{0}(n\!-\!m\!-\!l) f_{0}(m) \right\} }_{g_{0}(n-l)}\alpha_{N}^{k \left( n-l+1 \right)}c_{k}(l)
\end{align}
\noindent where ${g}_{0}(n)$ represents the  scaled convolution of the prototype analysis filter $h_{0}(n)$ and the prototype synthesis filter $f_{0}(n)$, of length $2N-1$, i.e. 
\begin{align}
     g_{0}(n) =  \frac{1}{D} \sum_{m=0}^{N-1} h_{0}(n-m) f_{0}(m)  \ .
\end{align}
Alternatively, \eqref{eq:LTI_filter_for_distortion_function_DFT_FB_subband} can be written as
\begin{align} \label{eq:PTWOLA_subband_intermediate_distoortion_fcn1}
    \hat{t}(n) &= \sum_{l=0}^{T-1} g_{0}(n-l) \underbrace{\left\{\sum_{k=0}^{N-1} \alpha_{N}^{k \left( n-l+1 \right)}c_{k}(l)\right\}}_{\tilde{c}^{l}(n-l)} \nonumber \\ 
    &= \sum_{l=0}^{T-1} g_{0}(n-l) \tilde{c}^{l}(n-l)
\end{align}
where $\tilde{c}^{l}(n)$ are time-domain subband filter coefficients, obtained as the (circularly shifted) $N$-point \gls{IDFT} of the collection of $l^{th}$ filter coefficients for all subband filters, i.e.
\begin{align}\label{eq:periodic_nature_subband_low_orlder}
	\tilde{c}^{l}(n) \triangleq \sum_{k=0}^{N-1} c_{k}(l) \alpha_{N}^{k (n+1)}  
\end{align}
which is periodic with period $N$:  $\tilde{{c}}^{l}(n+N) = \tilde{{c}}^{l}(n) \quad  \forall l$ . \\ \noindent
The result in \eqref{eq:PTWOLA_subband_intermediate_distoortion_fcn1} can also be written as
\begin{align} \label{eq:distortion_function_sample_and_vector_form}
     \mathbf{\hat{t}}   =   \sum_{l=0}^{T-1}  \left\{ \begin{bmatrix} \mathbf{0}_{1,l}     &   \mathbf{g}_{0}     &   \mathbf{0}_{1,T-l-1} \end{bmatrix} \odot \begin{bmatrix} \mathbf{0}_{1,l}     &   \tilde{\mathbf{c}}^{l}     &    \mathbf{0}_{1,T-l-1} \end{bmatrix} \right\}
\end{align}
where, the symbol $\odot$ denotes the Hadamard (element-wise) product, \linebreak \noindent $\mathbf{g}_{0} \triangleq \begin{bmatrix} g_{0}(0), \ g_{0}(1),  \ \cdots,  \ g_{0}(2N-2) \end{bmatrix}$ and \linebreak \noindent $\tilde{\mathbf{c}}^{l} \triangleq \begin{bmatrix}
    \tilde{{c}}^{l}(0), \ \cdots, \ \tilde{{c}}^{l}(N-2),\ \tilde{{c}}^{l}(0), \ \cdots,  \ \tilde{{c}}^{l}(N-1)
\end{bmatrix}$, are vectors of length $2N-1$.
The two periodic images of $\tilde{c}^{l}(n)$ in $\tilde{\mathbf{c}}^{l}$, represent the circular convolution effect in the distortion function $\mathbf{\hat{t}}$. It is noted that conventional \gls{WOLA} has $T=0$, and hence the summation in \eqref{eq:distortion_function_sample_and_vector_form} is reduced to one term.

\subsection{Aliasing Transfer Functions}\label{WOLA_subband_filtering_Sec:Low_order_filter_Subsec:Aliasing_transfer_fcn}
Similar to \eqref{eq:hat_t_g0_n_sums}, each alias transfer function $\hat{A}_{m}(z)$, $\left(m \, = \, 1 \, \cdots \, D-1\right)$,  in \eqref{eq:overall_transfer_fun_low_order_subband} corresponds to time-domain filter coefficients given as
\begin{align}\label{eq:LTI_filter_for_aliasing_transfer_function_DFT_FB_subband}
	\hat{a}_{m}(n) \! = \! \frac{1}{D}\sum_{k = 0}^{N-1} \left(h_{0}(n) \alpha_{N}^{nk} \alpha_{D}^{nm}\right) \! \star \! \left(f_{0}(n) \alpha_{N}^{nk}\alpha_{N}^{k}\right) \!  \star \!  \left(c_{k}(n)\alpha_{D}^{nm} \right)
\end{align}
Similar to the derivation in Section \ref{WOLA_subband_filtering_Sec:Low_order_filter_Subsec:Distortion_transfer_fcn}, it can be shown that the expression in \eqref{eq:LTI_filter_for_aliasing_transfer_function_DFT_FB_subband} can be simplified to
\begin{align}
	\hat{a}_{m}(n) &\!=\! \resizebox{0.85\hsize}{!}{%
		$ \sum_{l=0}^{T-1}\! \left(\underbrace{\frac{1}{D} \sum_{p} h_{0}(p) f_{0}(n-l-p) \alpha_{D}^{\left(p+l\right) m}}_{\textstyle
			\begin{array}{c}{\psi}_{m}(n-l) \alpha_{D}^{lm}\end{array}} \underbrace{\sum_{k=0}^{N-1}  \alpha_{N}^{(n+1)k} c_{k}(l) \alpha_{N}^{-lk}}_{\textstyle
			\begin{array}{c}\tilde{{c}}^{l}(n-l)\end{array}}\right) $}\nonumber \\
	&= \sum_{l=0}^{T-1} \alpha_{D}^{lm} \  {\psi}_{m}(n-l) \ \tilde{{c}}^{l}(n-l) 
\end{align}
and
\begin{align}\label{eq:LTI_filter_for_aliasing_transfer_function_DFT_FB_subband_expanded}
 \mathbf{\hat{a}}_{m} = \sum_{l=0}^{T-1}\alpha_{D}^{lm}  \begin{bmatrix} \mathbf{0}_{1,l} \! & \! \boldsymbol{\psi}_{m} \!&\!  \mathbf{0}_{1,T-l-1} \end{bmatrix}  \odot \begin{bmatrix} \mathbf{0}_{1,l} \!&\!  \tilde{\mathbf{c}}^{l} \!&\!  \mathbf{0}_{1,T-l-1} \end{bmatrix}
\end{align} 
where, $ \mathbf{\hat{a}}_{m}  \triangleq \begin{bmatrix} \hat{a}_{m}(0), \ \hat{a}_{m}(1),  \ \cdots,  \ \hat{a}_{m}(2N-2) \end{bmatrix}$ and ${\psi}_{m}(n)$ represents the scaled modulated convolution of the prototype analysis and synthesis filter, of length $2N-1$, i.e. 
\begin{align}
	{\psi}_{m}(n) = \frac{1}{D}\sum_{p=0}^{N-1} h_{0}(p) f_{0}(n-p) \alpha_{D}^{p m} \ .
\end{align} 
and finally $ \boldsymbol{\psi}_{m}  \triangleq \begin{bmatrix} {\psi}_{m}(0), \ {\psi}_{m}(1),  \ \cdots,  \ {\psi}_{m}(2N-2) \end{bmatrix}$. It is noted again that conventional \gls{WOLA} has $T=0$, and hence the summation in \eqref{eq:LTI_filter_for_aliasing_transfer_function_DFT_FB_subband_expanded} is reduced to one term.

%% file: WOLA_subband_domain_steady_state_solution.tex
\section{Steady-State Solution}\label{WOLA_subband_filtering_Sec:Steady_state_solution}
This section derives the steady-state \gls{MSE} solution for the generalized \gls{WOLA} filter bank based subband system identification.

For each subband, the subband filter coefficients are estimated independently by minimizing the following \gls{MSE} cost function
\begin{mini}
	{\mathbf{c}_{k}}{\mathbb{E}\left\{ \left|{\check{Y}_{k}(q)- Y_{k}(q)}\right|^{2} \right\}}{\label{eq:PTWOLA_optimization_MMSE_filter_coefficients}}{}
\end{mini}  
It can be verified that for a white stationary input signal, the steady-state solution for \eqref{eq:PTWOLA_optimization_MMSE_filter_coefficients} corresponds to the \gls{LS} solution of
\begin{align}\label{eq:WOLA_based_subband_processing_open_loop_LS_equation}
	H_{k}(z) C_{k}(z) \stackrel{\text{\gls{LS}}}{=} H_{k}(z) W(z)
\end{align}

For a uniform \gls{DFT} modulated filter bank, the corresponding time-domain relation can be written as
\begin{align}\label{eq:steady_state_sol_uDFT_TD_relation}
	\left(h_{0}(n) \alpha_{N}^{nk}\right) \star  c_{k}(n) \stackrel{\text{\gls{LS}}}{=} \left(h_{0}(n) \alpha_{N}^{nk}\right) \star w(n) 
\end{align}
Defining, $\mathbf{w} \triangleq \begin{bmatrix}
	w(0) & w(1) & \cdots & w(L-1)& \mathbf{0}_{N-L,1}
\end{bmatrix}$ as the column vector of length $N$ representing the impulse response of the unknown \gls{LTI} system, \eqref{eq:WOLA_based_subband_processing_open_loop_LS_equation} can be expressed in matrix form as \eqref{eq:PTWOLA_subband_filter_converged_assumption_vectorized_raw}, shown at the top of the next page,
\begin{figure*}[ht!]
	\begin{align}\label{eq:PTWOLA_subband_filter_converged_assumption_vectorized_raw}
		\underbrace{\begin{bmatrix}
				h_{0}(0)                    & 0                     & \cdots & 0                     \\
				h_{0}(1)                  & h_{0}(0)                  & \cdots & 0                  \\
				\vdots & \ddots &          \ddots             & \vdots \\
				0                     & 0                     & \cdots & h_{0}(N-1) \\ \hdashline
				\multicolumn{4}{c}{\mathbf{0}_{N-T,T}}
		\end{bmatrix}}_{\mathbf{H}_{c}} \text{diag}\left( \dft{N}(k,0:T-1) \right) \mathbf{c}_{k}^{\trans} \stackrel{\text{\gls{LS}}}{=} 
		\underbrace{\begin{bmatrix}
				h_{0}(0)                    & 0                     & \cdots & 0                     \\
				h_{0}(1)                  & h_{0}(0)                  & \cdots & 0                  \\
				\vdots & \ddots &          \ddots             & \vdots \\
				0                     & 0                     & \cdots & h_{0}(N-1)                  
		\end{bmatrix}}_{\mathbf{H}_{w}} \text{diag}\left( \dft{N}(k,:) \right) \mathbf{w}^{\trans}
	\end{align}  
	\hrulefill
	\vspace{-.7\baselineskip}
\end{figure*}
where $\mathbf{H}_{c}$ and $\mathbf{H}_{w}$ are a Toeplitz matrix of size $(2N-1) \times T$ and $(2N-1) \times N$ respectively. 
The \gls{LS} solution for \eqref{eq:PTWOLA_subband_filter_converged_assumption_vectorized_raw} can then be written as
\begin{align}\label{eq:PTWOLA_defination_Z0_primary_expression}
	\text{diag}\left( \dft{N}(k,0:T-1) \right) \mathbf{c}_{k}^{\trans} = \underbrace{\left( \mathbf{H}_{c} \right)^{\pinv} \mathbf{H}_{w}}_{{\mathbf{Z}}_{0}} \text{diag}\left( \dft{N}(k,:) \right) \mathbf{w}^{\trans}
\end{align}
where $\dft{N}$ is the $N \times N$ \gls{DFT}-matrix and $\left( \mathbf{H}_{c} \right)^{\pinv} \triangleq \left( \left(\mathbf{H}_{c}\right)^{\trans} \mathbf{H}_{c}  \right)^{-1} \left(\mathbf{H}_{c}\right)^{\trans}$ represents the left Moore–Penrose inverse (pseudoinverse) of  $\mathbf{H}_{c}$. Therefore, ${\mathbf{Z}}_{0}$ is a matrix of size $T \times N$.

The $l^{th}$ filter coefficient of the $k^{th}$ subband filter ($\mathbf{c}_{k}$) in \eqref{eq:PTWOLA_defination_Z0_primary_expression}, is then given as
\begin{align}\label{eq:PTWOLA_focus_lth_coeff_kth_bin}
	\dft{N}(k,l) c_{k}(l) = {\mathbf{Z}}_{0}(l,:)\text{diag}\left( \dft{N}(k,:) \right)\mathbf{w}^{\trans}
\end{align}
Stacking the $l^{th}$ filter coefficients of all the subband filters, then leads to
\begin{align}
	\resizebox{1\hsize}{!}{%
		$
		\text{diag}\left( \dft{N}(l,:)\right) \begin{bmatrix}
			c_{0}(l) \\
			\vdots \\
			c_{N-1}(l)
		\end{bmatrix} = \begin{bmatrix}
			{\mathbf{Z}}_{0}(l,:) &\cdots &0\\
			\vdots&  \ddots & \vdots\\
			0& \cdots &  {\mathbf{Z}}_{0}(l,:)
		\end{bmatrix} \begin{bmatrix}
			\text{diag}\left( \dft{N}(0,:) \right) \\
			\vdots \\
			\text{diag}\left( \dft{N}(N-1,:) \right)
		\end{bmatrix}\mathbf{w}^{\trans}
		$ }%
\end{align}
which can be simplified as
\begin{align}\label{eq:PTWOLA_stacked_c_k_l_forall_k}
	\text{diag}\left( \dft{N}(l,:)\right) \begin{bmatrix}
		c_{0}(l) \\
		\vdots \\
		c_{N-1}(l)
	\end{bmatrix} = \dft{N} 
	\text{diag}\left( {\mathbf{Z}}_{0}(l,:) \right) \mathbf{w}^{\trans}
\end{align}
From \eqref{eq:periodic_nature_subband_low_orlder}, it follows that  $\begin{bmatrix} c_{0}(l) & \cdots & c_{N-1}(l) \end{bmatrix}^{\trans} = \dft{N} \left( \tilde{\tilde{\mathbf{c}}}^{l} \right)^{\trans}$ with $\tilde{\tilde{\mathbf{c}}}^{l} \triangleq \begin{bmatrix} \tilde{c}^{l}(0) & \cdots & \tilde{c}^{l}(N-1) \end{bmatrix}$ , and hence  \eqref{eq:PTWOLA_stacked_c_k_l_forall_k} becomes
\begin{align}
	\text{diag}\left( \dft{N}(l,:)\right) \dft{N} \left( \tilde{\tilde{\mathbf{c}}}^{l} \right)^{\trans} = \dft{N} \text{diag}\left( {\mathbf{Z}}_{0}(l,:) \right)\mathbf{w}^{\trans} \ .
\end{align} 
Finally, the time-domain representation of subband filters appearing in \eqref{eq:distortion_function_sample_and_vector_form} and \eqref{eq:LTI_filter_for_aliasing_transfer_function_DFT_FB_subband_expanded}, is then given as $\tilde{\mathbf{c}}^{l} = \left[ \begin{tabular}{c|c}
	$\tilde{\tilde{\mathbf{c}}}^{l}$     & $\tilde{\tilde{\mathbf{c}}}^{l}(0:N-2)$   
\end{tabular} \right]$ and
\begin{align}\label{eq:PTWOLA_relation_TD_filter_coeff_tilde_Z0}
	\left( \tilde{\tilde{\mathbf{c}}}^{l} \right)^{\trans} = \underbrace{\left( \text{diag}\left( \dft{N}(l,:)\right) \dft{N} \right)^{-1} \dft{N}}_{\boldsymbol{\mathcal{I}}_{N}^{l}} \text{diag}\left( {\mathbf{Z}}_{0}(l,:) \right)\mathbf{w}^{\trans}
\end{align}
where $\boldsymbol{\mathcal{I}}_{N}^{l}$ is the $l$-point circularly shifted identity matrix of size $N$, with the element in the $i^{\text{th}}$ row and $j^{\text{th}}$ column defined as
\begin{align}
\boldsymbol{\mathcal{I}}_{N}^{l}(i,j) = \delta_{((i+l) {\bmod} N), j}
\end{align}
and $\delta_{a,b}$ denotes the Kronecker delta, which equals 1 if $a = b$ and 0 otherwise.

\noindent The matrix ${\mathbf{Z}}_{0}$ depends only on the prototype analysis filter $\mathbf{h}_{0}$, presented in $\mathbf{H}_{w}$ and $\mathbf{H}_{c}$, as can be seen from \eqref{eq:PTWOLA_subband_filter_converged_assumption_vectorized_raw} and \eqref{eq:PTWOLA_defination_Z0_primary_expression}. Moreover, from \eqref{eq:PTWOLA_subband_filter_converged_assumption_vectorized_raw} it can be seen that  $\mathbf{H}_{c}$ is a submatrix of  $\mathbf{H}_{w}$, such that $\mathbf{H}_{c} =\mathbf{H}_{w}(:,0:T-1) $. Therefore,  $\mathbf{H}_{w}$ can be written as
\begin{align}\label{eq:PTWOLA_Hw_H0c_Hw0_definition}
	\mathbf{H}_{w} = \left[ \begin{array}{c|c}
		\mathbf{H}_{c}  & \mathbf{H}_{w}(:,T:N-1)
	\end{array} \right]
\end{align}
Substituting  \eqref{eq:PTWOLA_Hw_H0c_Hw0_definition} in  ${\mathbf{Z}}_{0}$ as defined in \eqref{eq:PTWOLA_defination_Z0_primary_expression} leads to
\begin{align}
		{\mathbf{Z}}_{0} = \left( \left(\mathbf{H}_{c}\right)^{\trans} \mathbf{H}_{c}  \right)^{-1} \left(\mathbf{H}_{c}\right)^{\trans}\left[ \begin{array}{c|c}
			\mathbf{H}_{c}  & \mathbf{H}_{w}(:,T:N-1)
		\end{array} \right]
\end{align}
Therefore, it can be observed that ${\mathbf{Z}}_{0}$ has the following structure
\begin{align}\label{eq:PTWOLA_appendix_Z0_matrix_structure_I_eta_0}
	{\mathbf{Z}}_{0} = 
	\left[ \begin{tabular}{c|c}
		$\mathbf{I}_{T}$     & ${\boldsymbol{\xi}}_{0}$     
	\end{tabular} \right]
\end{align}
where, ${\boldsymbol{\xi}}_{0} \triangleq \left( \left(\mathbf{H}_{c}\right)^{\trans} \mathbf{H}_{c}  \right)^{-1} \left(\mathbf{H}_{c}\right)^{\trans}\mathbf{H}_{w}(:,T:N-1)$ is a (non-sparse) matrix of size $\left( T \times N-T \right)$, only dependent on the  prototype analysis filter.

It is noted that the last $N-L$ components of $\mathbf{w}$ are zero, hence the last $N-L$ columns of $\mathbf{Z}_{0}$ do not play a role.

It can be observed from \eqref{eq:PTWOLA_appendix_Z0_matrix_structure_I_eta_0}  that as the subband filter order $T$ increases, the non-zero part of the $\mathbf{Z}_{0}$ matrix approaches an identity matrix. For $L=T$, $\tilde{\mathbf{c}}^{l} = \begin{bmatrix}
	w(l) & \mathbf{0}_{N-1,1}
\end{bmatrix}$. Furthermore, from \eqref{eq:PTWOLA_relation_TD_filter_coeff_tilde_Z0} and \eqref{eq:PTWOLA_appendix_Z0_matrix_structure_I_eta_0}, the maximum length of $\tilde{\mathbf{c}}^{l}$ is $L-T+1$, as will also be exemplified with simulations in Section \ref{WOLA_subband_filtering_Sec:Simulation_results_AEC} (Fig. \ref{fig:PTWOLA_L_512_and_55_C_l_rect_window}). 

%% file: WOLA_subband_domain_system_identification.tex
\section{Steady-State \Gls{MSE} Performance of Subband System Identification}\label{WOLA_subband_filtering_Sec:MSE_performance_analysis}
This section addresses the \gls{MMSE}-based subband system identification problem, deriving the overall steady-state \gls{MSE} performance of the generalized \gls{WOLA} filter bank. The analysis is framed within the context of \gls{ERLE}, which is commonly employed in  \gls{AEC} applications, but also serves as a more broadly applicable metric, and examines the impact of the subband filter order, the \gls{FIR} length of the unknown \gls{LTI} system, and the prototype analysis and synthesis filters on the system’s overall performance.

%% file: WOLA_subband_performance_analysis.tex
The normalized \gls{MSE} in \gls{AEC} is conventionally represented by the \gls{ERLE}, which provides a measure of the achieved echo suppression and is inversely related to the normalized \gls{MSE}. The \gls{ERLE} for the considered system is given as
\begin{align}\label{eq:PTWOLA_ERLE_primary_expression}
    \erl = \frac{ \E{\left|{y(n)}\right|^{2}}}{ \E{\left|{y(n) - \hat{y}(n)}\right|^{2}} } \propto \frac{1}{\text{MSE}} \ 
\end{align} 
where $\E{\cdot}$ represents the expected value operation.
Substituting \eqref{eq:general_LTI_input_output_equation_with_delay} and \eqref{eq:PTWOLA_estimated_TD_echo_signal} into \eqref{eq:PTWOLA_ERLE_primary_expression}, and for simplification, assuming the input signal is a zero mean white signal with variance $\sigma_{u}^{2}$, the \gls{ERLE} expression can be rewritten as a function of the impulse response of the unknown \gls{LTI} system, the distortion function and aliasing transfer functions, i.e.
\begin{align}\label{eq:PTWOLA_ERLE_expression_primary_gaussian_sig}
    \erl = \frac {\normsq{\mathbf{w}_{ext}} \sigma_{u}^{2} }{\left(  \normsq{\mathbf{\hat{t}}-\mathbf{w}_{ext}} + \sum_{m=1}^{D-1} \normsq{\mathbf{\hat{a}}_{m}}  \right)\sigma_{u}^{2} }
\end{align}
The  $\mathbf{w}_{ext}$ represents the impulse response of the unknown \gls{LTI} system, padded with zeros, i.e.
\begin{align}\label{eq:distortion_function_vector_form_equal_RIR_general}
	\mathbf{w}_{ext} = 
	\begin{bmatrix}
		\mathbf{0}_{1,N-1} & \mathbf{w} & \mathbf{0}_{1,T-1}
	\end{bmatrix}
\end{align}
with $\mathbf{w} \triangleq \begin{bmatrix}
	w(0) & w(1) & \cdots & w(L-1)& \mathbf{0}_{1,N-L}
\end{bmatrix}$  the row vector of length $N$ representing the impulse response of the unknown \gls{LTI} system. The zeros appended at the front account for the delay introduced by the subband filtering in the considered filter bank\footnote{The expression in \eqref{eq:distortion_function_vector_form_equal_RIR_general} does not account for the delay of $\lfloor \frac{N}{D}-1\rfloor D $ samples associated with the buffer for block processing, represented by $\tau$ in \eqref{eq:output_no_subband_processing}. For a \gls{WOLA} filter bank, this buffer-induced delay depends solely on the number of subbands $N$ and  the decimation (or interpolation) factor $D$, and is therefore not considered in this analysis.  The delay in \eqref{eq:distortion_function_vector_form_equal_RIR_general} pertains to the overall transfer function in  \eqref{eq:distortion_function_sample_and_vector_form}, which corresponds to the suppression of the first periodic image of $\tilde{c}^{l}(n-l)$ to avoid circular convolution effects. Alternatively, the definition of  $\mathbf{w}_{ext}$ can be adjusted to $\mathbf{w}_{ext} = 
	\begin{bmatrix}
		\mathbf{w} & \mathbf{0}_{1,2N+T-L-2}
\end{bmatrix}$, corresponding to the suppression of the second periodic image of  $\tilde{c}^{l}(n-l)$. A detailed explanation has been provided by the authors in \cite{10289725}.}, while the zeros at the end ensure the extended impulse response matches with the length of the distortion function.

Substituting \eqref{eq:PTWOLA_relation_TD_filter_coeff_tilde_Z0} into \eqref{eq:distortion_function_sample_and_vector_form} and \eqref{eq:LTI_filter_for_aliasing_transfer_function_DFT_FB_subband_expanded}, the expression for the overall distortion function and aliasing transfer functions can be expressed in terms of the unknown \gls{LTI} impulse response, as given in \eqref{eq:PTWOLA_tilde_t_tilde_am_Z0_form}, shown at the top of the next page.
\begin{figure*}[h]
\begin{subequations} \label{eq:PTWOLA_tilde_t_tilde_am_Z0_form}
	\begin{align}
		\left( \hat{\mathbf{t}} \right)^{\trans} &= \sum_{l=0}^{T-1}\left\{ \text{diag}\!\left( \begin{bmatrix} \mathbf{0}_{1\times l} \!\!&\! \mathbf{g}_{0} \!\!&\! \mathbf{0}_{1\times(T-l-1)} \end{bmatrix} \right)
		\left[ \begin{tabular}{c}
			$\mathbf{0}_{l\times1}$ \\ \hdashline
			$\boldsymbol{\mathcal{I}}_{N}^{l}(0:N-2,:) \text{diag}\left( \mathbf{Z}_{0}(l,:) \right) 
				\mathbf{w}^{\trans}$ \\ \hdashline $\boldsymbol{\mathcal{I}}_{N}^{l} \text{diag}\left( \mathbf{Z}_{0}(l,:) \right) 
				\mathbf{w}^{\trans}$ \\ \hdashline $\mathbf{0}_{(T-l-1)\times1}$
		\end{tabular} \right]
		\right\} \nonumber \\
		&= \underbrace{ \sum_{l=0}^{T-1}\left\{ 
			\text{diag}\left(
			\begin{bmatrix}
				\mathbf{0}_{1,l} & \mathbf{g}_0 & \mathbf{0}_{1,T-l-1}
			\end{bmatrix}
			\right)
			\left[
			\begin{array}{c}
				\mathbf{0}_{l \times (2N-1)} \\
				\cdashline{1-1}
				\mathbf{I}_{2N-1} \\
				\cdashline{1-1}
				\mathbf{0}_{(T-l-1) \times (2N-1)}
			\end{array}
			\right]
	\left[
\begin{array}{c}
			\boldsymbol{\mathcal{I}}_{N}^{l}(0:N-2, :) \\ 
				\cdashline{1-1} 
				\noalign{\vskip 2pt}
			\boldsymbol{\mathcal{I}}_{N}^{l}
	\end{array}
\right]
			\text{diag}(\mathbf{Z}_0(l,:))
			\right\} }_{\mathbf{G_{\xi}}}
		\mathbf{w}^{\trans} \label{eq:PTWOLA_tilde_t_tilde_am_Z0_form_tilde_t} \\
		\left( \mathbf{\hat{a}}_{m} \right)^{\trans} &= \sum_{l=0}^{T-1}\left\{ \alpha_{D}^{lm} \text{diag}\!\left( \begin{bmatrix} \mathbf{0}_{1\times l} \!\!&\! \boldsymbol{\psi}_{m} \!\!&\! \mathbf{0}_{1\times (T-l-1)} \end{bmatrix} \right)
			\left[ \begin{tabular}{c}
			$\mathbf{0}_{l\times1}$ \\ \hdashline
			$\boldsymbol{\mathcal{I}}_{N}^{l}(0:N-2,:) \text{diag}\left( \mathbf{Z}_{0}(l,:) \right) 
			\mathbf{w}^{\trans}$ \\ \hdashline $\boldsymbol{\mathcal{I}}_{N}^{l} \text{diag}\left( \mathbf{Z}_{0}(l,:) \right) 
			\mathbf{w}^{\trans}$ \\ \hdashline $\mathbf{0}_{(T-l-1)\times1}$
		\end{tabular} \right]
		\right\} \nonumber \\ 
		&=\underbrace{\sum_{l=0}^{T-1}\left\{ 
		\alpha_{D}^{lm} \text{diag}\!\left( \begin{bmatrix} \mathbf{0}_{1\times l} \!\!&\! \boldsymbol{\psi}_{m} \!\!&\! \mathbf{0}_{1\times(T-l-1)} \end{bmatrix} \right)
			\left[
			\begin{array}{c}
				\mathbf{0}_{l \times (2N-1)} \\
				\cdashline{1-1}
				\mathbf{I}_{2N-1} \\
				\cdashline{1-1}
				\mathbf{0}_{(T-l-1) \times (2N-1)}
			\end{array}
			\right]
			\left[
			\begin{array}{c}
				\boldsymbol{\mathcal{I}}_{N}^{l}(0:N-2, :) \\ 
				\cdashline{1-1} 
				\noalign{\vskip 2pt}
				\boldsymbol{\mathcal{I}}_{N}^{l}
			\end{array}
			\right]
			\text{diag}(\mathbf{Z}_0(l,:))
			\right\} }_{\boldsymbol{\Psi_{\xi_{m}}}}\mathbf{w}^{\trans} \label{eq:PTWOLA_tilde_t_tilde_am_Z0_form_tilde_am}
	\end{align}
\end{subequations}
\hrulefill
\vspace{-.7\baselineskip}
\end{figure*}

Finally, the analytical \gls{ERLE} can be expressed as a function of the prototype filters and the system impulse response. Taking the  \gls{ERLE} formulation from  \eqref{eq:PTWOLA_ERLE_expression_primary_gaussian_sig}, the matrix representations of  $\mathbf{\hat{t}} $ and $\mathbf{\hat{a}}_{m} $ from \eqref{eq:PTWOLA_tilde_t_tilde_am_Z0_form_tilde_t} and \eqref{eq:PTWOLA_tilde_t_tilde_am_Z0_form_tilde_am}, respectively, are substituted into the expression.
The distortion error term $ \normsq{\mathbf{\hat{t}}-\mathbf{w}_{ext}}$ becomes $\normsq{\mathbf{G^{\trans}_{\xi}} \mathbf{w}- \mathbf{w}_{ext}}$, and the aliasing error $\sum_{m=1}^{D-1} \normsq{\mathbf{\hat{a}}_{m}}$ becomes $\sum_{m=1}^{D-1} \normsq{\boldsymbol{\Psi^{\trans}_{\xi_{m}}} \mathbf{w}}$. Inserting these into \eqref{eq:PTWOLA_ERLE_expression_primary_gaussian_sig} yields the final expression for the analytical \gls{ERLE}: 
\begin{align}\label{eq:PTWOLA_ERLE_final_form_MSE_analysis}
    \erl = \frac{\normsq{\mathbf{w}_{ext}}\sigma_{u}^{2}}{\left(
    	 \normsq{\mathbf{G^{\trans}_{\xi}} \mathbf{w}- \mathbf{w}_{ext}}+\sum_{m=1}^{D-1} \normsq{\boldsymbol{\Psi^{\trans}_{\xi_{m}}} \mathbf{w}}\right)\sigma_{u}^{2} } \ .
\end{align}

%% file: WOLA_subband_PTEQ_WOLA.tex
\section{\Gls{PTWOLA}: Computationally Efficient Generalized \gls{WOLA}}\label{WOLA_subband_filtering_Sec:Per-Tone_WOLA_comp_eff_multi_tap_WOLA}

\noindent The preceding sections have established the theoretical advantages and superior modeling potential of the generalized \gls{WOLA} filter bank framework. However, despite its promise, the framework’s high computational complexity limits its practical implementation, particularly as the subband filter order increases. This section addresses the critical trade-off between performance and computational efficiency by first introducing the \gls{PTWOLA}, a novel architecture that preserves the performance of the generalized \gls{WOLA} while significantly reducing its computational complexity. A detailed analysis of computational complexity is then presented in Section \ref{WOLA_subband_filtering_SubSec:Per-Tone_WOLA_complexity_analysis}.

The large computational complexity of the generalized \gls{WOLA} filter bank compared to the conventional \gls{WOLA} filter bank,  stems from two reasons. First, although the subband error for the subband adaptive filtering is computed at the downsampled rate \eqref{eq:PTWOLA_optimization_MMSE_filter_coefficients}, the subband filtering is performed at the full-rate, before the downsampling operation. In conventional \gls{WOLA}, the subband filters are zeroth-order, and hence can be moved  after the downsampling operation, allowing the subband filtering operation at the downsampled rate.
This is not directly feasible with the higher-order subband filters in the generalized \gls{WOLA} filter bank.
Second, additional computational complexity stems from the computation of the input signals for the subband filters, as will be explained next.

In Fig. \ref{fig:generalized_WOLA_subband_system_identification_block_diagram}, the output of the $k^{th}$ subband filter $C_{k}(z)$, after the downsampling operation, can be formulated as
\begin{align}\label{eq:PTWOLA_downsampled_kth_subband_filter_output_equation}
	\check{Y}_{k}(q) &= \left[u(n) \star \left(h_{0}(n) \alpha_{N}^{nk}\right) \star c_{k}(n) \right]_{\downarrow D} \nonumber \\ 
	&= \sum_{l=0}^{T-1}c_{k}(l) \sum_{m = 0}^{N-1} h_{0}(m) \alpha_{N}^{mk} u(qD-m-l)
\end{align}
where $\left[\cdot\right]_{\downarrow D}$ represents the $D$-fold downsampling operation and $q$ represents the time index at the downsampled rate (frame index). This can be expressed in matrix form as \eqref{eq:STFT_out_kth_band_matrix_form}, shown at the top of the next page, where $\mathbf{\check{u}}_{k}\!\left[q\right]$ represents the input signal vector for the $k^{th}$ subband filter.
\begin{figure*}[hbt!]
    \begin{align} \label{eq:STFT_out_kth_band_matrix_form}
        \check{Y}_{k}(q) = \underbrace{\begin{bmatrix}
        \alpha_{N}^{0}\\
        \alpha_{N}^{k}\\
        \vdots \\
        \alpha_{N}^{(N-1)k} 
    \end{bmatrix}^{\trans} 
    \begin{bmatrix}
        h_{0}(0) &0&\cdots &0\\
        0 & h_{0}(1) & \cdots&0\\
        \vdots& & \ddots & \vdots\\
        0& \cdots & 0& h_{0}(N-1)
      \end{bmatrix}
      \underbrace{\begin{bmatrix}
        u(qD) & \cdots & u(qD-T+1) \\
        u(qD -1)& \cdots & u(qD-T)\\
        \vdots& \cdots & \vdots \\
        u(qD-N+1)& \cdots&  u(qD-N-T+2) 
      \end{bmatrix} }_{\mathbf{U}\!\left[q\right]}}_{\mathbf{\check{u}}_{k}\!\left[q\right] \; \triangleq \; \begin{bmatrix}\check{U}_{k}(qD), \cdots, \check{U}_{k}(qD-T+1)\end{bmatrix}}
      \underbrace{\begin{bmatrix}
        c_{k}(0)\\
        c_{k}(1)\\
        \vdots \\
        c_{k}(T-1) 
    \end{bmatrix}}_{\mathbf{c}^{\trans}_{k}}
    \end{align}
\hrulefill
	\vspace{-.7\baselineskip}
\end{figure*}

\noindent Given that the vector $\begin{bmatrix}
	\alpha_{N}^{0},
	\cdots, 
	\alpha_{N}^{(N-1)k} 
\end{bmatrix}$ represents the $k^\text{th}$ row of the \gls{IDFT} matrix, \eqref{eq:STFT_out_kth_band_matrix_form} can be reformulated as
\begin{align} \label{eq:T_IDFTs_PTEQ_WOLA_compact}
	\check{Y}_{k}(q) &= \mathbf{\check{u}}_{k}\!\left[q\right] \mathbf{c}_{k}^{\trans} = \underbrace{ \idft{N}(k,:)  \text{diag}\left( \mathbf{h}_{0} \right)}_{\triangleq \dft{N}^{win^{*}}(k,:)}  \mathbf{U}\!\left[q\right]  \mathbf{c}_{k}^{\trans} \nonumber \\
	&= \text{row}_{k} \underbrace{\left(\dft{N}^{win^{*}}  \mathbf{U}\!\left[q\right] \right)}_{T\text{-IDFT operations}}  \mathbf{c}_{k}^{\trans}
\end{align}
where, $\dft{N}^{win^{*}}$ represents the windowed \gls{IDFT} matrix $\dft{N}^{win^{*}} \triangleq \idft{N}  \text{diag}\left( \mathbf{h}_{0} \right)$. Therefore, from  \eqref{eq:T_IDFTs_PTEQ_WOLA_compact}  it can be observed that as opposed to performing a single \gls{IDFT} operation for each time frame (i.e. for each $q$) in conventional \gls{WOLA}, $T-$\gls{IDFT} operations seem to be necessary for each frame, i.e., an \gls{IDFT} operation for every column in the matrix $\mathbf{U}\!\left[q\right]$.

However, it is worth noting that the $T$-\gls{IDFT} operations can be computed efficiently for any general cosine window used as a prototype analysis filter, including the rectangular window as a special case. This section primarily focuses on the extensively used (root-)raised cosine windows as the prototype analysis filter. Nevertheless, the analysis and findings discussed herein can be extended to other general cosine windows.

A prototype analysis filter, represented by a root-raised cosine window of length $N$, can be written as
\begin{align}\label{eq:PTWOLA_generalised_root_raised_cosine_window_expression}
	h_{0}(n) = \left(\rho - \eta  \cos{\frac{2 \pi n}{N}}\right)^{0.5}, \forall n \in \left[0, N-1\right]
\end{align}
where, $\rho$ and $\eta$ are scalar parameters in the range $ \left[0,1\right]$, with $\eta \leq \rho$ to ensure the expression remains real-valued for all $n$.

Using the binomial expansion \((a + x)^\beta = a^\beta + \beta a^{\beta - 1} x + \frac{\beta(\beta - 1)}{2} a^{\beta - 2} x^2  + \cdots\), for $|x|<|a|$, followed by trigonometric power-reduction identities (e.g. \(\cos^2(x) = \frac{1}{2}(1 + \cos(2x))\), etc.), \eqref{eq:PTWOLA_generalised_root_raised_cosine_window_expression} can be expanded into the Fourier cosine series, i.e.,
\begin{align}\label{eq:PTWOLA_root-Hann_window_taylor_series_expansion}
	h_{0}(n) =& \resizebox{0.9\hsize}{!}{%
		$\underbrace{\left(\rho^{2} - \frac{\eta^2}{16}\right)}_{\gamma(0)} - \underbrace{\left(\rho \eta + \frac{3}{64} \rho \eta^{2}\right)}_{-\gamma(1)} \cos{\frac{2 \pi n}{N}} - \underbrace{\frac{\eta^{2}}{16}}_{-\gamma(2)} \cos{\frac{4 \pi n}{N}} 
		- \underbrace{\frac{1}{64}\rho \eta^{2}}_{-\gamma(3)}\cos{\frac{6 \pi n}{N}} + \cdots $ }\nonumber \\
	&= \sum_{r=0}^{\infty} \gamma(r) \cos{\frac{2 \pi r n}{N}} \quad \text{with}\ \left|\gamma(r)\right|>\left|\gamma(r+1)\right| 
\end{align}
Therefore, the \gls{IDFT} applied to the windowed $t^{th}$ column of $\mathbf{U}\!\left[q\right]$ in \eqref{eq:STFT_out_kth_band_matrix_form}, evaluated for subband $k$, can be expressed as
\begin{align}\label{eq:PTWOLA_generalised_cosine_window_final_equation}
	\check{U}_{k}&(qD-t) = \sum_{m = 0}^{N-1} \sum_{r=0}^{\infty} \gamma(r) \cos{\frac{2 \pi r m}{N}} u(qD-t-m) \alpha_{N}^{mk} \nonumber \\
	&=  \sum_{r=0}^{\infty} \frac{\gamma(r)}{2}  \left[ \begin{multlined} \overbrace{\sum_{m = 0}^{N-1} u(qD-t-m) \alpha_{N}^{\left(k-r\right)m}}^{\idft{N}\left((k-r) \% N, : \right) \cdot \mathbf{U}\!\left[q\right](:,t)} \\ + \underbrace{\sum_{m = 0}^{N-1} u(qD-t-m) \alpha_{N}^{\left(k+r\right)m}}_{\idft{N}\left((k+r) \% N, : \right) \cdot \mathbf{U}\!\left[q\right](:,t)} \end{multlined} \right] \nonumber \\ 
	& \resizebox{0.95\hsize}{!}{%
		$\approxeq \sum_{r=0}^{R} \frac{\gamma(r)}{2} \left[\idft{N}\left((k - r) \% N, : \right) + \idft{N}\left((k + r) \% N, : \right)\right] \mathbf{U}\!\left[q\right]\!(:,t) %
	$}
\end{align}
where, $\%$ represents the modulo operation and $\left[\idft{N}\left((k \pm r) \% N, : \right) \cdot\mathbf{U}\!\left[q\right]\!(:,t) \right]$ corresponds to the \gls{IDFT} of the non-windowed (or rectangular windowed with $h_{0}(n) = 1, \, \forall n \in \left[0, N-1\right]$) $t^{th}$ column of $\mathbf{U}\!\left[q\right]$ in \eqref{eq:STFT_out_kth_band_matrix_form}, evaluated for subbands $\left(k \pm r\right)  \% N$, and where  the infinite sum has been truncated to $R$, leading to $2R$ so-called ``cross-terms".
The number of cross-terms required for the computation of the windowed \gls{IDFT}, based on the non-windowed \gls{IDFT}, is dependent on the prototype analysis filter $h_{0}(n)$. For instance, utilizing a cosine window as the prototype analysis filter necessitates two cross-terms ($R=1$), while employing a root-Hann window as the prototype analysis filter achieves an acceptable approximation with $R\approxeq 4$, corresponding to 8 cross-terms.

\begingroup%
\makeatletter%
\providecommand\color[2][]{%
  \errmessage{(Inkscape) Color is used for the text in Inkscape, but the package 'color.sty' is not loaded}%
  \renewcommand\color[2][]{}%
}%
\providecommand\transparent[1]{%
  \errmessage{(Inkscape) Transparency is used (non-zero) for the text in Inkscape, but the package 'transparent.sty' is not loaded}%
  \renewcommand\transparent[1]{}%
}%
\providecommand\rotatebox[2]{#2}%
\newcommand*\fsize{\dimexpr\f@size pt\relax}%
\newcommand*\lineheight[1]{\fontsize{\fsize}{#1\fsize}\selectfont}%
\ifx\svgwidth\undefined%
  \setlength{\unitlength}{485.50236259bp}%
  \ifx\svgscale\undefined%
    \relax%
  \else%
    \setlength{\unitlength}{\unitlength * \real{\svgscale}}%
  \fi%
\else%
  \setlength{\unitlength}{\svgwidth}%
\fi%
\global\let\svgwidth\undefined%
\global\let\svgscale\undefined%
\makeatother%
\begin{figure*}[hbt!]
    \centering
\begin{picture}(1,0.29141992)%
  \lineheight{1}%
  \setlength\tabcolsep{0pt}%
  \put(0,0){\includegraphics[width=\unitlength,page=1]{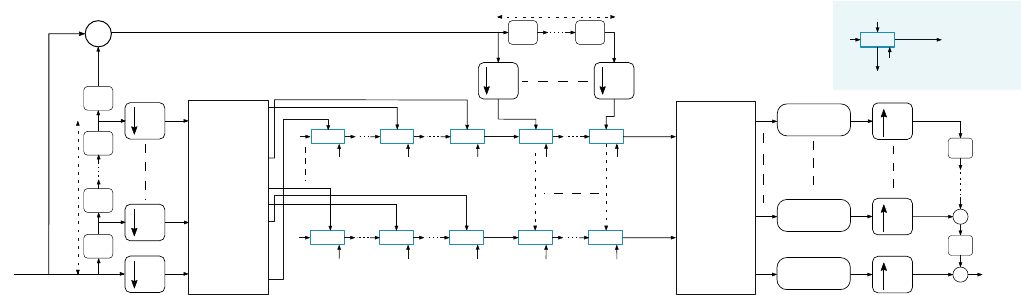}}%
  \put(0.09,0.0407552){\makebox(0,0)[lt]{\lineheight{1.25}\smash{\begin{tabular}[t]{l}$\Delta$\end{tabular}}}}%
  \put(0.503,0.25160437){\makebox(0,0)[lt]{\lineheight{1.25}\smash{\begin{tabular}[t]{l}$\Delta$\end{tabular}}}}%
  \put(0.570,0.25160437){\makebox(0,0)[lt]{\lineheight{1.25}\smash{\begin{tabular}[t]{l}$\Delta$\end{tabular}}}}%
  \put(0.934,0.138){\makebox(0,0)[lt]{\lineheight{1.25}\smash{\begin{tabular}[t]{l}$\Delta$\end{tabular}}}}%
  \put(0.934,0.0415){\makebox(0,0)[lt]{\lineheight{1.25}\smash{\begin{tabular}[t]{l}$\Delta$\end{tabular}}}}%
  \put(0.21,0.08934525){\makebox(0,0)[lt]{\lineheight{1.25}\smash{\begin{tabular}[t]{l}$\idft{N}$\end{tabular}}}}%
  \put(0.691,0.0884625){\makebox(0,0)[lt]{\lineheight{1.25}\smash{\begin{tabular}[t]{l}$\dft{N}$\end{tabular}}}}%
  \put(0.055,0.08840258){\makebox(0,0)[lt]{\lineheight{1.25}\smash{\begin{tabular}[t]{l}$N$\end{tabular}}}}%
  \put(0.598,0.20363857){\makebox(0,0)[lt]{\lineheight{1.25}\smash{\begin{tabular}[t]{l}$D$\end{tabular}}}}%
  \put(0.14,0.16404995){\makebox(0,0)[lt]{\lineheight{1.25}\smash{\begin{tabular}[t]{l}$D$\end{tabular}}}}%
  \put(0.87,0.01206677){\makebox(0,0)[lt]{\lineheight{1.25}\smash{\begin{tabular}[t]{l}$D$\end{tabular}}}}%
  \put(0.87,0.06942963){\makebox(0,0)[lt]{\lineheight{1.25}\smash{\begin{tabular}[t]{l}$D$\end{tabular}}}}%
  \put(0.87,0.16372801){\makebox(0,0)[lt]{\lineheight{1.25}\smash{\begin{tabular}[t]{l}$D$\end{tabular}}}}%
  \put(0.818,0.24829994){\makebox(0,0)[lt]{\lineheight{1.25}\smash{\begin{tabular}[t]{l}$a$\end{tabular}}}}%
  \put(0.854,0.27314545){\makebox(0,0)[lt]{\lineheight{1.25}\smash{\begin{tabular}[t]{l}$b$\end{tabular}}}}%
  \put(0.852,0.202){\makebox(0,0)[lt]{\lineheight{1.25}\smash{\begin{tabular}[t]{l}$b$\end{tabular}}}}%
  \put(0.867,0.22102829){\makebox(0,0)[lt]{\lineheight{1.25}\smash{\begin{tabular}[t]{l}$c$\end{tabular}}}}%
  \put(0.925,0.24443253){\makebox(0,0)[lt]{\lineheight{1.25}\smash{\begin{tabular}[t]{l}$a+b\cdot c$\end{tabular}}}}%
  \put(0.52,0.27639445){\makebox(0,0)[lt]{\lineheight{1.25}\smash{\begin{tabular}[t]{l}$T-1$\end{tabular}}}}%
  \put(0.93271793,0.0134){\makebox(0,0)[lt]{\lineheight{1.25}\smash{\begin{tabular}[t]{l}$+$\end{tabular}}}}%
  \put(0.089,0.251){\makebox(0,0)[lt]{\lineheight{1.25}\smash{\begin{tabular}[t]{l}$+$\end{tabular}}}}%
  \put(0.93273993,0.0707){\makebox(0,0)[lt]{\lineheight{1.25}\smash{\begin{tabular}[t]{l}$+$\end{tabular}}}}%
  \put(0.14,0.01238875){\makebox(0,0)[lt]{\lineheight{1.25}\smash{\begin{tabular}[t]{l}$D$\end{tabular}}}}%
  \put(0,0.028){\makebox(0,0)[lt]{\lineheight{1.25}\smash{\begin{tabular}[t]{l}$u(n)$\end{tabular}}}}%
  \put(0.483,0.20363857){\makebox(0,0)[lt]{\lineheight{1.25}\smash{\begin{tabular}[t]{l}$D$\end{tabular}}}}%
  \put(0.14,0.06357242){\makebox(0,0)[lt]{\lineheight{1.25}\smash{\begin{tabular}[t]{l}$D$\end{tabular}}}}%
  \put(0.28205904,0.14668712){\makebox(0,0)[lt]{\lineheight{1.25}\smash{\begin{tabular}[t]{l}{\footnotesize $0$}\end{tabular}}}}%
  \put(0.328,0.1212122){\makebox(0,0)[lt]{\lineheight{1.25}\smash{\begin{tabular}[t]{l}{\footnotesize$v_{0}(0)$}\end{tabular}}}}%
  \put(0.397,0.1212122){\makebox(0,0)[lt]{\lineheight{1.25}\smash{\begin{tabular}[t]{l}{\footnotesize $v_{0}(R)$}\end{tabular}}}}%
  \put(0.458,0.1212122){\makebox(0,0)[lt]{\lineheight{1.25}\smash{\begin{tabular}[t]{l}{\footnotesize $v_{0}(2R)$}\end{tabular}}}}%
  \put(0.528,0.1212122){\makebox(0,0)[lt]{\lineheight{1.25}\smash{\begin{tabular}[t]{l}{\footnotesize $v_{0}\!(\!2R\!\!+\!\!1\!)$}\end{tabular}}}}%
  \put(0.598,0.1212122){\makebox(0,0)[lt]{\lineheight{1.25}\smash{\begin{tabular}[t]{l}{\footnotesize $v_{0}$}\!{\footnotesize $($}{\tiny 2R+T-1\!}{\footnotesize $)$}\end{tabular}}}}%
  \put(0.63515428,0.137){\makebox(0,0)[lt]{\lineheight{1.25}\smash{\begin{tabular}[t]{l}{\footnotesize $\check{Y}_{0}$}\end{tabular}}}}%
  \put(0.63515428,0.072){\makebox(0,0)[lt]{\lineheight{1.25}\smash{\begin{tabular}[t]{l}{\footnotesize $\check{Y}_{\!\frac{N}{2}}$}\end{tabular}}}}%
  \put(0.28205904,0.05610968){\makebox(0,0)[lt]{\lineheight{1.25}\smash{\begin{tabular}[t]{l}{\footnotesize $0$}\end{tabular}}}}%
  \put(0.96202173,0.0273765){\makebox(0,0)[lt]{\lineheight{1.25}\smash{\begin{tabular}[t]{l}$\hat{y}(n)$\end{tabular}}}}%
  \put(0.775,0.16715725){\makebox(0,0)[lt]{\lineheight{1.25}\smash{\begin{tabular}[t]{l}$f_{0}(0)$\end{tabular}}}}%
  \put(0.763,0.072){\makebox(0,0)[lt]{\lineheight{1.25}\smash{\begin{tabular}[t]{l}$f_{0}(\!N\!\!-\!\!2 \!)$\end{tabular}}}}%
  \put(0.763,0.015){\makebox(0,0)[lt]{\lineheight{1.25}\smash{\begin{tabular}[t]{l}$f_{0}(\!N\!\!-\!\!1\!)$\end{tabular}}}}%
  \put(0.5945,0.022){\makebox(0,0)[lt]{\lineheight{1.25}\smash{\begin{tabular}[t]{l}{\footnotesize $v_{\!\frac{N}{2}}$}\!{\footnotesize $($}{\tiny 2R+T-1\!}{\footnotesize $)$}\end{tabular}}}}%
  \put(0.52,0.022){\makebox(0,0)[lt]{\lineheight{1.25}\smash{\begin{tabular}[t]{l}{\footnotesize $v_{\!\frac{N}{2}}\!(\!2R\!\!+\!\!1\!)$}\end{tabular}}}}%
  \put(0.45,0.022){\makebox(0,0)[lt]{\lineheight{1.25}\smash{\begin{tabular}[t]{l}{\footnotesize $v_{\!\frac{N}{2}}\!(2R)$}\end{tabular}}}}%
  \put(0.38,0.022){\makebox(0,0)[lt]{\lineheight{1.25}\smash{\begin{tabular}[t]{l}{\footnotesize $v_{\!\frac{N}{2}}\!(R)$}\end{tabular}}}}%
  \put(0.312,0.022){\makebox(0,0)[lt]{\lineheight{1.25}\smash{\begin{tabular}[t]{l}{\footnotesize $v_{\!\frac{N}{2}}\!(0)$}\end{tabular}}}}%
  \put(0.42,0.2){\makebox(0,0)[lt]{\lineheight{1.25}\smash{\begin{tabular}[t]{l}{\footnotesize ${U}^{rec}_{R}$}\end{tabular}}}}%
  \put(0.50,0.175){\makebox(0,0)[lt]{\lineheight{1.25}\smash{\begin{tabular}[t]{l}{\footnotesize $\Delta u(0)$}\end{tabular}}}}%
  \put(0.605,0.175){\makebox(0,0)[lt]{\lineheight{1.25}\smash{\begin{tabular}[t]{l}{\footnotesize $\Delta u($}{\tiny \!-T+2\!}{\footnotesize $)$}\end{tabular}}}}%
  \put(0.3928812,0.17){\makebox(0,0)[lt]{\lineheight{1.25}\smash{\begin{tabular}[t]{l}{\footnotesize ${U}^{rec}_{0}$}\end{tabular}}}}%
  \put(0.32733337,0.164){\makebox(0,0)[lt]{\lineheight{1.25}\smash{\begin{tabular}[t]{l}{\footnotesize ${U}^{rec}_{-R}$}\end{tabular}}}}%
  \put(0.32843791,0.0715){\makebox(0,0)[lt]{\lineheight{1.25}\smash{\begin{tabular}[t]{l}{\footnotesize ${U}^{rec}_{\!\frac{N}{2}\!-\!R}$}\end{tabular}}}}%
  \put(0.394,0.075){\makebox(0,0)[lt]{\lineheight{1.25}\smash{\begin{tabular}[t]{l}{\footnotesize ${U}^{rec}_{\!\frac{N}{2}}$}\end{tabular}}}}%
  \put(0.46220465,0.075){\makebox(0,0)[lt]{\lineheight{1.25}\smash{\begin{tabular}[t]{l}{\footnotesize ${U}^{rec}_{\!\frac{N}{2}\!+\!R}$}\end{tabular}}}}%
  \put(0.09,0.18703466){\makebox(0,0)[lt]{\lineheight{1.25}\smash{\begin{tabular}[t]{l}$\Delta$\end{tabular}}}}%
  \put(0.09,0.14226409){\makebox(0,0)[lt]{\lineheight{1.25}\smash{\begin{tabular}[t]{l}$\Delta$\end{tabular}}}}%
  \put(0.09,0.08556838){\makebox(0,0)[lt]{\lineheight{1.25}\smash{\begin{tabular}[t]{l}$\Delta$\end{tabular}}}}%
  \put(0.107,0.235){\makebox(0,0)[lt]{\lineheight{1.25}\smash{\begin{tabular}[t]{l}-\end{tabular}}}}%
\end{picture}%
\caption{\Acrfull{PTWOLA} filter bank.}
\label{fig:WOLA_subband_processing_blck_diag_PTWOLA}
\end{figure*}
\endgroup%

Comparing  the expression for $\check{U}_{k}(qD-t)$ in \eqref{eq:STFT_out_kth_band_matrix_form} and \eqref{eq:T_IDFTs_PTEQ_WOLA_compact} with \eqref{eq:PTWOLA_generalised_cosine_window_final_equation}, it can be observed that
\begin{equation}\label{eq:Relation_windowed_IDFT_gamma_IDFT}
	\dft{N}^{win^{*}}(k,:) = \pmb{\gamma}_R \, \idft{N}\left((k-R:k+R) \% N, : \right)
\end{equation}
where,  $\pmb{\gamma}_R$ is a symmetric vector containing the $\gamma(r)$ coefficients given as
\begin{align}
	\pmb{\gamma}_R \triangleq \begin{bmatrix}
		\gamma(R), \! & \gamma(R-1), \! & \cdots, \! & \gamma(0), \! & \gamma(1), \! & \cdots, \! & \gamma(R)
	\end{bmatrix}
\end{align}
Consequently, the windowed \gls{IDFT} of the $t^{th}$ column of $\mathbf{U}\!\left[q\right]$, evaluated for subband $k$, can be expressed as a weighted linear combination of the non-windowed \gls{IDFT} of the $t^{th}$ column of $\mathbf{U}\!\left[q\right]$, evaluated for subbands $k \pm r  \ (r \, = \, 0 \, , \, \cdots \, , \, R)$.

Substituting \eqref{eq:Relation_windowed_IDFT_gamma_IDFT} in \eqref{eq:T_IDFTs_PTEQ_WOLA_compact} then leads to
\begin{equation}\label{eq:T_IDFTs_PTEQ_WOLA_compact_with_gamma}
	\check{Y}_{k}(q) = \pmb{\gamma}_R \, \idft{N}\left((k-R:k+R) \% N, : \right) \, \mathbf{U}\!\left[q\right] \, \mathbf{c}_{k}^{\trans}
\end{equation}

For further simplification, \eqref{eq:T_IDFTs_PTEQ_WOLA_compact_with_gamma} can be reformulated as \eqref{eq:STFT_out_kth_band_matrix_form_gamma_vector_u}, where matrix $\boldsymbol{\mathcal{F}}_{k}$ reflects the (computationally complex) $T$-non-windowed \gls{IDFT} operations for the $k^{th}$ subband.
\begin{figure*}
	\begin{align}\label{eq:STFT_out_kth_band_matrix_form_gamma_vector_u}
		\check{Y}_{k}(q) \! = \! \mathbf{c}_{k} \!
		\underbrace{\left[\begin{array}{p{1.25em}p{1.25em}cp{1.25em}}
			\cline{1-1}
			\multicolumn{1}{|c|}{\pmb{\gamma}_R} 	& $\cdots$        	&           	& $\mathbf{0}$                      \\ \cline{1-2}
			$\mathbf{0}$    					& \multicolumn{1}{|c|}{\pmb{\gamma}_R}	&  & $\mathbf{0}$               \\ \cline{2-2}
			$\vdots$                           	&                	& \ddots      	&                                   \\ \cline{4-4} 
			$\mathbf{0}$               			& $\cdots$      	&  				& \multicolumn{1}{|c|}{\pmb{\gamma}_R} \\ \cline{4-4} 
		\end{array}\right]}_{\textstyle
		\begin{array}{c}\overline{\pmb{\gamma}^{T}_{R}}\end{array}}
			\underbrace{\left[\begin{array}{ccccccc}
				\cline{1-3}
				\multicolumn{3}{|c|}{\idft{N}\left((k-R:k+R) \% N, : \right)} &  &  & \cdots & \mathbf{0} \\ \cline{1-4}
				\multicolumn{1}{c|}{0} & \multicolumn{3}{c|}{\idft{N}\left((k-R:k+R) \% N, : \right)} &  & \cdots & \mathbf{0} \\ \cline{2-4}
				\vdots &  &  & \ddots &  &  & \vdots \\ \cline{5-7} 
				0 &  & \cdots & \multicolumn{1}{c|}{} & \multicolumn{3}{c|}{\idft{N}\left((k-R:k+R) \% N, : \right)} \\ \cline{5-7} 
			\end{array} \right]}_{\textstyle
			\begin{array}{c}\boldsymbol{\mathcal{F}}_{k}\end{array}} \!\!
			\underbrace{\begin{bmatrix}
					u(qD)\\
					u(qD -1)\\
					\vdots \\
					u(qD\!\!-\!\!N\!\!-\!\!T\!\!+\!\!2)
			\end{bmatrix}}_{\textstyle
			\begin{array}{c}\mathbf{u}\left[q\right]\end{array}}
	\end{align}
	\vspace{-.7\baselineskip}
\end{figure*}

Fortunately, the structure in the matrix $\boldsymbol{\mathcal{F}}_{k}$ can be exploited to reduce the required number of non-windowed \gls{IDFT} operations to just one for each frame, by decomposing it as a product of a lower block triangular matrix and a sparse matrix, as shown in \eqref{eq:PTWOLA_windowed_2R_1_cross_T_IDFTs_difference_terms_matrix}, 
\begin{figure*}[h]
	\begin{align}\label{eq:PTWOLA_windowed_2R_1_cross_T_IDFTs_difference_terms_matrix}
		\resizebox{1\hsize}{!}{%
			$\boldsymbol{\mathcal{F}}_{k} =  \underbrace{\left[\begin{array}{c:cccc}
					\left(\alpha_{N}^{k} \mathds{I}_{R}\right)^{-0} & \mathbf{0}_{(2R+1)\times 1} & \mathbf{0}_{(2R+1)\times 1} & \cdots & \mathbf{0}_{(2R+1)\times 1} \\
					\left(\alpha_{N}^{k} \mathds{I}_{R}\right)^{-1} & -\text{diag}\! \left(\left(\alpha_{N}^{k} \mathds{I}_{R}\right)^{-1}\right) & \mathbf{0}_{(2R+1)\times 1} & \cdots & \mathbf{0}_{(2R+1)\times 1} \\
					\vdots &  & \ddots &  & \vdots \\
					\left(\alpha_{N}^{k} \mathds{I}_{R}\right)^{-(T-1)} & -\text{diag} \! \left(\left(\alpha_{N}^{k} \mathds{I}_{R}\right)^{-(T-1)}\right) & -\text{diag}\! \left(\left(\alpha_{N}^{k} \mathds{I}_{R}\right)^{-(T-2)}\right) & \cdots & -\text{diag}\! \left(\left(\alpha_{N}^{k} \mathds{I}_{R}\right)^{-1}\right) 
				\end{array} \right]}_{\textstyle
				\begin{array}{c}\left(\overline{\mathds{L}_{k}}\right)_{T(2R+1) \times (2R+T)}\end{array}}
			\underbrace{\left[\begin{array}{cc|c}
				\multicolumn{2}{c | }{\smash{\raisebox{0.1\normalbaselineskip}{$\left( \idft{N}\left( (k-R:k+R) {\bmod} N,: \right) \right)$}}} &  \mathbf{0}_{2R+1\times (T-1)}  \\
				\hline 
				\multicolumn{1}{c | }{\mathbf{I}_{T-1}} & \mathbf{0}_{(T-1) \times N-T} & -\mathbf{I}_{T-1}
			\end{array}\right]}_{\textstyle
			\begin{array}{c}{ \overline{\mathbb{F}_{k}}}\end{array}} $}
	\end{align}
	\vspace{-.7\baselineskip}
\end{figure*}
where $\mathds{I}_{R}$ is a diagonal matrix given as
\begin{align}
	\mathds{I}_{R} =  \begin{bmatrix}
		\alpha_{N}^{-R} & 0 & \cdots & 0 \\
		0 & \alpha_{N}^{-R+1} & \cdots & 0 \\
		\vdots &  & \ddots & 0 \\
		0 & \cdots &  & \alpha_{N}^{R} 
	\end{bmatrix} 
\end{align}

With \eqref{eq:PTWOLA_windowed_2R_1_cross_T_IDFTs_difference_terms_matrix}, expression  \eqref{eq:STFT_out_kth_band_matrix_form_gamma_vector_u} can be written as
\begin{equation}
	\check{Y}_{k}(q) = \mathbf{c}_{k} \:  \overline{\pmb{\gamma}^{T}_{R}} \:  \overline{\mathds{L}_{k}} \: \overline{\mathbb{F}_{k}} \: \mathbf{u}\!\left[q\right]
\end{equation}
where both the $\overline{\pmb{\gamma}^{T}_{R}}$ and $ \overline{\mathds{L}_{k}}$  are independent of the input signal $u(n)$.

Finally, to further reduce the computational complexity, the matrix $\left(\overline{\pmb{\gamma}^{T}_{R}}  \:    \overline{\mathds{L}_{k}}\right)$  can be integrated into the subband adaptive filter coefficients $\mathbf{c}_{k}$, i.e.  $\left(\mathbf{c}_{k}  \:   \overline{\pmb{\gamma}^{T}_{R}}  \:    \overline{\mathds{L}_{k}}\right)$ is replaced  by $\mathbf{v}_{k}$, where $\mathbf{v}_{k}$ then represents the modified subband adaptive filter coefficients for subband $k$. This integration allows each subband $k$ in \gls{PTWOLA} to have a filter input with one direct term and $2R$ cross-terms (represented by the first $2R+1$ rows of the matrix $\overline{\mathbb{F}_{k}}$), which are calculated from the output of a single \gls{IDFT} operation, plus $T-1$ so-called difference terms \cite{Per_tone_equalizer_katleen} (represented by the last $T-1$ rows of the matrix $\overline{\mathbb{F}_{k}}$). The simplified expression can be formulated as \eqref{eq:PTWOLA_cross_diff_terms_final_form_modified_PTEQ_coeffs}
\begin{figure*}[h]
    \begin{align}\label{eq:PTWOLA_cross_diff_terms_final_form_modified_PTEQ_coeffs}
     \check{Y}_{k}(q) &= \mathbf{v}_{k} \left[\begin{array}{cc|c}
        \multicolumn{2}{c | }{\smash{\raisebox{0.1\normalbaselineskip}{$\left( \idft{N}\left( (k-R:k+R) {\%} N,: \right) \right)$}}} &  \mathbf{0}_{2R+1\times (T-1)}  \\
       \hline 
       \multicolumn{1}{c | }{\mathbf{I}_{T-1}} & \mathbf{0}_{(T-1) \times N-T} & -\mathbf{I}_{T-1}
   \end{array}\right] \begin{bmatrix}
    u(qD) \\ u(qD-1) \\ \vdots \\ u(qD-N-T+2)
   \end{bmatrix}
   &=\mathbf{v}_{k} \begin{bmatrix}
   			{U^{rec}}_{(k-R){\%} N}(qD) \\
   			\vdots \\
   			{U^{rec}}_{(k+R){\%} N}(qD) \\
   			\Delta u(q)\\
   			\vdots\\
   			\Delta u(q-T+2) 
   		\end{bmatrix}
\end{align}
\hrulefill
\end{figure*}
where, $\mathbf{v}_{k} \!\triangleq \!\begin{bmatrix} v_{k}(0) & \!\!\! \cdots & \!\!\!v_{k}(2R+T-1) \end{bmatrix}$ are the modified subband adaptive filter coefficients for \gls{PTWOLA}. 

In conclusion, the  $T$ windowed \gls{IDFT} operations for each $q$ in  \eqref{eq:STFT_out_kth_band_matrix_form}, can be replaced by a single non-windowed \gls{IDFT} operation, accompanied by $T-1$ difference terms along with $2R$ cross-terms, as illustrated in \eqref{eq:PTWOLA_cross_diff_terms_final_form_modified_PTEQ_coeffs}, shown at the top of the next page, where $U^{rec}_{k}(qD)$ represents the non-windowed \gls{IDFT} as defined in \eqref{eq:PTWOLA_generalised_cosine_window_final_equation} i.e., $U^{rec}_{k}(qD) \triangleq \idft{N}\left((k) \% N, : \right)  \mathbf{U}\!\left[q\right]\!(:,1)$  and $\Delta u(q-j)  \triangleq u(qD-j)-u(qD-j-N)$ is a so-called difference term \cite{Per_tone_equalizer_katleen}.

A comparison between \eqref{eq:PTWOLA_cross_diff_terms_final_form_modified_PTEQ_coeffs} for \gls{PTWOLA} and \eqref{eq:STFT_out_kth_band_matrix_form} for the  generalized \gls{WOLA} filter bank reveals that the \gls{PTWOLA} implementation necessitates an additional $2R$ subband adaptive filter coefficients; i.e. each subband adaptive filter has $T+2R$ taps instead of $T$ taps. Nonetheless, in the simulations presented in Section~\ref{subsection:WOLA_based_subband_PTWOLA_Performance_Results}, the number of difference terms (or cross-terms) in \gls{PTWOLA} is intentionally reduced to ensure the same number of subband adaptive filter coefficients across both implementations. This adjustment enables a fair comparison. Under these conditions, it will be subsequently demonstrated that \gls{PTWOLA} performance remains comparable to that of the generalized \gls{WOLA} filter bank for identical total subband adaptive filter lengths.

The structure of \gls{PTWOLA}, with a single non-windowed \gls{IDFT} operation and $T-1$ difference terms along with $2R$ cross-terms, is depicted in Fig. \ref{fig:WOLA_subband_processing_blck_diag_PTWOLA}.

\subsection{Computational Complexity}\label{WOLA_subband_filtering_SubSec:Per-Tone_WOLA_complexity_analysis}
This subsection presents an analysis of the computational complexity of the generalized \gls{WOLA} filter bank and \gls{PTWOLA}. The computational complexity is evaluated in terms of the number of real multiplications and real additions required to process a single frame. The following standard conventions are adopted in the analysis.  A complex-by-complex multiplication operation is counted as $4$ real multiplications and $2$ real additions, and a real-by-complex multiplication operation is counted as $2$ real multiplications. Similarly, both complex-plus-complex and real-plus-complex addition operations are counted as $2$ real additions, since the latter is typically implemented using a complex adder by treating the real input as having zero imaginary part. Finally, an $N$-point  (I)\gls{DFT} operation, based on the standard radix-2 algorithm, is assumed to require $2N \log_{2}(N)$ real multiplications and $3N \log_{2}(N)$ real additions.

\subsubsection{Generalized \gls{WOLA} filter bank}
The first stage of the generalized \gls{WOLA} filter bank based subband processing involves the computation of the term $\left(\dft{N}^{win^{*}}  \mathbf{U}\!\left[q\right] \right)$ as shown in  \eqref{eq:T_IDFTs_PTEQ_WOLA_compact}. This involves a windowing operation with $N$ real multiplications followed by an $N$-point \gls{IDFT} applied to each of the  $T$ columns of the input matrix $\mathbf{U}\!\left[q\right]$. Therefore,  this transform stage requires $TN(2 \log_{2}(N)+1)$ real multiplications and $3TN \log_{2}(N)$ real additions.

The second stage performs subband filtering through a complex-by-complex vector dot product between two $T$-element  vectors, evaluated for $N/2$ unique subbands, followed by $T-1$ complex additions per subband. Hence the subband filtering stage requires $2NT$ real multiplications and $N(2T-1)$ real additions. 

Overall, the computational complexity is dominated by the $T$ separate $N$-point \gls{IDFT}  operations per frame. Therefore, the asymptotic complexity of the generalized \gls{WOLA} filter bank is  $\mathcal{O}(TN\log_{2}(N))$.\\

\subsubsection{\gls{PTWOLA}}
To allow a fair comparison with the generalized \gls{WOLA} filter bank, the subsequent analysis assumes that the total number of filter taps in \gls{PTWOLA} is identical to that of generalized \gls{WOLA}, with $2R+1$  cross-terms and  $T-(2R+1)$  difference terms. 

The first stage of  \gls{PTWOLA} involves computing the input samples for the subband filtering stage.  This consists of a single $N$-point \gls{IDFT} applied to the first column of the input data matrix $U[q]$, along with the computation of $T-(2R+1)$ real difference terms. Therefore, this stage requires  $2N \log_{2}(N)$ real multiplications and $3N \log_{2}(N) + (T-2R-1)$ real additions.

The second stage comprises subband filtering. It involves a complex-by-complex vector dot product between two $2R-1$-element vectors for the cross-terms, and a real-by-complex vector dot product between two $T-(2R+1)$-element vectors for the difference terms, both computed over $N/2$ unique subbands. The resulting $T$ products are then summed, requiring $T-1$ complex additions. Hence, the subband filtering stage requires $N(2R+T+1)$ real multiplications and $N(2R+T)$ real additions.

In this case, the computational complexity is primarily dictated by the single $N$-point \gls{IDFT} operation and the subband filtering stage. Therefore, the asymptotic complexity of  \gls{PTWOLA} is $\mathcal{O}(N \log_{2}(N) + N(T+R))$. In typical scenarios where $N\gg T$,  this simplifies to $\mathcal{O}(N\log_{2}(N))$,  which is equivalent to the computational complexity of conventional \gls{WOLA}.

The  computational complexities for the generalized \gls{WOLA} filter bank and the \gls{PTWOLA} are summarized in the Table \ref{tab:genwola_ptwola_complexity}.
\begin{table}[ht]
	\caption{Number of floating-point  operations required by generalized \gls{WOLA} and \gls{PTWOLA} based subband filtering.}
	\label{tab:genwola_ptwola_complexity}
	\resizebox{\linewidth}{!}{%
		\begin{tabular}{|l|c|c|}
			\hline
			\textbf{Metric} & \textbf{Generalized \gls{WOLA}} & \textbf{\gls{PTWOLA}} \\
			\hline
			Real Mult. & $TN(2 \log_{2}(N)+3)$ & $2N\log_{2}(N) + N(2R+T+1)$ \\
			\hline
			Real Add. & $TN(3 \log_{2}(N)+2)-N$ & \parbox[c]{3.5cm}{$3N \log_{2}(N)+(T-2R-1)  +N(2R+T)$} \\
			\hline
			Asymptotic & $\mathcal{O}(TN\log_{2}(N))$ & $\mathcal{O}(N\log_{2}(N))$ (approx.) \\
			\hline
		\end{tabular}%
	}
\end{table}

%% file: WOLA_subband_results.tex
\section{Simulation Results}\label{WOLA_subband_filtering_Sec:Simulation_results_AEC}
\subsection{Acoustic Echo Cancellation (\gls{AEC})}
For the evaluation of the subband system identification, single channel \gls{AEC} is considered as a reference use-case scenario with the system distance and the 
\gls{ERLE} as  performance metrics. 
In this context, the \gls{RIR}, represented by $w(n)$, describes the acoustic path, including all echoes and reverberation, from the loudspeaker to the microphone 
The system distance provides a measure of resemblance between the estimated distortion function $\mathbf{\hat{t}}$ and the actual \gls{RIR} $\mathbf{w}_{ext}$, and is  defined in decibels as $10 \log_{10}\left(\frac{\normsq{\mathbf{w}_{ext}-\mathbf{\hat{t}}}}{\normsq{\mathbf{w}_{ext}}}\right)$. The \gls{ERLE} provides a measure of the achieved echo suppression. 
For such \gls{AEC} application in \eqref{eq:general_LTI_input_output_equation}, $y(n)$ represents the microphone signal, $u(n)$ represents the far-end loudspeaker signal and $w(n)$ represents the \gls{RIR} between the loudspeaker and the microphone. Moreover, $\mathbf{c}_{k}$ (or $\mathbf{v}_{k}$) represents the (converged) coefficients of the estimated low-order \gls{RIR} in subband $k$.

\subsection{Simulation Scenario}
A \gls{LEM} system is considered with a single microphone and a single loudspeaker. A non-stationary stochastic process with exponential decay,  modeled after Polack’s framework \cite{polac1988}, is employed to generate the  \gls{RIR} of length $L=512$ between the loudspeaker and microphone, i.e.
\begin{align}
	w(n) = \sqcup_{L}(n)\Omega(n)\exp^{-\kappa n}
\end{align}
where, $\sqcup_{L}$ is a rectangular function such that $ \sqcup_{L}(n) = 1 \text{ for } n = 0 \cdots L-1$, $\Omega(n)$ represents a zero-mean white Gaussian noise with unit variance and  $\kappa$ is the decay parameter of the exponential function, related to the reverberation time $T_{60}$ by $\kappa = 3 \ln(10)/T_{60}$. 
The far-end and near-end signals are both zero-mean white Gaussian noise signals, with an average \gls{EBR} of $20\, \text{dB}$. A \gls{WOLA} filter bank is used with \gls{DFT} size of $N=1024$ and with $50\%$  overlap (i.e., $D = 512$). Adaptation of subband coefficients $\mathbf{c}_{k}$ (or $\mathbf{v}_{k}$) is performed using the \gls{RLS} algorithm \cite{haykin2014adaptive}. For the simulations, the \gls{RLS} algorithm was operated with a forgetting factor of $\lambda=0.999$, and initialized with a covariance matrix $P(0) = 100 \cdot \mathbf{I}_{T}$ and all-zero filter coefficients. 
The simulation is performed at a sampling frequency of $16\, \text{kHz}$.

\scalebox{0}{%
	\begin{tikzpicture}
		\begin{axis}[hide axis]
			\addplot [
				color={rgb,1:red,0.34667;green,0.53600;blue,0.69067}, dashed, mark=o,	
				forget plot
				]
				(0,0);\label{fig_caption:Rectangular_prototype_synthesis_filter_norm}
			\addplot [
				color={rgb,1:red,0.34667;green,0.53600;blue,0.69067}, mark=o,	
				forget plot
				]
				(0,0);\label{fig_caption:Rectangular_prototype_synthesis_filter_distortion}
			\addplot [
				color={rgb,1:red,0.91529;green,0.28157;blue,0.28784},dashed, mark=+,	
				forget plot
				]
				(0,0);\label{fig_caption:Cosine_prototype_synthesis_filter_norm}
			\addplot [
				color={rgb,1:red,0.91529;green,0.28157;blue,0.28784}, mark=+,	
				forget plot
				]
				(0,0);\label{fig_caption:Cosine_prototype_synthesis_filter_distortion}
			\addplot [
				color={rgb,1:red,0.44157;green,0.74902;blue,0.43216},dashed, mark=asterisk,		
				forget plot
				]
				(0,0);\label{fig_caption:Root-Hann_prototype_synthesis_filter_norm}
				color={rgb,1:red,0.44157;green,0.74902;blue,0.43216}, mark=asterisk,		
				forget plot
				]
				(0,0);\label{fig_caption:Root-Hann_prototype_synthesis_filter_distortion}
			\addplot [
				color=black, dashed,	
				forget plot
				]
				(0,0);\label{fig_caption:General_prototype_synthesis_filter_norm}
			\addplot [
				color=black,	
				forget plot
				]
				(0,0);\label{fig_caption:General_prototype_synthesis_filter_distortion}
			\addplot [
				color=black, dotted,	
						forget plot
				]
				(0,0);\label{fig_caption:General_prototype_synthesis_filter_distortion_reversed}
		\end{axis}
	\end{tikzpicture}%
}

\subsection{Prototype Filters and Time-Domain Subband Filters}
\begin{figure*}
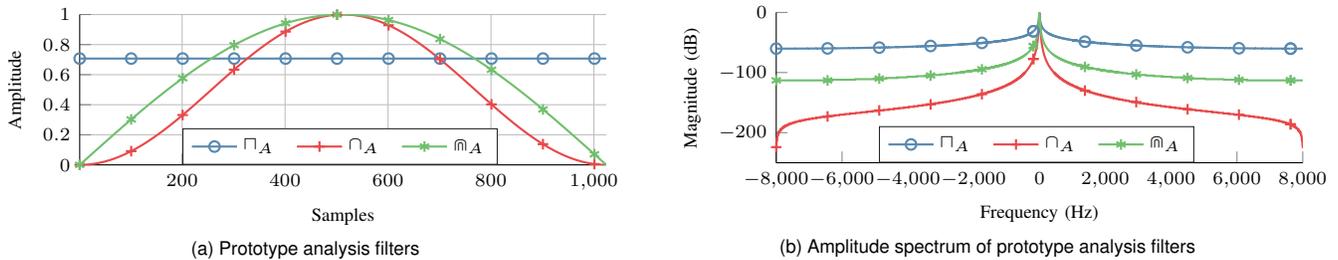

	\captionsetup[subfigure]{width=3.4in}
	\captionsetup[subfigure]{captionskip=-4pt}
	\captionsetup[subfigure]{font={scriptsize}}
	\begin{minipage}{0.49\textwidth}
	\subfloat[{\scriptsize Prototype analysis filters}]{\label{fig:prototype_analysis filters_time}}{%
	\setlength\fheight{2cm}
	\setlength\fwidth{7cm}
	\input{figures/PTWOLA_Results_ana_windows_time.tex}
	}
\end{minipage}\hfil
\begin{minipage}{0.49\textwidth}
\subfloat[{\scriptsize Amplitude spectrum of prototype analysis filters}]{\label{fig:prototype_analysis filters_freq}}{%
\setlength\fheight{2cm}
\setlength\fwidth{7cm}
\input{figures/PTWOLA_Results_ana_windows_freq.tex}
}
\end{minipage}
	\caption{Prototype analysis filters and their corresponding amplitude spectrum (envelope).}
	\label{fig:prototype_analysis_filters}
\end{figure*}

\begin{figure*}
	\captionsetup[subfigure]{width=3.4in}
	\captionsetup[subfigure]{captionskip=-4pt}
	\captionsetup[subfigure]{font={scriptsize}}
	\begin{minipage}{0.49\textwidth}
	\subfloat[{\scriptsize Minimum norm prototype synthesis filters}]{\label{fig:min_norm_prototype_synthesis filters_time}}{%
	\setlength\fheight{2cm}
	\setlength\fwidth{7cm}
	\input{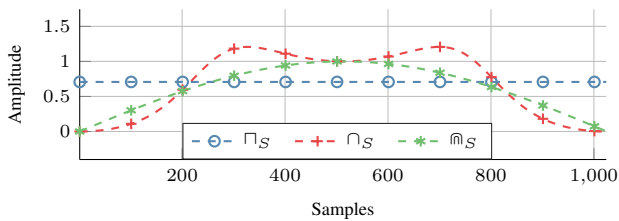}
	}
\end{minipage}\hfil
\begin{minipage}{0.49\textwidth}
\subfloat[{\scriptsize Minimum distortion prototype synthesis filters}]{\label{fig:min_distortion_prototype_synthesis filters_time}}{%
\setlength\fheight{2cm}
\setlength\fwidth{7cm}
\input{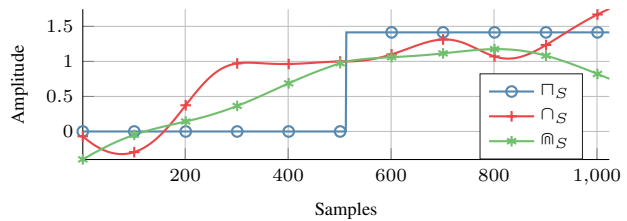}
}
\end{minipage}
	\caption{Minimum norm (\ref{fig_caption:General_prototype_synthesis_filter_norm}) and minimum-distortion (\ref{fig_caption:General_prototype_synthesis_filter_distortion}) prototype synthesis filters of prototype analysis filters shown in Fig. \ref{fig:prototype_analysis_filters}.}
	\label{fig:prototype_synthesis_filters}
\end{figure*}

\begin{figure}
	\setlength\fheight{2.5cm}
\setlength\fwidth{7cm}
	\input{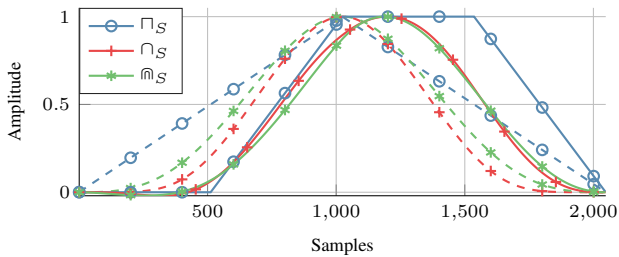}
	\caption{Scaled convolution of the considered prototype analysis filters with their respective minimum-norm (\ref{fig_caption:General_prototype_synthesis_filter_norm}) and minimum-distortion (\ref{fig_caption:General_prototype_synthesis_filter_distortion}) prototype synthesis filters.}
	\label{fig:scaled_convolution_prototype_filters}
\end{figure}

For the performance evaluation, three windows were considered as prototype analysis filters; namely the rectangular window (\rectwin{A}), the cosine window (\cosinewin{A}) and the root-Hann window (\roothannwin{A}), as illustrated in Fig. \ref{fig:prototype_analysis_filters}. This selection ensures  a broad evaluation across filters with distinct spectral characteristics. The rectangular and cosine window serve as representative extremes in terms of frequency selectivity, while the widely-used root-Hann window is included both for its relevance in practical applications  and its effectiveness in highlighting the trade-offs inherent in the proposed \gls{PTWOLA} implementation (number of implemented cross-terms versus performance).
The amplitude spectrum of the considered prototype analysis filters is depicted in Fig. \ref{fig:prototype_analysis filters_freq}. It can be observed that the cosine window exhibits the highest degree of frequency selectivity, while the rectangular window filter exhibits the lowest degree of frequency selectivity. 
For each prototype analysis filter, a corresponding perfect reconstruction prototype synthesis filter was designed, as shown in Fig \ref{fig:prototype_synthesis_filters}. The design was based on the (conventional) minimum-norm solution and on the recently proposed minimum-distortion solution presented in \cite{10289725}. 
It can be observed that the rectangular window as prototype analysis filter and its corresponding minimum-distortion prototype synthesis filter, as depicted in Fig. \ref{fig:prototype_analysis filters_time} and Fig. \ref{fig:min_distortion_prototype_synthesis filters_time}, respectively, exemplify the prototype filters used in the \gls{OLS} method.  
Fig.  \ref{fig:scaled_convolution_prototype_filters} shows the (scaled) convolution $g_{0}(n)$ for the considered prototype analysis and synthesis filters. 

\begin{figure*}
	\captionsetup[subfigure]{width=3.4in}
	\captionsetup[subfigure]{captionskip=-3pt}
	\captionsetup[subfigure]{font={scriptsize}}
	\begin{minipage}{0.49\textwidth}
	\subfloat[{\scriptsize $\sum_{l}\left|\tilde{{c}}^{l}(n)\right|$ for $T=15$ and $L=512$}]{\label{fig:PTWOLA_L_512_C_l_rect_window}}{%
	\setlength\fheight{2cm}
	\setlength\fwidth{7cm}
	\input{figures/PTWOLA_L_512_C_l_rect_window_sum_mag2db.tex}
	}
\end{minipage}\hfil
\begin{minipage}{0.49\textwidth}
\subfloat[{\scriptsize $\sum_{l} \left| \tilde{{c}}^{l}(n)\right|$ for $T=15$ and $L=55$}]{\label{fig:PTWOLA_L_55_C_l_rect_window}}{%
\setlength\fheight{2cm}
\setlength\fwidth{7cm}
\input{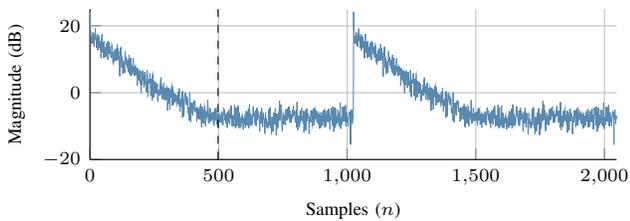}
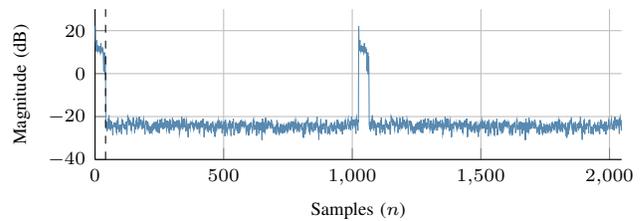
}
\end{minipage}
	\caption{$\sum_{l}\left|\tilde{{c}}^{l}(n)\right|$ values in dB with rectangular prototype analysis filter and $T=15$ for \gls{RIR} length of $L=512$ and $L=55$. The vertical line (\ref{fig_caption:General_prototype_synthesis_filter_norm}) marks the sample point at $n=L$.}
	\label{fig:PTWOLA_L_512_and_55_C_l_rect_window}
\end{figure*}

Finally, Fig. \ref{fig:PTWOLA_L_512_and_55_C_l_rect_window} presents the values of $\sum_{l} \left| \tilde{{c}}^{l}(n)\right|$ for $n = 0 \cdots 2N-2$, derived from the converged subband filter coefficients as defined in \eqref{eq:periodic_nature_subband_low_orlder} . For brevity, results are only shown for the rectangular prototype analysis filter with a subband filter length of $T=15$. However, without loss of generality, the same conclusions can be drawn for different prototype analysis filters and subband filter lengths. Moreover, to highlight the effect of the \gls{RIR} length $L$, the values are presented for  $L=512$ and $L=15$, as shown in Fig. \ref{fig:PTWOLA_L_512_C_l_rect_window} and \ref{fig:PTWOLA_L_55_C_l_rect_window},  respectively. As can be observed from Fig. \ref{fig:PTWOLA_L_512_and_55_C_l_rect_window}, a significant portion of the power of $\tilde{{c}}^{l}(n)$ is contained approximately between samples $n=0 \cdots L-T$  (and between $n = N-1 \cdots  N+L-T-1$ due to periodicity as derived in \eqref{eq:periodic_nature_subband_low_orlder}). Therefore, the maximum length of $\tilde{\mathbf{c}}^{l}$ is $L-T+1$, which substantiates the theoretical derivations presented in \eqref{eq:PTWOLA_relation_TD_filter_coeff_tilde_Z0} and \eqref{eq:PTWOLA_appendix_Z0_matrix_structure_I_eta_0}.

\subsection{Performance Evaluation: Generalized \gls{WOLA}}
\begin{figure*}
	\captionsetup[subfigure]{width=3.4in}
	\captionsetup[subfigure]{captionskip=-4pt}
	\captionsetup[subfigure]{font={scriptsize}}
	\begin{minipage}{0.49\textwidth}
	\subfloat[{\scriptsize System distance curves}]{\label{fig:System_distance_SWDFT_L_512}}{%
	\setlength\fheight{3cm}
	\setlength\fwidth{7cm}
	\input{figures/PTWOLA_Results_Sys_Distance_L512_SWDFT.tex}
	}
\end{minipage}\hfil
\begin{minipage}{0.49\textwidth}
\subfloat[{\scriptsize \gls{ERLE} curves}]{\label{fig:ERLE_SWDFT_L_512}}{%
\setlength\fheight{3cm}
\setlength\fwidth{7cm}
\input{figures/PTWOLA_Results_L512_SWDFT.tex}
}
\end{minipage}
\caption{System distance and \gls{ERLE} achieved by subband processing with different subband filter lengths and considered prototype analysis filters with their respective minimum-norm (\ref{fig_caption:General_prototype_synthesis_filter_norm}) and minimum-distortion (\ref{fig_caption:General_prototype_synthesis_filter_distortion}) prototype synthesis filters.}
\label{fig:System_dist_and_ERLE_SWDFT_L_512}
\end{figure*}
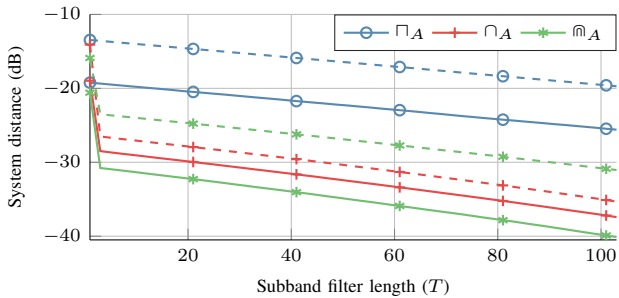
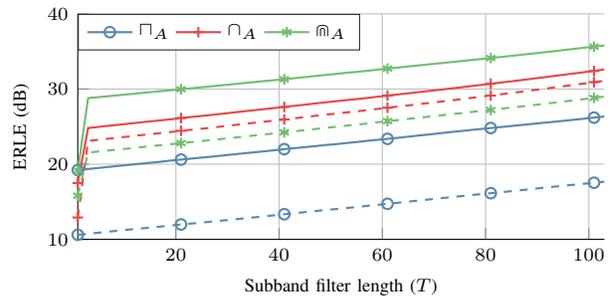

This subsection presents the experimental results that corroborate the theoretical conclusions drawn in Section \ref{WOLA_subband_filtering_Sec:MSE_performance_analysis} for the considered \gls{AEC} scenario. The system distance and \gls{ERLE} curves are respectively shown in  Fig. \ref{fig:System_distance_SWDFT_L_512} and \ref{fig:ERLE_SWDFT_L_512}, for different prototype filters and subband filter lengths ($T$). It is worth noting that the subband filter length of $T=1$, with different prototype analysis filters and their corresponding minimum-norm prototype synthesis filters, represents the conventional \gls{WOLA} method.

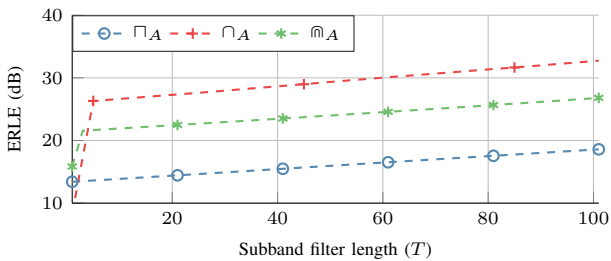
\begin{figure}
	\setlength\fheight{2.5cm}
\setlength\fwidth{7cm}
	\input{figures/PTWOLA_L_512_ERLE_MSE_analysis_norm_windows.tex}
	\caption{\gls{ERLE} calculated using \eqref{eq:PTWOLA_ERLE_final_form_MSE_analysis} for the prototype analysis filters shown in Fig. \ref{fig:prototype_analysis_filters} with their respective minimum-norm prototype synthesis filters.}
	\label{fig:ERLE_plot_MSE_analysis}
\end{figure}

It is observed that the system distance and the \gls{ERLE} performance for all the prototype filters improve as the length of the subband filters  ($T$) increases. This observation aligns with the findings derived in Section \ref{WOLA_subband_filtering_Sec:MSE_performance_analysis}, as presented in Fig. \ref{fig:ERLE_plot_MSE_analysis}. The figure illustrates the \gls{ERLE}, calculated using  \eqref{eq:PTWOLA_ERLE_final_form_MSE_analysis}, 
for the prototype analysis filters shown in Fig. \ref{fig:prototype_analysis_filters}, along with their respective minimum-norm (\ref{fig_caption:General_prototype_synthesis_filter_norm}) prototype synthesis filters.

Notably, the system distances achieved by the minimum-distortion prototype synthesis filters are smaller than those achieved by the minimum-norm synthesis filters for all prototype analysis filters and for all subband filter lengths, as shown in Fig. \ref{fig:System_distance_SWDFT_L_512}. 
In the open-loop subband system identification problem highlighted in \eqref{eq:WOLA_based_subband_processing_open_loop_LS_equation}, the converged filter coefficients are independent of the synthesis filter bank. Therefore, they remain unaltered under the same set of system conditions, such as prototype analysis filter, input signals, \gls{RIR}, filter length, etc., regardless of the prototype synthesis filter used. Hence, the system distance in Fig. \ref{fig:System_distance_SWDFT_L_512}, illustrates that, for a given prototype analysis filter and corresponding converged subband filter coefficients, the minimum-distortion prototype synthesis filters transform the subband filter coefficients into a distortion function that models the actual \gls{RIR} more accurately than the minimum-norm prototype synthesis filters. This improvement in system distance is partly due to the ability of minimum-distortion prototype synthesis filters to better suppress the circular convolution effect in the distortion function $\mathbf{\tilde{t}}$, which arises from the two periodic images in  $\tilde{\mathbf{c}}^{l}$. 

As shown in  Fig. \ref{fig:scaled_convolution_prototype_filters}, the amplitude of the $g_{0}(n)$ samples corresponding to the (first) periodic image in $\tilde{\mathbf{c}}^{l}$ is significantly lower for minimum-distortion prototype synthesis filters than for minimum-norm prototype synthesis filters. Therefore, from \eqref{eq:distortion_function_sample_and_vector_form}, this leads a to better suppression of the circular convolution effect in the distortion function $\mathbf{\tilde{t}}$, consequently resulting in a lower system distance. 
This improvement in achieved system distance also improves the overall \gls{AEC} performance of the system, as evident in Fig. \ref{fig:ERLE_SWDFT_L_512}, where the \gls{ERLE} achieved by the minimum-distortion prototype synthesis filters consistently outperforms the \gls{ERLE} achieved by the minimum-norm prototype synthesis filters for all prototype analysis filters and subband filter lengths. 

Finally, the rectangular prototype analysis filter with the corresponding minimum-distortion prototype synthesis filter, despite representing the prototype filters for the \gls{OLS} method, which facilitates perfect suppression of circular convolution effect, has its overall performance restricted by the poor frequency selectivity of the rectangular prototype analysis filter (Fig. \ref{fig:prototype_analysis filters_freq}), leading to a poor convergence of subband filters and motivates the use of prototype filters with better frequency selectivity.

\subsection{Performance Evaluation: \gls{PTWOLA}} \label{subsection:WOLA_based_subband_PTWOLA_Performance_Results}
While generalized \gls{WOLA} has demonstrated notable performance improvements, its high computational complexity, dominated by  $T$  windowed \gls{IDFT} operations per time frame, limits its practical application. To address this issue, Section~\ref{WOLA_subband_filtering_Sec:Per-Tone_WOLA_comp_eff_multi_tap_WOLA} proposes a computationally efficient \gls{PTWOLA} method, which reduces the  $T$  windowed \gls{IDFT} operations per frame to a single non-windowed \gls{IDFT} operation,  $T-1$  difference terms, and  $2R(+1)$  cross-terms ($+1$ direct term) per frame. The number of cross-terms depends on the prototype analysis filter (see \eqref{eq:PTWOLA_generalised_cosine_window_final_equation}).
This section compares the performance of \gls{PTWOLA} with generalized \gls{WOLA}, with identical total subband adaptive filter lengths, for both minimum-norm and minimum-distortion prototype synthesis filters.
\scalebox{0}{%
	\begin{tikzpicture}
		\begin{axis}[hide axis]
			\addplot [
				color={rgb,1:red,1;green,0.54902;blue,0.10196}, dashed, mark=square, mark options={solid, color={rgb,1:red,1;green,0.54902;blue,0.10196}},	
				forget plot
				]
				(0,0);\label{fig_caption:Rect_win_PTWOLA_R0}
			\addplot [
				color={rgb,1:red,0.49020;green,0.18039;blue,0.56078}, dashed, mark=diamond,	mark options={solid, color={rgb,1:red,0.49020;green,0.18039;blue,0.56078}},
				forget plot
				]
				(0,0);\label{fig_caption:Cos_win_PTWOLA_R1}
			\addplot [
				color={rgb,1:red,0.69020;green,0.40000;blue,0.23922},dashed, mark=star,	mark options={solid, color={rgb,1:red,0.69020;green,0.40000;blue,0.23922}},
				forget plot
				]
				(0,0);\label{fig_caption:Root_hann_win_PTWOLA_R2}
			\addplot [
				color={rgb,1:red,1;green,0.07451;blue,0.65098},dashed, mark=x,	mark options={solid, color={rgb,1:red,1;green,0.07451;blue,0.65098}},
				forget plot
				]
				(0,0);\label{fig_caption:Root_hann_win_PTWOLA_R4}	
			\addplot [
				color=black,dotted, line width=0.8pt,
				forget plot
				]
				(0,0);\label{fig_caption:PTWOLA_const_nf=51}
		\end{axis}
	\end{tikzpicture}%
}

\begin{figure*}
	\captionsetup[subfigure]{width=3.4in}
	\captionsetup[subfigure]{captionskip=-0pt}
	\captionsetup[subfigure]{font={scriptsize}}
	\begin{minipage}{0.49\textwidth}
	\subfloat[{\scriptsize \gls{ERLE} achieved with minimum-norm prototype synthesis filters.}]{\label{fig:ERLE_comp_L512_SWDFTvsconstR_norm}}{%
	\setlength\fheight{3cm}
	\setlength\fwidth{7cm}
	\input{figures/PTWOLA_Results_L512_SWDFTvsconstR_norm.tex}
	}
\end{minipage}\hfil
\begin{minipage}{0.49\textwidth}
\subfloat[{\scriptsize \gls{ERLE} achieved with minimum-distortion prototype synthesis filters.}]{\label{fig:ERLE_comp_L512_SWDFTvsconstR_distortion}}{%
\setlength\fheight{3cm}
\setlength\fwidth{7cm}
\input{figures/PTWOLA_Results_L512_SWDFTvsconstR_des.tex}
}
\end{minipage}
	\caption{\gls{ERLE} achieved by \gls{PTWOLA} (\ref{fig_caption:General_prototype_synthesis_filter_norm}) and  generalized \gls{WOLA} (\ref{fig_caption:General_prototype_synthesis_filter_distortion}) for $L=512$ with different total subband filter lengths and constant number of cross-terms ($R$) for prototype filters shown in Fig. \ref{fig:prototype_analysis_filters} and Fig. \ref{fig:prototype_synthesis_filters}. The cross-terms for different prototype filters are: \rectwin{A} $R=0$ (\ref{fig_caption:Rect_win_PTWOLA_R0}), \cosinewin{A} $R=1$ (\ref{fig_caption:Cos_win_PTWOLA_R1}), \roothannwin{A} $R=2$ (\ref{fig_caption:Root_hann_win_PTWOLA_R2}), \roothannwin{A} $R=4$ (\ref{fig_caption:Root_hann_win_PTWOLA_R4}).}
	\label{fig:ERLE_comp_L512_SWDFTvsconstR}
\end{figure*}
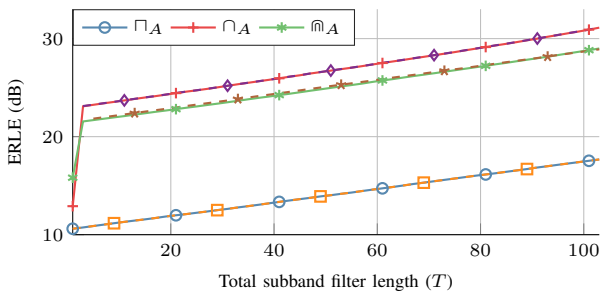
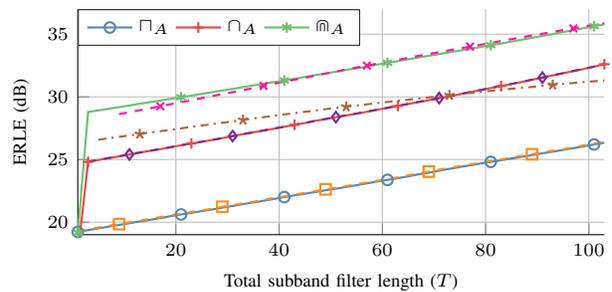

Fig. \ref{fig:ERLE_comp_L512_SWDFTvsconstR} presents a performance comparison between the \gls{PTWOLA} with $\mathbf{v}^{\trans}_{k}$ and the generalized \gls{WOLA} with $\mathbf{c}^{\trans}_{k}$, considering the same total filter lengths and a consistent number of cross-terms $2R$, necessary to represent each prototype analysis filter. Consequently, given a total filter length of $T$, the  generalized \gls{WOLA} necessitates $T$ windowed \gls{IDFT} inputs, whereas the \gls{PTWOLA} implementation requires $2R$$(+1)$ cross($+$direct)-terms and $\acute{T} = T-(2R+1)$  difference terms, as highlighted in  \eqref{eq:PTWOLA_cross_diff_terms_final_form_modified_PTEQ_coeffs}. 

As demonstrated in Fig. \ref{fig:ERLE_comp_L512_SWDFTvsconstR}, with a suitable number of cross-terms ($2R$), as defined in Section  \ref{WOLA_subband_filtering_Sec:Per-Tone_WOLA_comp_eff_multi_tap_WOLA}, the performance exhibited by the  \gls{PTWOLA} is comparable to the performance of the generalized \gls{WOLA} for identical total filter lengths ($T$). Moreover, \gls{PTWOLA} delivers this comparable performance by utilizing a smaller or equal number of difference terms, (which represents the sliding \glspl{IDFT} in  generalized \gls{WOLA}) with $\acute{T} = T-(2R+1)  \leq T$. This is because, unlike the  generalized \gls{WOLA}, where any prototype analysis window implies having constant symmetric subband cross-term filter coefficients ($\gamma(-r) = \gamma(r)$ in \eqref{eq:PTWOLA_generalised_cosine_window_final_equation}), \gls{PTWOLA} does not adhere to this symmetry constraint. This grants \gls{PTWOLA} extra degrees of freedom for identical total filter lengths. This has been meticulously observed in the converged subband filter coefficients for the cross-terms in \gls{PTWOLA}, which indeed in  general are non-symmetric.

Finally, while the required number of single-sided cross-terms ($R$) to represent the windowing operation is finite for the rectangular ($R=0$) and the cosine ($R=1$) prototype analysis filters, root-Hann prototype filters demand an infinite number of cross-terms, as shown in \eqref{eq:PTWOLA_root-Hann_window_taylor_series_expansion}. Consequently, the inter-subband aliasing effect diminishes as the number of cross-terms for the root-Hann prototype analysis filter is increased. However, for a fixed subband filter length ($T$), the increase in number of cross-terms reduces the available difference terms ($\acute{T} = T-(2R+1)$), thereby amplifying the distortion error. 
Therefore, the necessary number of cross-terms ($R$) for \gls{PTWOLA} with the root-Hann prototype analysis filter to achieve comparable performance to generalized \gls{WOLA} depends on whether the inter-subband aliasing or distortion error is more dominant. This is exemplified in Fig. \ref{fig:ERLE_comp_L512_SWDFTvsconstR_norm}, where \gls{PTWOLA} matches the performance of  generalized \gls{WOLA} with $R = 2$ for a minimum-norm prototype synthesis filter. However, as shown in Fig. \ref{fig:ERLE_comp_L512_SWDFTvsconstR_distortion}, \gls{PTWOLA} requires a larger number of cross-terms, $R = 4$, to achieve the performance of  generalized \gls{WOLA}. This discrepancy arises because, for the root-Hann prototype analysis filter, the minimum-norm synthesis filter is subject to significant distortion error. Therefore, performance is limited by the distortion errors and \gls{PTWOLA} requires more difference terms to offset it and match the performance of the generalized \gls{WOLA}. Meanwhile, with the minimum-distortion synthesis filter, the distortion errors are confined, and \gls{PTWOLA} performance is limited by the inter-subband aliasing, thereby requiring additional cross-terms to match the performance of the generalized \gls{WOLA} filter bank.

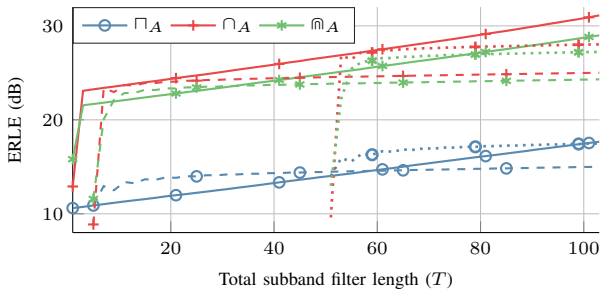
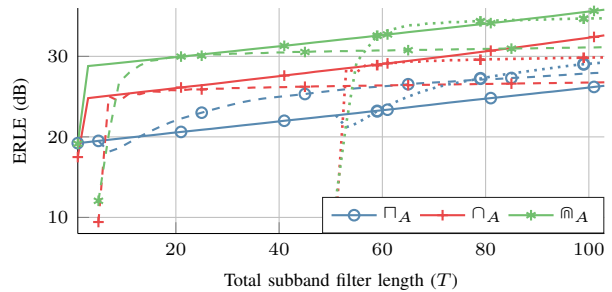
\begin{figure*}
	\captionsetup[subfigure]{width=3.4in}
	\captionsetup[subfigure]{captionskip=-0pt}
	\captionsetup[subfigure]{font={scriptsize}}
	\begin{minipage}{0.49\textwidth}
	\subfloat[{\scriptsize \gls{ERLE} achieved with minimum-norm prototype synthesis filters}]{\label{fig:ERLE_comp_L512_SWDFTvsconstnf_norm}}{%
	\setlength\fheight{3cm}
	\setlength\fwidth{7cm}
	\input{figures/PTWOLA_Results_L512_SWDFTvsconstnf_norm.tex}
	}
\end{minipage}\hfil
\begin{minipage}{0.49\textwidth}
\subfloat[{\scriptsize \gls{ERLE} achieved with minimum-distortion prototype synthesis filters}]{\label{fig:ERLE_comp_L512_SWDFTvsconstnf_distortion}}{%
\setlength\fheight{3cm}
\setlength\fwidth{7cm}
\input{figures/PTWOLA_Results_L512_SWDFTvsconstnf_des.tex}
}
\end{minipage}
	\caption{\gls{ERLE} achieved by \gls{PTWOLA} ($\acute{T}=5$ (\ref{fig_caption:General_prototype_synthesis_filter_norm}), $\acute{T}=51$ (\ref{fig_caption:PTWOLA_const_nf=51})) and generalized \gls{WOLA} (\ref{fig_caption:General_prototype_synthesis_filter_distortion}) for $L=512$ with different total subband filter lengths and constant number of difference terms ($\acute{T}$) for  prototype filters shown in Fig. \ref{fig:prototype_analysis_filters} and Fig. \ref{fig:prototype_synthesis_filters}.}
	\label{fig:ERLE_comp_L512_SWDFTvsconstnf}
\end{figure*}

Fig. \ref{fig:ERLE_comp_L512_SWDFTvsconstnf} presents a similar performance analysis for \gls{PTWOLA} with a fixed number of difference terms ($\acute{T})$. This implies that for a specific total filter length ($T$), the number of single-sided cross-terms is given by $R=\frac{T-\acute{T}-1}{2}$. For the sake of brevity and clarity, the performance analysis focuses on $\acute{T} = 5$ and $\acute{T} = 51$. However, it should be noted that these findings can be extrapolated to other values of $\acute{T}$, thus maintaining the generality of the observations.

For the cosine prototype analysis filter, minimal performance improvement is seen beyond $R=1$, particularly when using the minimum-norm synthesis filter. This trend is attributed to the frequency selectivity of the cosine prototype analysis filter (Fig. \ref{fig:prototype_analysis filters_freq}), where the magnitude spectrum for the cosine prototype analysis filter drops significantly beyond $R=1$, limiting its impact on the inter-subband aliasing error. 
With the rectangular prototype analysis filter, which exhibits poor frequency selectivity (Fig. \ref{fig:prototype_analysis filters_freq}), performance is largely dictated by the inter-subband aliasing error. Consequently, adding cross-terms improves performance, even outperforming the  generalized \gls{WOLA} for the same filter length. However, this performance improvement plateaus earlier for the minimum-norm synthesis filter (Fig. \ref{fig:ERLE_comp_L512_SWDFTvsconstnf_norm}) compared to the minimum-distortion synthesis filter (Fig. \ref{fig:ERLE_comp_L512_SWDFTvsconstnf_distortion}). This is due to the higher distortion error for the rectangular prototype analysis filter with the minimum-norm synthesis filter compared to the minimum-distortion synthesis filter (Fig. \ref{fig:scaled_convolution_prototype_filters}). Hence, as the inter-subband aliasing error decreases with an increased number of cross-terms, the distortion error dominates the aliasing error much earlier for the minimum-norm synthesis filter than for the minimum-distortion synthesis filter.
Lastly, a similar pattern can be observed for the root-Hann prototype analysis filter, where an increase in the number of cross-terms results in a lower inter-subband aliasing error. Nevertheless, this performance enhancement becomes minimal with an increasing number of cross-terms of $R$, as their contribution to inter-subband aliasing drops significantly, as shown in Fig. \ref{fig:prototype_analysis filters_freq}.

Furthermore, the analysis presented in Fig. \ref{fig:ERLE_comp_L512_SWDFTvsconstnf}  also serves as a direct comparison with state-of-the-art cross-band filtering methods from the literature \cite{avargel2006performance, 4156182}. These methods are structurally equivalent to a special case of the \gls{PTWOLA} framework where the number of difference terms is minimal ($\acute{T}=1$) and performance is improved by adding cross-terms ($R>0$). Fig. \ref{fig:ERLE_comp_L512_SWDFTvsconstnf}  illustrates precisely this concept, showing how the \gls{ERLE} evolves as the number of cross-terms is increased for a fixed number of difference terms.

%% file: figures/PTWOLA_L_512_C_l_rect_window_sum_mag2db.tex
\definecolor{mycolor1}{rgb}{0.34667,0.53600,0.69067}%
\begin{tikzpicture}[font=\scriptsize]
\begin{axis}[%
  width=\fwidth,
  height=\fheight,
  at={(0,0)},
scale only axis,
xmin=0,
xmax=2047,
xlabel={Samples ($n$)},
ylabel={Magnitude (dB)},
ymin=-20,
ymax=25,
axis background/.style={fill=white},
axis x line*=bottom,
axis y line*=left,
xmajorgrids,
ymajorgrids
]
\addplot [color=mycolor1, forget plot]
  table[row sep=crcr]{%
1	24.2152658340388\\
2	18.9732992184289\\
3	17.9893948414987\\
4	16.235504088376\\
5	14.9383774520707\\
6	16.7443878815893\\
7	15.7691740920373\\
8	14.5929563286864\\
9	16.884465709092\\
10	14.6574831708371\\
11	18.0731007717205\\
12	17.4706156843247\\
13	15.6970288404305\\
14	16.6696668187634\\
15	16.7127315521734\\
16	13.7904045868231\\
17	15.9737789991134\\
18	17.6691385178252\\
19	16.6038826502055\\
20	17.0806134405114\\
21	19.2013970036352\\
22	16.0143078416773\\
23	16.725670862163\\
24	15.6491887451479\\
25	15.7840677189604\\
26	13.4817832780626\\
27	15.1337191972658\\
28	17.0906274218435\\
29	16.8804295901114\\
30	12.7206962347386\\
31	17.3706131232834\\
32	15.1365797098253\\
33	14.3667164462187\\
34	13.5711933849088\\
35	14.0344146552421\\
36	16.8047776942458\\
37	13.5280289587065\\
38	15.3983327076078\\
39	15.3495717237738\\
40	13.6342160807469\\
41	13.2490347568901\\
42	15.3838034136635\\
43	10.3434182796515\\
44	13.1884240496556\\
45	16.3004102768634\\
46	16.9304495072367\\
47	17.5255519197999\\
48	15.061895328534\\
49	13.0448211438308\\
50	11.9289732633938\\
51	13.0186356378531\\
52	14.9954324452771\\
53	14.0080850618869\\
54	10.3953990347766\\
55	15.3917900715262\\
56	12.7311074471895\\
57	12.2034639556235\\
58	12.8112756772934\\
59	10.751238318924\\
60	13.4230818716566\\
61	13.4920471544779\\
62	12.8295978026457\\
63	10.7563532157794\\
64	14.6963457668857\\
65	12.217594850786\\
66	14.3788239235443\\
67	13.8859590657207\\
68	15.1420052530787\\
69	11.290219617166\\
70	9.87643747051281\\
71	15.2335695743797\\
72	15.0931114155801\\
73	13.6570978147367\\
74	14.5750798319365\\
75	14.7035130899023\\
76	11.7024329375217\\
77	11.7222142952031\\
78	13.7677467705269\\
79	12.2023921869308\\
80	12.0222713126177\\
81	10.9393609864185\\
82	10.5364458535393\\
83	10.5741622724085\\
84	10.6707347104836\\
85	14.2633421554749\\
86	13.9219977861729\\
87	11.4995472965278\\
88	12.2624547985862\\
89	9.12942015933526\\
90	9.68934751733995\\
91	8.49826700521516\\
92	10.3563274484963\\
93	11.6319489402847\\
94	9.4410263591998\\
95	12.4123286280135\\
96	12.1655421848194\\
97	11.9242691785786\\
98	9.72669172816943\\
99	12.6333537503864\\
100	9.53058486717058\\
101	13.2090131568443\\
102	11.1838629171612\\
103	10.4786588530886\\
104	10.0941492780267\\
105	10.6836628784255\\
106	14.5892386454079\\
107	13.3846260607929\\
108	9.76759459379411\\
109	13.3524457227086\\
110	11.6482961455633\\
111	12.4061627184917\\
112	10.1505906559496\\
113	10.8039400072146\\
114	8.96174410995158\\
115	9.10071684001672\\
116	4.89042569745509\\
117	11.9802670491242\\
118	11.4133465830714\\
119	13.3033217306963\\
120	11.0723905271223\\
121	13.7712015124487\\
122	9.41898418432223\\
123	9.07452885732923\\
124	7.43030833349397\\
125	10.3756000900025\\
126	10.0503955948916\\
127	10.0197987224298\\
128	5.70720742818049\\
129	7.74785055170721\\
130	6.91888633813318\\
131	6.82707445419999\\
132	9.32958496650982\\
133	8.169918863403\\
134	6.39813991495764\\
135	5.97838996533214\\
136	10.1577798792753\\
137	12.3562589385407\\
138	9.233963157408\\
139	10.3917817856214\\
140	10.2455910013354\\
141	10.7287484951569\\
142	8.28424894788604\\
143	9.2243976808594\\
144	8.23093909635965\\
145	8.28227544261516\\
146	7.59293996569221\\
147	9.89614543964855\\
148	7.07556483006286\\
149	8.85965922283134\\
150	6.17528090634019\\
151	6.94955330958751\\
152	11.2459370152412\\
153	6.82456430441256\\
154	5.72124658565706\\
155	6.78870684916329\\
156	6.74204756130291\\
157	5.22711743618877\\
158	10.4144360138684\\
159	5.01267838540787\\
160	5.93548493706857\\
161	7.3278835947154\\
162	6.27419143644168\\
163	5.40263308179313\\
164	8.08049566775472\\
165	5.46116724143835\\
166	7.18393063591635\\
167	8.62605874938238\\
168	10.2660464059224\\
169	8.04720012576855\\
170	6.51589382217661\\
171	6.52648380214756\\
172	5.52736995333837\\
173	5.01941375278588\\
174	7.56280074036749\\
175	7.84289195877465\\
176	5.14843982413853\\
177	5.77627621378298\\
178	5.30961451066576\\
179	7.82186262136039\\
180	6.68263332809549\\
181	6.58061372973305\\
182	6.38620737742224\\
183	7.43410574893256\\
184	2.08984979889716\\
185	4.91308635956266\\
186	5.54569800298222\\
187	7.19725196023398\\
188	6.30355509004839\\
189	4.02321413937562\\
190	6.04188577240646\\
191	3.87905414001834\\
192	4.04762673464581\\
193	4.09698721891747\\
194	6.74689234107442\\
195	2.64652872610834\\
196	3.62814660727114\\
197	3.29706921938545\\
198	2.71605487766129\\
199	-0.0689652215104489\\
200	3.05000805610789\\
201	6.36710400145857\\
202	5.33093829108853\\
203	2.29973181532411\\
204	6.16453236260857\\
205	6.00171611899936\\
206	6.28797712517157\\
207	5.9774264617744\\
208	3.77676418112822\\
209	5.02555948225059\\
210	1.53162036684786\\
211	3.8191952657351\\
212	3.58090998476659\\
213	6.151395483072\\
214	5.08938007353492\\
215	5.34032082472837\\
216	3.30572303829644\\
217	2.0789839918871\\
218	3.37417156975214\\
219	2.77656023654009\\
220	0.194583317860252\\
221	3.3513911849056\\
222	6.72784214793513\\
223	5.3704457738347\\
224	4.55750436837424\\
225	-1.24088532914405\\
226	-0.124346126775811\\
227	3.50980637218587\\
228	3.50062920027679\\
229	2.88766500401177\\
230	3.96177641781471\\
231	2.83499794322054\\
232	3.57452467389266\\
233	5.82158508945248\\
234	4.04360920189589\\
235	2.78083775743304\\
236	-0.966324693574279\\
237	-0.327678069258545\\
238	-0.455764062515315\\
239	4.07455306567638\\
240	3.36379765880679\\
241	4.02952646080185\\
242	-2.27139529803348\\
243	5.04733286616514\\
244	5.43487743202179\\
245	0.823985614966225\\
246	3.82175081625985\\
247	1.34122778566782\\
248	-1.39374181346816\\
249	1.79586501233883\\
250	3.45951961016017\\
251	1.24798041064138\\
252	3.24537941698306\\
253	1.42468365610915\\
254	-0.322109440864678\\
255	1.76380190432202\\
256	2.00183440667803\\
257	3.03105520889559\\
258	-0.235451687006802\\
259	2.23706558572807\\
260	1.89123796262103\\
261	1.76297417575712\\
262	2.57438639214948\\
263	-2.39707558673826\\
264	-2.45709704974976\\
265	0.179055489310314\\
266	0.141198161189633\\
267	0.94605566350363\\
268	0.130198786738128\\
269	-0.117036790696034\\
270	-0.158456471845706\\
271	-0.9153727994359\\
272	2.48799851563446\\
273	-0.488161695166403\\
274	3.24135579216889\\
275	3.79062150575148\\
276	3.32120443206496\\
277	4.54943003700711\\
278	0.36572210195446\\
279	0.707680587107553\\
280	1.17499890277589\\
281	-0.0572607351424387\\
282	-1.96680341413041\\
283	-0.0761810224424853\\
284	1.78638318024642\\
285	-0.130873567450479\\
286	0.822230101444925\\
287	2.94455201911416\\
288	-0.00186150761178735\\
289	0.28058460267541\\
290	-2.22371937099151\\
291	0.637943129757678\\
292	0.820737679792151\\
293	1.31923809975839\\
294	-0.533391134563924\\
295	0.793657703337207\\
296	-2.02415978649587\\
297	-1.89954292218221\\
298	1.61389388429362\\
299	1.20789010103217\\
300	-1.93297617770473\\
301	-1.33483990106409\\
302	-5.40816828694477\\
303	-1.85752804239235\\
304	0.858502871529552\\
305	1.48199480859346\\
306	-1.04755448266365\\
307	-2.78887975459299\\
308	-3.29969287232333\\
309	1.12814756277414\\
310	0.56880491257275\\
311	-5.02063598188735\\
312	0.769516682830477\\
313	0.0706937018096264\\
314	-1.33284609497416\\
315	-0.667299308260449\\
316	0.13678224616992\\
317	1.5892533134855\\
318	2.00090762093895\\
319	-0.0762133301966883\\
320	-3.78750430708277\\
321	0.509660029999033\\
322	1.65244082807803\\
323	-0.273432121559675\\
324	-3.61715060370443\\
325	0.168020189192076\\
326	-1.30744934642911\\
327	-0.276508100509198\\
328	-2.16577622947618\\
329	-2.93493432601805\\
330	-0.409528359012069\\
331	-1.79408858125015\\
332	-2.65690052322106\\
333	-0.480023231305546\\
334	-2.57699171654317\\
335	-0.979368705527487\\
336	-0.927775880893026\\
337	-2.39085644206471\\
338	-2.2112239006558\\
339	-3.36494088061897\\
340	-2.85855545095947\\
341	-1.51157002647224\\
342	0.188768514312951\\
343	-2.03545886119777\\
344	-0.162683082966518\\
345	-3.06129716180135\\
346	-2.6790710179677\\
347	1.31295975744741\\
348	-3.59837537595995\\
349	-2.22193877598694\\
350	-5.59815628380928\\
351	-3.25853478324614\\
352	-3.47772770613182\\
353	-2.81714503379937\\
354	-3.28162351921562\\
355	-1.91906177137334\\
356	-3.19955997606995\\
357	-1.21633309788873\\
358	0.224473969185703\\
359	-3.28700470677233\\
360	-7.48555948101105\\
361	0.449416076177133\\
362	-4.91465709194946\\
363	-3.44331200416449\\
364	-3.13177916021921\\
365	-6.95892639061234\\
366	-6.09503125517772\\
367	-1.73629558605015\\
368	-1.26029263021794\\
369	-0.165960786449586\\
370	-1.25327455408817\\
371	-0.0396300536177672\\
372	-1.97419636808702\\
373	-1.66119997600149\\
374	-0.927257380726564\\
375	-3.65778697936779\\
376	-1.79762754704959\\
377	-1.86580482080993\\
378	-2.46110193683549\\
379	-1.88953325493463\\
380	-2.60983226385165\\
381	-6.06521285450669\\
382	-4.97830608191198\\
383	-4.80115020576099\\
384	-4.75133481255186\\
385	-4.69744556841794\\
386	-8.38354439085582\\
387	-3.89543695563698\\
388	0.101222404535536\\
389	-3.01957761332397\\
390	-4.02550583990008\\
391	-3.01230986009999\\
392	-8.80588048638009\\
393	-6.36128407356672\\
394	-3.2420315130198\\
395	-4.07213659695015\\
396	-3.04256543938801\\
397	-4.25342748967459\\
398	-6.10522949079049\\
399	-3.21706792449368\\
400	-6.89741663646107\\
401	-4.23935728334386\\
402	-4.56789450965248\\
403	-2.51068618421962\\
404	-3.32337131200618\\
405	-3.29306230556693\\
406	-2.77719022903973\\
407	-3.16187347465725\\
408	-2.58615911425832\\
409	-5.71171077341289\\
410	-5.09731890505511\\
411	-5.74631163462592\\
412	-5.73104629023648\\
413	-4.99991767226139\\
414	-4.86860742323132\\
415	-8.61307166379659\\
416	-6.42858565975236\\
417	-5.79749708067684\\
418	-4.35008104560141\\
419	-4.95522181224367\\
420	-6.75843087842706\\
421	-7.26661658013188\\
422	-7.94440448125402\\
423	-4.86566661923958\\
424	-5.59804864363385\\
425	-5.1564727413974\\
426	-5.41032037768799\\
427	-4.06542508463056\\
428	-7.01352536776397\\
429	-8.01167870632466\\
430	-6.57342839171181\\
431	-3.27929727422556\\
432	-5.85178755241912\\
433	-8.42011265695817\\
434	-10.325458996419\\
435	-4.67145793458307\\
436	-4.50757901581353\\
437	-6.43174033050108\\
438	-9.00413975588371\\
439	-4.17955266472614\\
440	-8.10839434832882\\
441	-7.85788037744818\\
442	-9.25758898676088\\
443	-5.47946359404161\\
444	-5.59860705025148\\
445	-7.47846767655284\\
446	-4.06877130668099\\
447	-4.37735137641044\\
448	-8.24965036104341\\
449	-9.97358297138271\\
450	-6.84685168815546\\
451	-6.34003740776435\\
452	-5.87134215607123\\
453	-6.28520026471587\\
454	-6.35513114061694\\
455	-7.31869510268598\\
456	-6.8253617098685\\
457	-3.65416690801414\\
458	-2.2261821435559\\
459	-4.33172348075937\\
460	-4.71911731597115\\
461	-9.22737070491747\\
462	-6.52034930689991\\
463	-9.07862567733267\\
464	-6.38879641521806\\
465	-8.26177618753766\\
466	-6.57267370526274\\
467	-4.60484266487896\\
468	-6.34131666168843\\
469	-8.51244099596348\\
470	-6.41473096613603\\
471	-7.96085465346924\\
472	-7.39212954209784\\
473	-7.75358157667196\\
474	-7.00532684910023\\
475	-7.65575424033404\\
476	-6.04358107938568\\
477	-5.98717398157935\\
478	-7.97032549932201\\
479	-7.6599369593765\\
480	-6.40684826909468\\
481	-5.64775583579677\\
482	-9.8463838779335\\
483	-9.0298530401589\\
484	-7.50982296247491\\
485	-4.69517005818183\\
486	-7.61313331647577\\
487	-7.50602243904495\\
488	-8.15446778867714\\
489	-5.30905713327532\\
490	-5.61631186375838\\
491	-5.23823594797256\\
492	-8.46966114827519\\
493	-9.53820761468831\\
494	-8.77034744034922\\
495	-7.66719896104394\\
496	-6.55998117314577\\
497	-6.32714123123084\\
498	-7.9145721012143\\
499	-10.4557848684992\\
500	-7.8069715646916\\
501	-7.09923559757862\\
502	-8.43894456223455\\
503	-6.36829559842011\\
504	-6.52576091504489\\
505	-8.93217963126627\\
506	-7.06148417859535\\
507	-8.22577411452405\\
508	-7.53737482280862\\
509	-8.97452411568244\\
510	-6.49017537778182\\
511	-6.31966365773759\\
512	-5.56725271565118\\
513	-7.06946977746694\\
514	-6.74115462312951\\
515	-9.62376059905546\\
516	-7.75773658352119\\
517	-9.6061941636474\\
518	-7.26870325447145\\
519	-6.32736268613998\\
520	-4.4307935893738\\
521	-5.27498913560952\\
522	-6.34114743450056\\
523	-5.22540834483385\\
524	-6.22564170241089\\
525	-8.30501964808342\\
526	-7.74434008353256\\
527	-7.33922058574625\\
528	-8.48283130188119\\
529	-9.17565085802998\\
530	-7.33383516726503\\
531	-6.1497591094448\\
532	-6.90127399438392\\
533	-10.2319699170113\\
534	-10.3469840980905\\
535	-5.81648694912288\\
536	-6.60259122545162\\
537	-8.94642263151683\\
538	-7.60239923384438\\
539	-5.26672740066003\\
540	-7.97953318324352\\
541	-7.1434572606438\\
542	-6.26513570948639\\
543	-9.43640581517758\\
544	-7.91504082332794\\
545	-5.25430083288673\\
546	-4.59297623561766\\
547	-6.72033293004688\\
548	-7.7348964050434\\
549	-3.84770374336153\\
550	-5.18377438524408\\
551	-7.62952631538651\\
552	-9.39799522140261\\
553	-10.0577038455777\\
554	-8.67386973778953\\
555	-6.74646096327698\\
556	-7.27510635594175\\
557	-9.03697463115047\\
558	-9.88058612014698\\
559	-8.28545004258083\\
560	-7.05374485455892\\
561	-6.38684686068739\\
562	-3.65047002339661\\
563	-4.11133758672627\\
564	-10.0975938387187\\
565	-9.21653268530341\\
566	-10.5844542174486\\
567	-7.26919977711491\\
568	-4.45560405432288\\
569	-7.84823350600046\\
570	-9.6181720398745\\
571	-9.86847110021367\\
572	-8.67926150502974\\
573	-8.42284680582616\\
574	-9.74207399056098\\
575	-5.8044592492075\\
576	-3.90762792052699\\
577	-8.32284485062222\\
578	-7.17096426181794\\
579	-7.07718161231469\\
580	-10.5874016514557\\
581	-6.01002668964154\\
582	-6.94438491564997\\
583	-10.4019797547665\\
584	-8.37364537781769\\
585	-7.5789709097867\\
586	-9.39639854376087\\
587	-9.91643913258115\\
588	-9.17070586970066\\
589	-8.1539217497173\\
590	-8.74879198638912\\
591	-7.16384960079967\\
592	-6.77971553027291\\
593	-8.50589072911493\\
594	-9.15935177595733\\
595	-7.51771176133478\\
596	-10.9440219447582\\
597	-8.35968152555924\\
598	-7.81041265175456\\
599	-9.15706255496745\\
600	-8.47827726934108\\
601	-8.78326053079052\\
602	-7.23052764469276\\
603	-7.65759109264848\\
604	-10.447965140814\\
605	-8.69542170853152\\
606	-7.7334890371921\\
607	-5.48333128611934\\
608	-6.42522602850535\\
609	-7.75736616502472\\
610	-9.76144883378411\\
611	-7.15126351743057\\
612	-5.68193446506242\\
613	-9.6587602730359\\
614	-8.88689488380184\\
615	-6.93820522921203\\
616	-4.7388744074538\\
617	-10.6030190299009\\
618	-6.99686629903262\\
619	-7.97497995561108\\
620	-3.50545424488316\\
621	-5.67089582236611\\
622	-7.71377552312965\\
623	-9.77899854273292\\
624	-9.60899728783357\\
625	-6.25996699755502\\
626	-7.06645423164041\\
627	-7.25072099991328\\
628	-7.975476309326\\
629	-9.16273060901991\\
630	-9.52726327014452\\
631	-8.87181384415977\\
632	-6.8614491546021\\
633	-6.73895319946233\\
634	-5.03809121876004\\
635	-4.11744159211955\\
636	-6.11874954640985\\
637	-6.65736246290761\\
638	-6.34504580736472\\
639	-7.35174585669139\\
640	-7.55532909794465\\
641	-8.80226738715928\\
642	-8.78383046081109\\
643	-9.5876450015371\\
644	-5.30425111157144\\
645	-7.03337370475058\\
646	-9.96533040650552\\
647	-7.78276367226166\\
648	-5.93187835790914\\
649	-4.50414070645539\\
650	-5.08172611464556\\
651	-8.17300238328457\\
652	-5.83437878080017\\
653	-5.71598730500015\\
654	-7.47657029265661\\
655	-6.41035849876615\\
656	-6.59917931624055\\
657	-6.67924708999237\\
658	-8.00194483322211\\
659	-10.2075781996128\\
660	-11.6232655841994\\
661	-7.49725289426275\\
662	-8.52139247946373\\
663	-5.78094704960883\\
664	-8.67298593201301\\
665	-9.2502069726191\\
666	-7.41803816843322\\
667	-4.5307313782771\\
668	-5.0101054397767\\
669	-6.14127376499615\\
670	-9.13267679364985\\
671	-10.3460021489219\\
672	-9.76534828383692\\
673	-7.20707669852919\\
674	-6.09519915026838\\
675	-7.17656493382932\\
676	-7.62223282191215\\
677	-7.96541777398613\\
678	-8.11666171064597\\
679	-11.9773315660352\\
680	-12.1431376787835\\
681	-10.6793458308736\\
682	-6.26069129274732\\
683	-6.49468755845485\\
684	-8.4153295910707\\
685	-8.2139405110386\\
686	-9.31590109288011\\
687	-8.59662079919122\\
688	-10.3807007590243\\
689	-8.92904676986081\\
690	-9.56501485002096\\
691	-8.71488737652095\\
692	-10.5503690382607\\
693	-9.51963787740109\\
694	-5.42732533468207\\
695	-4.98744691841986\\
696	-9.75568101840406\\
697	-8.35674524897279\\
698	-4.79974933775481\\
699	-6.82974030879389\\
700	-4.6057182721403\\
701	-4.79870138445157\\
702	-8.00943719992152\\
703	-10.5940945282271\\
704	-9.72436137864579\\
705	-10.1123653900954\\
706	-12.35897201618\\
707	-8.63151036272416\\
708	-7.25070094633076\\
709	-8.90124097194932\\
710	-8.33169917366249\\
711	-7.86459204400566\\
712	-11.5433794718594\\
713	-8.43783484813111\\
714	-6.9776605635841\\
715	-5.99480429561867\\
716	-7.20359030775288\\
717	-6.30812845824767\\
718	-9.53193235067128\\
719	-7.95081936210157\\
720	-4.79551076770307\\
721	-4.74445686642777\\
722	-8.1645677534426\\
723	-8.64117284874566\\
724	-12.6763977786938\\
725	-10.4394739526066\\
726	-7.31077024145162\\
727	-6.27099781443994\\
728	-7.05347339455504\\
729	-9.13910158036975\\
730	-10.4445381979193\\
731	-10.1889201255691\\
732	-8.41751720778351\\
733	-6.4111226104531\\
734	-6.0088735432537\\
735	-8.13556867285159\\
736	-4.70831146994444\\
737	-2.44536139785255\\
738	-6.06290636682089\\
739	-8.11434418659332\\
740	-7.93120296939824\\
741	-4.58569867238747\\
742	-3.54789576090669\\
743	-6.4432186594065\\
744	-5.88343404932005\\
745	-5.71161709322481\\
746	-9.38264357038612\\
747	-10.9401861313457\\
748	-11.4952923243685\\
749	-8.75322834557213\\
750	-7.27739160511048\\
751	-6.99414748121213\\
752	-6.01569766320322\\
753	-7.31392904420038\\
754	-9.48026947817219\\
755	-7.42047812560583\\
756	-9.32760676984073\\
757	-7.15200205153485\\
758	-7.41484462047556\\
759	-7.62390134922524\\
760	-9.82861697409385\\
761	-6.5296489004304\\
762	-5.00382021969409\\
763	-6.63286219224158\\
764	-6.58635866710274\\
765	-7.03958911237292\\
766	-8.25373239094369\\
767	-4.69531964926509\\
768	-7.0167890826122\\
769	-7.85113857563468\\
770	-6.4014231047383\\
771	-6.68801026319833\\
772	-6.43186100324552\\
773	-4.11396949099555\\
774	-5.16320699326269\\
775	-6.25991076993311\\
776	-7.201976403619\\
777	-8.90630901817228\\
778	-8.89790461420262\\
779	-5.52198875024466\\
780	-8.37768927582652\\
781	-6.38182192718302\\
782	-8.78102719652708\\
783	-8.11703236286195\\
784	-6.54571970324535\\
785	-7.52082607651481\\
786	-6.94302653395886\\
787	-6.3490152215054\\
788	-5.24268298702905\\
789	-5.15911486308273\\
790	-7.07932159045122\\
791	-3.99551016932561\\
792	-5.58721034934704\\
793	-5.96665408873275\\
794	-6.80927035938089\\
795	-6.05705302627517\\
796	-5.51354621214724\\
797	-7.82768465922942\\
798	-8.88384489763373\\
799	-7.69797091987588\\
800	-8.9164981510026\\
801	-8.16303172074366\\
802	-5.24980634734043\\
803	-10.3585457357691\\
804	-6.73709725390097\\
805	-2.8116104991943\\
806	-6.08558941948085\\
807	-7.1399778320427\\
808	-11.1470177918076\\
809	-8.58130124955817\\
810	-8.34259037803527\\
811	-6.87784661087434\\
812	-9.73789098629141\\
813	-9.83347855783371\\
814	-9.84180824975332\\
815	-10.0112706100914\\
816	-9.18926994998604\\
817	-5.43874141713729\\
818	-8.52567988353234\\
819	-8.35751775691058\\
820	-7.3645861218188\\
821	-7.15170851261033\\
822	-5.36007110666717\\
823	-4.94885531205161\\
824	-8.19565464797167\\
825	-8.12363665821664\\
826	-7.8521751549504\\
827	-8.35721899266678\\
828	-7.47333258388353\\
829	-7.89015850350885\\
830	-5.47988701934007\\
831	-6.68603565317446\\
832	-11.9275701959458\\
833	-9.54423179645491\\
834	-9.16621492397899\\
835	-6.11763087636411\\
836	-5.81385709017304\\
837	-8.97166910872922\\
838	-8.31361314820289\\
839	-11.2715930181559\\
840	-8.09647051594451\\
841	-6.98781132525188\\
842	-6.32567749928452\\
843	-4.77809455609927\\
844	-7.32987571513397\\
845	-8.55981193111485\\
846	-6.87944760087896\\
847	-6.26848328844741\\
848	-6.38210027638944\\
849	-6.41283107677243\\
850	-6.88964027215308\\
851	-7.73201928141935\\
852	-12.3713187152129\\
853	-10.3200091227359\\
854	-7.86189401400286\\
855	-8.02482101200956\\
856	-7.22397234361651\\
857	-8.09514756722146\\
858	-8.83161546092249\\
859	-5.27118710985672\\
860	-7.12462783048588\\
861	-6.18159285383305\\
862	-6.13363001941158\\
863	-5.57622658574243\\
864	-9.95296996701544\\
865	-9.67985679280033\\
866	-7.99705509832233\\
867	-5.14253625702875\\
868	-8.06558241192836\\
869	-4.69882063898379\\
870	-5.48039991094323\\
871	-9.83987651430845\\
872	-6.92260309405644\\
873	-4.96270603385061\\
874	-4.69260048742073\\
875	-6.17967201671079\\
876	-6.73744806647614\\
877	-7.43758663020659\\
878	-7.92795195179773\\
879	-7.10682086672429\\
880	-6.62255550069591\\
881	-5.43667448194825\\
882	-7.34302174715254\\
883	-7.72207040776138\\
884	-8.16884464872822\\
885	-6.10576424233831\\
886	-8.79194138464265\\
887	-8.1486989987839\\
888	-9.4540720663498\\
889	-7.48062697612401\\
890	-6.51992662079939\\
891	-7.03599147256798\\
892	-8.9149296757166\\
893	-10.7667034843701\\
894	-9.013258128629\\
895	-8.59933976413962\\
896	-5.9242896504979\\
897	-10.4075662805101\\
898	-9.63439172259357\\
899	-8.04207464941025\\
900	-8.48650385048644\\
901	-6.86516142422248\\
902	-8.8108091124648\\
903	-8.78007554584751\\
904	-9.80480001654129\\
905	-5.80350347528476\\
906	-8.63074513677252\\
907	-7.97397029417925\\
908	-5.36411796967103\\
909	-5.2473419042266\\
910	-8.51294941615661\\
911	-7.5338845170861\\
912	-8.80604417356274\\
913	-10.3620558005113\\
914	-7.24588714552976\\
915	-5.53878764798804\\
916	-5.86631655217132\\
917	-6.8988707233052\\
918	-10.0467963078116\\
919	-6.90356271799294\\
920	-7.99963497636556\\
921	-8.60600007417465\\
922	-5.49498915013821\\
923	-6.41561393392016\\
924	-6.63900174063556\\
925	-8.82621283783257\\
926	-8.15069275208486\\
927	-7.34374804881882\\
928	-4.822351743538\\
929	-7.91885525858511\\
930	-8.4694154905429\\
931	-6.73220716969627\\
932	-9.12506320800823\\
933	-10.9592191262007\\
934	-7.39691423676199\\
935	-8.39845177586136\\
936	-6.70171762753365\\
937	-4.48880669447432\\
938	-5.06345542531497\\
939	-6.3821608454311\\
940	-6.54414069398292\\
941	-7.03755591669645\\
942	-9.8708676118739\\
943	-8.49985955237265\\
944	-4.10465247290605\\
945	-7.30691100005669\\
946	-8.4328344423563\\
947	-9.62295681625984\\
948	-11.6662556547593\\
949	-9.74337611072113\\
950	-5.75765436976286\\
951	-4.47141607170959\\
952	-5.66718960733162\\
953	-10.5180749353504\\
954	-4.48936225762942\\
955	-5.41841379344883\\
956	-9.72064128576455\\
957	-7.57934834856677\\
958	-8.04563504273195\\
959	-7.31109247433199\\
960	-7.71770449287299\\
961	-9.17048640152683\\
962	-8.0924824874543\\
963	-8.2927853744733\\
964	-5.13298430201817\\
965	-6.87466329426296\\
966	-6.71256507513451\\
967	-6.41164060402439\\
968	-9.81602602416274\\
969	-9.07367634825281\\
970	-10.0654227612383\\
971	-8.36788915664818\\
972	-7.15420377831376\\
973	-3.06657125844206\\
974	-8.84467824184944\\
975	-10.8037316070016\\
976	-9.12932675430261\\
977	-7.68664455341733\\
978	-7.58458297002677\\
979	-9.61709684442249\\
980	-7.54965730726653\\
981	-7.12755265823813\\
982	-4.21039780591337\\
983	-4.77401679671809\\
984	-6.10390579626492\\
985	-8.64164288508978\\
986	-7.66834945955364\\
987	-8.17390559862351\\
988	-4.31492122393461\\
989	-6.53964959364788\\
990	-8.65142315396929\\
991	-7.16529720831537\\
992	-8.13985844156986\\
993	-3.71107842628329\\
994	-7.31767115884425\\
995	-10.0204776633984\\
996	-6.35746294187445\\
997	-9.00088000828173\\
998	-5.89892283837311\\
999	-8.80974560366836\\
1000	-7.50961761527269\\
1001	-7.51097213500681\\
1002	-7.99793224063634\\
1003	-8.18155707316217\\
1004	-9.98940244239746\\
1005	-7.57204845944852\\
1006	-7.28121099511528\\
1007	-6.15012346044209\\
1008	-6.12804204562682\\
1009	-5.8606789413589\\
1010	-6.99425349199695\\
1011	-7.06884912527246\\
1012	-7.58666697955355\\
1013	-15.5320088488968\\
1014	-5.67101987150883\\
1015	-7.43957934408027\\
1016	-5.91055766718349\\
1017	-5.07450367337425\\
1018	-7.78947756840856\\
1019	-7.3026059974974\\
1020	-9.29150517574895\\
1021	-7.32068742883887\\
1022	-7.50664473280819\\
1023	-6.78649389392905\\
1024	-4.93163034865469\\
1025	24.2152658340388\\
1026	18.9732992184286\\
1027	17.9893948414987\\
1028	16.2355040883758\\
1029	14.9383774520705\\
1030	16.7443878815897\\
1031	15.7691740920373\\
1032	14.5929563286864\\
1033	16.8844657090919\\
1034	14.6574831708369\\
1035	18.0731007717205\\
1036	17.4706156843251\\
1037	15.6970288404307\\
1038	16.6696668187635\\
1039	16.7127315521732\\
1040	13.7904045868231\\
1041	15.9737789991131\\
1042	17.6691385178252\\
1043	16.6038826502053\\
1044	17.0806134405115\\
1045	19.2013970036355\\
1046	16.0143078416774\\
1047	16.7256708621632\\
1048	15.6491887451487\\
1049	15.7840677189604\\
1050	13.4817832780624\\
1051	15.1337191972662\\
1052	17.0906274218436\\
1053	16.8804295901114\\
1054	12.7206962347385\\
1055	17.3706131232832\\
1056	15.1365797098255\\
1057	14.3667164462186\\
1058	13.5711933849091\\
1059	14.0344146552422\\
1060	16.8047776942458\\
1061	13.5280289587068\\
1062	15.3983327076074\\
1063	15.3495717237737\\
1064	13.634216080747\\
1065	13.2490347568899\\
1066	15.3838034136634\\
1067	10.3434182796521\\
1068	13.1884240496558\\
1069	16.3004102768634\\
1070	16.9304495072365\\
1071	17.5255519197999\\
1072	15.0618953285341\\
1073	13.0448211438308\\
1074	11.9289732633938\\
1075	13.0186356378531\\
1076	14.995432445277\\
1077	14.0080850618873\\
1078	10.3953990347766\\
1079	15.3917900715262\\
1080	12.7311074471899\\
1081	12.2034639556228\\
1082	12.8112756772935\\
1083	10.751238318924\\
1084	13.4230818716563\\
1085	13.4920471544778\\
1086	12.8295978026458\\
1087	10.7563532157796\\
1088	14.6963457668858\\
1089	12.2175948507857\\
1090	14.3788239235444\\
1091	13.8859590657207\\
1092	15.1420052530786\\
1093	11.2902196171665\\
1094	9.87643747051314\\
1095	15.2335695743796\\
1096	15.0931114155801\\
1097	13.6570978147367\\
1098	14.5750798319366\\
1099	14.7035130899027\\
1100	11.7024329375215\\
1101	11.722214295203\\
1102	13.7677467705268\\
1103	12.2023921869309\\
1104	12.0222713126177\\
1105	10.9393609864189\\
1106	10.5364458535393\\
1107	10.5741622724086\\
1108	10.6707347104839\\
1109	14.263342155475\\
1110	13.9219977861726\\
1111	11.4995472965279\\
1112	12.2624547985867\\
1113	9.12942015933554\\
1114	9.68934751733993\\
1115	8.49826700521486\\
1116	10.3563274484954\\
1117	11.6319489402847\\
1118	9.44102635919965\\
1119	12.4123286280133\\
1120	12.1655421848199\\
1121	11.9242691785787\\
1122	9.72669172816912\\
1123	12.6333537503864\\
1124	9.5305848671705\\
1125	13.2090131568446\\
1126	11.183862917161\\
1127	10.4786588530889\\
1128	10.0941492780265\\
1129	10.6836628784254\\
1130	14.5892386454082\\
1131	13.3846260607933\\
1132	9.76759459379435\\
1133	13.3524457227086\\
1134	11.6482961455636\\
1135	12.4061627184916\\
1136	10.1505906559503\\
1137	10.8039400072144\\
1138	8.96174410995189\\
1139	9.10071684001609\\
1140	4.89042569745664\\
1141	11.9802670491242\\
1142	11.4133465830712\\
1143	13.3033217306955\\
1144	11.0723905271224\\
1145	13.7712015124487\\
1146	9.41898418432252\\
1147	9.07452885732915\\
1148	7.43030833349412\\
1149	10.3756000900027\\
1150	10.0503955948917\\
1151	10.0197987224302\\
1152	5.70720742818015\\
1153	7.74785055170759\\
1154	6.91888633813205\\
1155	6.82707445419992\\
1156	9.32958496650946\\
1157	8.16991886340313\\
1158	6.39813991495824\\
1159	5.97838996533202\\
1160	10.1577798792755\\
1161	12.356258938541\\
1162	9.23396315740702\\
1163	10.3917817856211\\
1164	10.2455910013351\\
1165	10.7287484951575\\
1166	8.28424894788513\\
1167	9.22439768085952\\
1168	8.23093909635972\\
1169	8.28227544261472\\
1170	7.59293996569224\\
1171	9.89614543964873\\
1172	7.07556483006305\\
1173	8.85965922283203\\
1174	6.17528090634023\\
1175	6.9495533095874\\
1176	11.2459370152412\\
1177	6.82456430441299\\
1178	5.72124658565718\\
1179	6.78870684916344\\
1180	6.74204756130251\\
1181	5.22711743618839\\
1182	10.4144360138684\\
1183	5.01267838540722\\
1184	5.9354849370685\\
1185	7.32788359471573\\
1186	6.27419143644128\\
1187	5.40263308179319\\
1188	8.08049566775454\\
1189	5.4611672414376\\
1190	7.18393063591656\\
1191	8.62605874938176\\
1192	10.2660464059224\\
1193	8.04720012576954\\
1194	6.51589382217784\\
1195	6.52648380214795\\
1196	5.52736995333862\\
1197	5.01941375278736\\
1198	7.56280074036749\\
1199	7.84289195877473\\
1200	5.14843982413827\\
1201	5.77627621378363\\
1202	5.30961451066608\\
1203	7.82186262136026\\
1204	6.68263332809548\\
1205	6.58061372973282\\
1206	6.38620737742158\\
1207	7.43410574893245\\
1208	2.08984979889655\\
1209	4.91308635956274\\
1210	5.5456980029816\\
1211	7.19725196023365\\
1212	6.30355509004778\\
1213	4.02321413937485\\
1214	6.04188577240664\\
1215	3.87905414001794\\
1216	4.04762673464632\\
1217	4.09698721891758\\
1218	6.74689234107484\\
1219	2.64652872610856\\
1220	3.62814660727108\\
1221	3.29706921938571\\
1222	2.71605487766087\\
1223	-0.0689652215121091\\
1224	3.05000805610807\\
1225	6.3671040014583\\
1226	5.33093829108846\\
1227	2.29973181532478\\
1228	6.16453236260923\\
1229	6.00171611899955\\
1230	6.28797712517158\\
1231	5.97742646177392\\
1232	3.77676418112789\\
1233	5.02555948224991\\
1234	1.53162036684861\\
1235	3.8191952657351\\
1236	3.58090998476686\\
1237	6.15139548307178\\
1238	5.08938007353429\\
1239	5.3403208247288\\
1240	3.30572303829659\\
1241	2.07898399188661\\
1242	3.37417156975142\\
1243	2.7765602365408\\
1244	0.194583317860195\\
1245	3.35139118490515\\
1246	6.72784214793579\\
1247	5.37044577383488\\
1248	4.5575043683742\\
1249	-1.24088532914617\\
1250	-0.124346126775705\\
1251	3.50980637218577\\
1252	3.50062920027697\\
1253	2.88766500400996\\
1254	3.96177641781445\\
1255	2.83499794322088\\
1256	3.57452467389289\\
1257	5.8215850894523\\
1258	4.04360920189588\\
1259	2.78083775743371\\
1260	-0.966324693574231\\
1261	-0.327678069259437\\
1262	-0.455764062517115\\
1263	4.07455306567613\\
1264	3.36379765880703\\
1265	4.02952646080143\\
1266	-2.27139529803188\\
1267	5.04733286616506\\
1268	5.43487743202249\\
1269	0.8239856149664\\
1270	3.82175081626142\\
1271	1.34122778566882\\
1272	-1.39374181346834\\
1273	1.79586501233891\\
1274	3.45951961016038\\
1275	1.24798041064083\\
1276	3.2453794169826\\
1277	1.42468365610963\\
1278	-0.322109440864047\\
1279	1.76380190432172\\
1280	2.00183440667817\\
1281	3.03105520889546\\
1282	-0.23545168700648\\
1283	2.23706558572823\\
1284	1.89123796262083\\
1285	1.76297417575817\\
1286	2.57438639214903\\
1287	-2.39707558673806\\
1288	-2.45709704974935\\
1289	0.179055489309519\\
1290	0.141198161191298\\
1291	0.946055663504316\\
1292	0.130198786739131\\
1293	-0.117036790694574\\
1294	-0.158456471845629\\
1295	-0.915372799437112\\
1296	2.48799851563428\\
1297	-0.488161695168458\\
1298	3.24135579216882\\
1299	3.79062150575167\\
1300	3.32120443206466\\
1301	4.54943003700729\\
1302	0.365722101955081\\
1303	0.707680587106881\\
1304	1.17499890277542\\
1305	-0.0572607351425027\\
1306	-1.96680341413088\\
1307	-0.0761810224415446\\
1308	1.78638318024729\\
1309	-0.13087356745037\\
1310	0.822230101445286\\
1311	2.94455201911367\\
1312	-0.00186150761222235\\
1313	0.280584602676104\\
1314	-2.22371937099287\\
1315	0.6379431297598\\
1316	0.820737679792493\\
1317	1.31923809975923\\
1318	-0.533391134564633\\
1319	0.793657703337207\\
1320	-2.0241597864954\\
1321	-1.89954292217884\\
1322	1.613893884293\\
1323	1.20789010103289\\
1324	-1.93297617770359\\
1325	-1.33483990106472\\
1326	-5.40816828694285\\
1327	-1.85752804239311\\
1328	0.85850287153013\\
1329	1.481994808593\\
1330	-1.04755448266477\\
1331	-2.78887975459146\\
1332	-3.29969287232271\\
1333	1.12814756277367\\
1334	0.568804912571268\\
1335	-5.02063598189211\\
1336	0.76951668283089\\
1337	0.0706937018103572\\
1338	-1.33284609497394\\
1339	-0.667299308258011\\
1340	0.136782246170127\\
1341	1.58925331348643\\
1342	2.00090762093894\\
1343	-0.0762133301955832\\
1344	-3.78750430708344\\
1345	0.509660029998975\\
1346	1.65244082807938\\
1347	-0.273432121559997\\
1348	-3.61715060370462\\
1349	0.168020189190927\\
1350	-1.30744934642833\\
1351	-0.276508100508281\\
1352	-2.16577622947468\\
1353	-2.93493432601923\\
1354	-0.409528359011763\\
1355	-1.79408858125004\\
1356	-2.65690052322101\\
1357	-0.480023231306746\\
1358	-2.57699171654473\\
1359	-0.979368705526109\\
1360	-0.927775880894442\\
1361	-2.39085644206323\\
1362	-2.21122390065687\\
1363	-3.36494088061891\\
1364	-2.85855545095833\\
1365	-1.51157002647206\\
1366	0.188768514314021\\
1367	-2.03545886119704\\
1368	-0.162683082966458\\
1369	-3.06129716180347\\
1370	-2.67907101796924\\
1371	1.31295975744636\\
1372	-3.59837537596137\\
1373	-2.22193877598905\\
1374	-5.59815628380813\\
1375	-3.25853478324443\\
1376	-3.47772770613114\\
1377	-2.81714503380023\\
1378	-3.28162351921659\\
1379	-1.91906177137224\\
1380	-3.19955997606997\\
1381	-1.21633309789009\\
1382	0.22447396918418\\
1383	-3.28700470677788\\
1384	-7.4855594810081\\
1385	0.449416076177943\\
1386	-4.91465709195048\\
1387	-3.44331200416519\\
1388	-3.13177916022037\\
1389	-6.95892639061762\\
1390	-6.09503125517722\\
1391	-1.73629558604975\\
1392	-1.26029263021865\\
1393	-0.16596078644639\\
1394	-1.25327455408649\\
1395	-0.0396300536176732\\
1396	-1.97419636808722\\
1397	-1.66119997600112\\
1398	-0.927257380725807\\
1399	-3.65778697936788\\
1400	-1.79762754705003\\
1401	-1.86580482081072\\
1402	-2.46110193683666\\
1403	-1.88953325493568\\
1404	-2.60983226385063\\
1405	-6.06521285450817\\
1406	-4.97830608191335\\
1407	-4.80115020576039\\
1408	-4.7513348125527\\
1409	-4.69744556841501\\
1410	-8.3835443908556\\
1411	-3.89543695563605\\
1412	0.101222404536171\\
1413	-3.01957761332401\\
1414	-4.02550583990011\\
1415	-3.01230986009912\\
1416	-8.80588048638126\\
1417	-6.36128407356906\\
1418	-3.24203151301997\\
1419	-4.07213659695102\\
1420	-3.04256543938641\\
1421	-4.25342748967498\\
1422	-6.10522949079215\\
1423	-3.21706792449271\\
1424	-6.89741663646122\\
1425	-4.23935728334352\\
1426	-4.56789450965178\\
1427	-2.51068618422154\\
1428	-3.3233713120069\\
1429	-3.29306230556427\\
1430	-2.77719022903888\\
1431	-3.16187347465573\\
1432	-2.58615911425764\\
1433	-5.71171077341578\\
1434	-5.09731890505289\\
1435	-5.74631163462385\\
1436	-5.73104629023422\\
1437	-4.99991767226234\\
1438	-4.86860742323155\\
1439	-8.61307166379648\\
1440	-6.42858565975131\\
1441	-5.79749708067832\\
1442	-4.35008104560063\\
1443	-4.95522181224289\\
1444	-6.75843087842796\\
1445	-7.26661658013255\\
1446	-7.94440448125809\\
1447	-4.86566661923769\\
1448	-5.5980486436326\\
1449	-5.15647274139723\\
1450	-5.41032037768647\\
1451	-4.06542508462913\\
1452	-7.01352536776752\\
1453	-8.01167870632656\\
1454	-6.57342839171013\\
1455	-3.2792972742253\\
1456	-5.85178755242075\\
1457	-8.42011265696091\\
1458	-10.3254589964216\\
1459	-4.67145793458363\\
1460	-4.50757901581795\\
1461	-6.43174033050377\\
1462	-9.00413975588559\\
1463	-4.17955266472704\\
1464	-8.1083943483245\\
1465	-7.85788037744724\\
1466	-9.25758898675263\\
1467	-5.47946359403651\\
1468	-5.59860705025243\\
1469	-7.47846767655084\\
1470	-4.06877130668093\\
1471	-4.377351376409\\
1472	-8.24965036104329\\
1473	-9.97358297138639\\
1474	-6.84685168815638\\
1475	-6.34003740776468\\
1476	-5.87134215607234\\
1477	-6.28520026471158\\
1478	-6.35513114061628\\
1479	-7.31869510268626\\
1480	-6.82536170986889\\
1481	-3.65416690801706\\
1482	-2.2261821435576\\
1483	-4.3317234807608\\
1484	-4.71911731596798\\
1485	-9.22737070491746\\
1486	-6.52034930690103\\
1487	-9.07862567733779\\
1488	-6.38879641521742\\
1489	-8.2617761875361\\
1490	-6.57267370526278\\
1491	-4.6048426648752\\
1492	-6.34131666169084\\
1493	-8.51244099596756\\
1494	-6.41473096613716\\
1495	-7.96085465346889\\
1496	-7.39212954209561\\
1497	-7.75358157667122\\
1498	-7.00532684909674\\
1499	-7.65575424033349\\
1500	-6.04358107938364\\
1501	-5.98717398157601\\
1502	-7.97032549932136\\
1503	-7.65993695937744\\
1504	-6.40684826909407\\
1505	-5.64775583579736\\
1506	-9.84638387793779\\
1507	-9.02985304016173\\
1508	-7.50982296247002\\
1509	-4.6951700581836\\
1510	-7.61313331647639\\
1511	-7.50602243904653\\
1512	-8.15446778868171\\
1513	-5.30905713327715\\
1514	-5.61631186376278\\
1515	-5.23823594797489\\
1516	-8.46966114828097\\
1517	-9.53820761468435\\
1518	-8.77034744034689\\
1519	-7.6671989610444\\
1520	-6.55998117314792\\
1521	-6.3271412312293\\
1522	-7.91457210121257\\
1523	-10.455784868496\\
1524	-7.80697156470227\\
1525	-7.09923559757614\\
1526	-8.43894456223423\\
1527	-6.36829559842325\\
1528	-6.52576091504732\\
1529	-8.93217963126405\\
1530	-7.06148417859772\\
1531	-8.22577411452356\\
1532	-7.53737482281596\\
1533	-8.97452411567804\\
1534	-6.4901753777831\\
1535	-6.31966365773776\\
1536	-5.56725271565149\\
1537	-7.06946977746233\\
1538	-6.74115462312994\\
1539	-9.62376059904723\\
1540	-7.75773658352706\\
1541	-9.60619416364743\\
1542	-7.26870325447232\\
1543	-6.32736268613744\\
1544	-4.43079358937429\\
1545	-5.27498913561116\\
1546	-6.34114743449741\\
1547	-5.22540834483118\\
1548	-6.22564170241394\\
1549	-8.30501964808567\\
1550	-7.74434008353382\\
1551	-7.33922058575008\\
1552	-8.48283130188469\\
1553	-9.17565085802649\\
1554	-7.33383516726477\\
1555	-6.14975910943811\\
1556	-6.90127399438455\\
1557	-10.2319699170171\\
1558	-10.3469840980932\\
1559	-5.81648694911989\\
1560	-6.60259122545568\\
1561	-8.94642263152119\\
1562	-7.60239923384277\\
1563	-5.26672740066216\\
1564	-7.97953318324191\\
1565	-7.14345726064536\\
1566	-6.2651357094853\\
1567	-9.43640581518109\\
1568	-7.9150408233319\\
1569	-5.2543008328821\\
1570	-4.59297623561988\\
1571	-6.72033293004904\\
1572	-7.73489640504416\\
1573	-3.84770374336043\\
1574	-5.18377438524758\\
1575	-7.62952631538958\\
1576	-9.39799522139756\\
1577	-10.0577038455756\\
1578	-8.67386973778766\\
1579	-6.7464609632788\\
1580	-7.27510635594359\\
1581	-9.03697463115063\\
1582	-9.88058612015014\\
1583	-8.28545004258063\\
1584	-7.05374485455817\\
1585	-6.38684686068499\\
1586	-3.65047002339742\\
1587	-4.11133758672715\\
1588	-10.0975938387228\\
1589	-9.21653268531299\\
1590	-10.5844542174471\\
1591	-7.26919977710978\\
1592	-4.4556040543246\\
1593	-7.84823350599916\\
1594	-9.61817203987116\\
1595	-9.86847110021312\\
1596	-8.67926150503329\\
1597	-8.42284680583025\\
1598	-9.74207399056516\\
1599	-5.8044592492102\\
1600	-3.90762792052824\\
1601	-8.32284485062418\\
1602	-7.17096426181843\\
1603	-7.07718161231175\\
1604	-10.5874016514578\\
1605	-6.01002668964367\\
1606	-6.94438491564673\\
1607	-10.4019797547668\\
1608	-8.37364537782277\\
1609	-7.57897090978488\\
1610	-9.39639854375974\\
1611	-9.91643913257976\\
1612	-9.17070586969602\\
1613	-8.15392174971858\\
1614	-8.7487919863836\\
1615	-7.16384960080075\\
1616	-6.77971553027259\\
1617	-8.50589072911285\\
1618	-9.15935177595513\\
1619	-7.51771176133289\\
1620	-10.9440219447464\\
1621	-8.35968152556721\\
1622	-7.81041265175858\\
1623	-9.157062554973\\
1624	-8.47827726934152\\
1625	-8.7832605307905\\
1626	-7.2305276446926\\
1627	-7.65759109264445\\
1628	-10.4479651408101\\
1629	-8.69542170852633\\
1630	-7.7334890371911\\
1631	-5.48333128612585\\
1632	-6.42522602850765\\
1633	-7.75736616502543\\
1634	-9.7614488337792\\
1635	-7.15126351742656\\
1636	-5.68193446506099\\
1637	-9.65876027303091\\
1638	-8.88689488380844\\
1639	-6.93820522920891\\
1640	-4.73887440745454\\
1641	-10.603019029897\\
1642	-6.99686629903032\\
1643	-7.97497995561225\\
1644	-3.50545424487776\\
1645	-5.67089582237386\\
1646	-7.71377552312796\\
1647	-9.77899854273395\\
1648	-9.60899728783563\\
1649	-6.25996699755295\\
1650	-7.0664542316392\\
1651	-7.25072099991325\\
1652	-7.97547630933477\\
1653	-9.16273060901328\\
1654	-9.52726327014241\\
1655	-8.8718138441604\\
1656	-6.86144915460131\\
1657	-6.73895319946206\\
1658	-5.03809121875941\\
1659	-4.11744159211716\\
1660	-6.11874954640968\\
1661	-6.65736246290439\\
1662	-6.3450458073624\\
1663	-7.35174585668757\\
1664	-7.55532909794535\\
1665	-8.80226738716052\\
1666	-8.78383046081009\\
1667	-9.58764500154701\\
1668	-5.30425111157187\\
1669	-7.03337370474996\\
1670	-9.9653304065161\\
1671	-7.78276367226226\\
1672	-5.93187835790974\\
1673	-4.50414070645318\\
1674	-5.08172611464391\\
1675	-8.17300238328054\\
1676	-5.83437878080088\\
1677	-5.71598730499747\\
1678	-7.47657029266018\\
1679	-6.4103584987644\\
1680	-6.59917931624267\\
1681	-6.679247089988\\
1682	-8.00194483322446\\
1683	-10.2075781996187\\
1684	-11.6232655842035\\
1685	-7.49725289426335\\
1686	-8.5213924794612\\
1687	-5.78094704960806\\
1688	-8.67298593201348\\
1689	-9.25020697262085\\
1690	-7.41803816843307\\
1691	-4.53073137827588\\
1692	-5.01010543977258\\
1693	-6.14127376498828\\
1694	-9.13267679364873\\
1695	-10.3460021489234\\
1696	-9.7653482838425\\
1697	-7.20707669853013\\
1698	-6.09519915026807\\
1699	-7.17656493383296\\
1700	-7.62223282191473\\
1701	-7.96541777398455\\
1702	-8.1166617106476\\
1703	-11.9773315660259\\
1704	-12.1431376787806\\
1705	-10.6793458308781\\
1706	-6.2606912927486\\
1707	-6.494687558455\\
1708	-8.41532959107229\\
1709	-8.21394051103609\\
1710	-9.31590109288161\\
1711	-8.59662079918789\\
1712	-10.3807007590253\\
1713	-8.9290467698637\\
1714	-9.5650148500183\\
1715	-8.71488737651731\\
1716	-10.5503690382642\\
1717	-9.51963787740017\\
1718	-5.42732533468469\\
1719	-4.98744691841587\\
1720	-9.75568101841437\\
1721	-8.35674524897199\\
1722	-4.79974933775153\\
1723	-6.82974030879001\\
1724	-4.60571827213727\\
1725	-4.79870138445056\\
1726	-8.00943719992606\\
1727	-10.5940945282284\\
1728	-9.72436137863597\\
1729	-10.1123653900933\\
1730	-12.3589720161733\\
1731	-8.63151036272398\\
1732	-7.25070094633042\\
1733	-8.90124097195191\\
1734	-8.33169917366208\\
1735	-7.86459204400039\\
1736	-11.5433794718596\\
1737	-8.43783484813121\\
1738	-6.9776605635807\\
1739	-5.99480429562113\\
1740	-7.20359030775536\\
1741	-6.30812845825105\\
1742	-9.53193235067473\\
1743	-7.95081936210445\\
1744	-4.79551076770778\\
1745	-4.74445686643021\\
1746	-8.1645677534364\\
1747	-8.64117284874343\\
1748	-12.6763977786933\\
1749	-10.4394739526102\\
1750	-7.31077024144923\\
1751	-6.27099781443937\\
1752	-7.05347339455649\\
1753	-9.13910158036126\\
1754	-10.444538197915\\
1755	-10.1889201255712\\
1756	-8.4175172077821\\
1757	-6.41112261045118\\
1758	-6.00887354324892\\
1759	-8.13556867284471\\
1760	-4.70831146994537\\
1761	-2.44536139785341\\
1762	-6.0629063668182\\
1763	-8.11434418658936\\
1764	-7.93120296939592\\
1765	-4.58569867238726\\
1766	-3.54789576090276\\
1767	-6.44321865940777\\
1768	-5.88343404931594\\
1769	-5.71161709322657\\
1770	-9.38264357038835\\
1771	-10.9401861313478\\
1772	-11.4952923243731\\
1773	-8.75322834557715\\
1774	-7.27739160511206\\
1775	-6.99414748121287\\
1776	-6.01569766320205\\
1777	-7.31392904420346\\
1778	-9.48026947817827\\
1779	-7.42047812560634\\
1780	-9.32760676984038\\
1781	-7.15200205153888\\
1782	-7.41484462047314\\
1783	-7.62390134922509\\
1784	-9.82861697410283\\
1785	-6.52964890043197\\
1786	-5.00382021969571\\
1787	-6.63286219224116\\
1788	-6.58635866709253\\
1789	-7.03958911237107\\
1790	-8.25373239094479\\
1791	-4.69531964926231\\
1792	-7.01678908260719\\
1793	-7.85113857563233\\
1794	-6.40142310474032\\
1795	-6.68801026319868\\
1796	-6.43186100324472\\
1797	-4.11396949099401\\
1798	-5.16320699326329\\
1799	-6.25991076992947\\
1800	-7.20197640361945\\
1801	-8.90630901817092\\
1802	-8.89790461419973\\
1803	-5.52198875025024\\
1804	-8.37768927582705\\
1805	-6.38182192718733\\
1806	-8.78102719653024\\
1807	-8.11703236286048\\
1808	-6.54571970324722\\
1809	-7.52082607650816\\
1810	-6.94302653395719\\
1811	-6.3490152215031\\
1812	-5.24268298703437\\
1813	-5.15911486308619\\
1814	-7.07932159045348\\
1815	-3.99551016932588\\
1816	-5.58721034934908\\
1817	-5.96665408873334\\
1818	-6.80927035938774\\
1819	-6.05705302628151\\
1820	-5.51354621214992\\
1821	-7.82768465923482\\
1822	-8.88384489763206\\
1823	-7.69797091988034\\
1824	-8.91649815100458\\
1825	-8.16303172074409\\
1826	-5.24980634734225\\
1827	-10.3585457357582\\
1828	-6.73709725389952\\
1829	-2.81161049919597\\
1830	-6.08558941948347\\
1831	-7.13997783203636\\
1832	-11.1470177918072\\
1833	-8.58130124955877\\
1834	-8.34259037803301\\
1835	-6.87784661086879\\
1836	-9.73789098629112\\
1837	-9.83347855783737\\
1838	-9.84180824976064\\
1839	-10.0112706100862\\
1840	-9.18926994998314\\
1841	-5.43874141713787\\
1842	-8.52567988353419\\
1843	-8.3575177569161\\
1844	-7.36458612181856\\
1845	-7.15170851260566\\
1846	-5.36007110666462\\
1847	-4.94885531205415\\
1848	-8.19565464797033\\
1849	-8.12363665821702\\
1850	-7.85217515494653\\
1851	-8.35721899266787\\
1852	-7.47333258388218\\
1853	-7.89015850351622\\
1854	-5.47988701933616\\
1855	-6.68603565317629\\
1856	-11.9275701959486\\
1857	-9.54423179645762\\
1858	-9.1662149239827\\
1859	-6.11763087636201\\
1860	-5.81385709017625\\
1861	-8.97166910872709\\
1862	-8.31361314820467\\
1863	-11.2715930181505\\
1864	-8.09647051594226\\
1865	-6.98781132525108\\
1866	-6.32567749928286\\
1867	-4.77809455610034\\
1868	-7.32987571513406\\
1869	-8.55981193110847\\
1870	-6.87944760086385\\
1871	-6.26848328844615\\
1872	-6.38210027638695\\
1873	-6.41283107677018\\
1874	-6.88964027215813\\
1875	-7.73201928141863\\
1876	-12.371318715205\\
1877	-10.320009122735\\
1878	-7.86189401400131\\
1879	-8.0248210120086\\
1880	-7.22397234361858\\
1881	-8.09514756722679\\
1882	-8.83161546092482\\
1883	-5.27118710985344\\
1884	-7.12462783048799\\
1885	-6.1815928538378\\
1886	-6.1336300194123\\
1887	-5.57622658574133\\
1888	-9.95296996701783\\
1889	-9.67985679280747\\
1890	-7.99705509832036\\
1891	-5.14253625702733\\
1892	-8.06558241192252\\
1893	-4.69882063898553\\
1894	-5.48039991094263\\
1895	-9.83987651430976\\
1896	-6.92260309405722\\
1897	-4.96270603385146\\
1898	-4.69260048742007\\
1899	-6.17967201671192\\
1900	-6.73744806647049\\
1901	-7.43758663020813\\
1902	-7.92795195180122\\
1903	-7.10682086672363\\
1904	-6.62255550069398\\
1905	-5.43667448194828\\
1906	-7.34302174714965\\
1907	-7.722070407762\\
1908	-8.16884464872912\\
1909	-6.10576424234004\\
1910	-8.79194138464761\\
1911	-8.14869899878323\\
1912	-9.45407206635119\\
1913	-7.48062697612592\\
1914	-6.5199266208009\\
1915	-7.03599147256765\\
1916	-8.91492967571529\\
1917	-10.7667034843743\\
1918	-9.01325812863041\\
1919	-8.59933976413441\\
1920	-5.92428965049651\\
1921	-10.4075662805229\\
1922	-9.63439172259162\\
1923	-8.04207464941186\\
1924	-8.48650385049093\\
1925	-6.86516142422695\\
1926	-8.8108091124639\\
1927	-8.78007554584582\\
1928	-9.80480001654094\\
1929	-5.8035034752863\\
1930	-8.63074513676264\\
1931	-7.9739702941749\\
1932	-5.36411796967238\\
1933	-5.24734190423077\\
1934	-8.5129494161603\\
1935	-7.5338845170807\\
1936	-8.80604417355574\\
1937	-10.3620558005131\\
1938	-7.24588714552347\\
1939	-5.53878764798658\\
1940	-5.86631655217133\\
1941	-6.89887072330377\\
1942	-10.0467963078052\\
1943	-6.90356271798855\\
1944	-7.99963497636702\\
1945	-8.60600007417261\\
1946	-5.49498915013351\\
1947	-6.4156139339162\\
1948	-6.63900174063321\\
1949	-8.82621283782698\\
1950	-8.15069275208085\\
1951	-7.34374804881944\\
1952	-4.82235174354018\\
1953	-7.91885525858167\\
1954	-8.46941549054895\\
1955	-6.73220716969869\\
1956	-9.12506320800363\\
1957	-10.9592191261994\\
1958	-7.39691423675954\\
1959	-8.39845177586304\\
1960	-6.70171762753183\\
1961	-4.48880669447273\\
1962	-5.06345542531484\\
1963	-6.38216084543317\\
1964	-6.54414069398039\\
1965	-7.03755591669965\\
1966	-9.87086761186909\\
1967	-8.49985955237221\\
1968	-4.10465247290209\\
1969	-7.30691100005008\\
1970	-8.43283444235364\\
1971	-9.62295681625991\\
1972	-11.666255654745\\
1973	-9.74337611072058\\
1974	-5.75765436976034\\
1975	-4.47141607170561\\
1976	-5.66718960733365\\
1977	-10.5180749353381\\
1978	-4.48936225763055\\
1979	-5.41841379345232\\
1980	-9.72064128576485\\
1981	-7.579348348566\\
1982	-8.04563504273011\\
1983	-7.31109247432775\\
1984	-7.71770449287576\\
1985	-9.17048640152296\\
1986	-8.09248248745374\\
1987	-8.29278537446599\\
1988	-5.13298430201757\\
1989	-6.87466329426485\\
1990	-6.71256507513042\\
1991	-6.41164060402272\\
1992	-9.81602602416558\\
1993	-9.07367634824785\\
1994	-10.0654227612544\\
1995	-8.36788915664958\\
1996	-7.1542037783166\\
1997	-3.06657125844294\\
1998	-8.8446782418452\\
1999	-10.8037316070038\\
2000	-9.12932675429902\\
2001	-7.68664455341626\\
2002	-7.58458297002165\\
2003	-9.6170968444254\\
2004	-7.54965730725969\\
2005	-7.127552658241\\
2006	-4.21039780591278\\
2007	-4.7740167967165\\
2008	-6.10390579626851\\
2009	-8.64164288510125\\
2010	-7.66834945954971\\
2011	-8.17390559862078\\
2012	-4.31492122393397\\
2013	-6.53964959364535\\
2014	-8.6514231539725\\
2015	-7.16529720831675\\
2016	-8.13985844155295\\
2017	-3.71107842628751\\
2018	-7.31767115884153\\
2019	-10.0204776633962\\
2020	-6.35746294187159\\
2021	-9.00088000828907\\
2022	-5.89892283837265\\
2023	-8.80974560367395\\
2024	-7.50961761526949\\
2025	-7.51097213500946\\
2026	-7.9979322406411\\
2027	-8.18155707316126\\
2028	-9.9894024424025\\
2029	-7.5720484594501\\
2030	-7.2812109951187\\
2031	-6.15012346044514\\
2032	-6.12804204562085\\
2033	-5.86067894135616\\
2034	-6.99425349200059\\
2035	-7.06884912527365\\
2036	-7.58666697955976\\
2037	-15.5320088488871\\
2038	-5.67101987150951\\
2039	-7.43957934407391\\
2040	-5.91055766718394\\
2041	-5.07450367337564\\
2042	-7.7894775684073\\
2043	-7.30260599750275\\
2044	-9.29150517574723\\
2045	-7.32068742883157\\
2046	-7.50664473280126\\
2047	-6.78649389393558\\
};
\draw [dashed] (axis cs:498,-20) -- (axis cs:498,25);
\end{axis}
\end{tikzpicture}%

%% file: figures/PTWOLA_Results_Sys_Distance_L512_SWDFT.tex
\definecolor{mycolor1}{rgb}{0.34667,0.53600,0.69067}%
\definecolor{mycolor2}{rgb}{0.91529,0.28157,0.28784}%
\definecolor{mycolor3}{rgb}{0.44157,0.74902,0.43216}%
\begin{tikzpicture}[font=\scriptsize]

\begin{axis}[%
  width=\fwidth,
  height=\fheight,
  at={(0,0)},
scale only axis,
xmin=1,
xmax=103,
xlabel={Subband filter length ($T$)},
ymin=-40.5,
ymax=-10,
ylabel={System distance (dB)},
axis background/.style={fill=white},
axis x line*=bottom,
axis y line*=left,
legend columns = -1,
legend style={at={(1,1)}, anchor=north east, legend cell align=left, align=left, draw=white!15!black},
xmajorgrids,
ymajorgrids,
mark repeat=10,
mark size=2.0pt
]
\addplot [line width=\plotlinewidth, color=mycolor1, dashed, mark=o, mark options={solid, mycolor1},forget plot]
  table[row sep=crcr]{%
1	-13.4587196065864\\
3	-13.5754035656404\\
5	-13.7011200564878\\
7	-13.8223474220223\\
9	-13.9462819109275\\
11	-14.0638872286439\\
13	-14.1865835494701\\
15	-14.3021951306689\\
17	-14.4225582331044\\
19	-14.5518638489009\\
21	-14.670489369051\\
23	-14.786428570941\\
25	-14.9118158964219\\
27	-15.0293868154256\\
29	-15.1550488223824\\
31	-15.2742276468197\\
33	-15.4000846438492\\
35	-15.5191535122633\\
37	-15.6392850467414\\
39	-15.763056638304\\
41	-15.8898101962954\\
43	-16.0111701499599\\
45	-16.1284115552996\\
47	-16.2502258287646\\
49	-16.3785502732181\\
51	-16.4933222139109\\
53	-16.6191886239435\\
55	-16.7406364426911\\
57	-16.8704204665378\\
59	-16.994065941414\\
61	-17.1118113212202\\
63	-17.235838086001\\
65	-17.3607517855996\\
67	-17.4909976514935\\
69	-17.6122313678983\\
71	-17.7397783137122\\
73	-17.8670014545814\\
75	-17.9929443882177\\
77	-18.1147802290531\\
79	-18.2342446651917\\
81	-18.3570598985673\\
83	-18.4768164210762\\
85	-18.6030669269071\\
87	-18.7226086974316\\
89	-18.8442186368036\\
91	-18.9737574834186\\
93	-19.0906490578651\\
95	-19.2170780697319\\
97	-19.3400731106065\\
99	-19.4642259547447\\
101	-19.5914731016378\\
103	-19.7158007613096\\
};

\addplot [line width=\plotlinewidth, color=mycolor2, dashed, mark=+, mark options={solid, mycolor2},forget plot]
  table[row sep=crcr]{%
1	-14.0790122472382\\
3	-26.5073195567979\\
5	-26.6795038917276\\
7	-26.8397475985811\\
9	-26.9924854923666\\
11	-27.1494198929576\\
13	-27.3130590260134\\
15	-27.4714865001324\\
17	-27.6237171316183\\
19	-27.7821093250745\\
21	-27.9414518774644\\
23	-28.1032849313458\\
25	-28.259625695607\\
27	-28.4257509535319\\
29	-28.5864579920236\\
31	-28.750347121733\\
33	-28.9217337920472\\
35	-29.0786547081155\\
37	-29.2446119602759\\
39	-29.407905498594\\
41	-29.5730130875218\\
43	-29.7453325809824\\
45	-29.9080824975072\\
47	-30.0917495468755\\
49	-30.2592464674069\\
51	-30.4369740920731\\
53	-30.6141310179815\\
55	-30.7847518990462\\
57	-30.9580699885882\\
59	-31.1358978746762\\
61	-31.3109390465971\\
63	-31.48727923889\\
65	-31.6633271110899\\
67	-31.8433282185703\\
69	-32.0265098152589\\
71	-32.2048578073333\\
73	-32.3854227775524\\
75	-32.5693594375306\\
77	-32.7537744020858\\
79	-32.9434428156013\\
81	-33.1276017747077\\
83	-33.3191648144781\\
85	-33.5125187546844\\
87	-33.7046169573766\\
89	-33.8987485048439\\
91	-34.0950179670604\\
93	-34.3010149227469\\
95	-34.4989304923481\\
97	-34.6971983754021\\
99	-34.8923755321484\\
101	-35.1016648203301\\
103	-35.3203981737969\\
};

\addplot [line width=\plotlinewidth, color=mycolor3, dashed, mark=asterisk, mark options={solid, mycolor3},forget plot]
  table[row sep=crcr]{%
1	-15.8906163797517\\
3	-23.5081935984187\\
5	-23.6479032973819\\
7	-23.7857555210089\\
9	-23.9235256048296\\
11	-24.0643896204391\\
13	-24.2077912817714\\
15	-24.3496546418825\\
17	-24.4921283922604\\
19	-24.6287341229843\\
21	-24.7703369860225\\
23	-24.9179331247024\\
25	-25.0570523218152\\
27	-25.2050618503672\\
29	-25.348088784424\\
31	-25.4911079770535\\
33	-25.6356254006705\\
35	-25.7836332182901\\
37	-25.9310878018384\\
39	-26.0685330052848\\
41	-26.2157155515009\\
43	-26.360637081385\\
45	-26.5083269822498\\
47	-26.6600806184157\\
49	-26.8072558516674\\
51	-26.9585516251616\\
53	-27.1150597489795\\
55	-27.2601424833791\\
57	-27.4108263359233\\
59	-27.5718599980639\\
61	-27.7255500635007\\
63	-27.8704893897552\\
65	-28.0297492264569\\
67	-28.1808150795808\\
69	-28.3372286673209\\
71	-28.487604527562\\
73	-28.6377954932022\\
75	-28.7909702637052\\
77	-28.9502658243557\\
79	-29.1044704830336\\
81	-29.2586111435594\\
83	-29.4238526666688\\
85	-29.5754288574613\\
87	-29.7371418033053\\
89	-29.8971756202706\\
91	-30.0592661431734\\
93	-30.2205606589745\\
95	-30.387260266831\\
97	-30.5537614325073\\
99	-30.7171835150868\\
101	-30.8796502408411\\
103	-31.0499906866465\\
};

\addplot [line width=\plotlinewidth, color=mycolor1, mark=o, mark options={solid, mycolor1}]
  table[row sep=crcr]{%
1	-19.2323061728234\\
3	-19.3541588964823\\
5	-19.4833280550243\\
7	-19.608502436152\\
9	-19.7325967239594\\
11	-19.8551805473424\\
13	-19.9810409238207\\
15	-20.0996702883644\\
17	-20.2243200880446\\
19	-20.3662335299628\\
21	-20.4841961855996\\
23	-20.5994595496521\\
25	-20.7223755337123\\
27	-20.8351994105552\\
29	-20.9672974437964\\
31	-21.0915720811586\\
33	-21.2208921359777\\
35	-21.3397362481112\\
37	-21.4615363329108\\
39	-21.5920463604585\\
41	-21.7242554955106\\
43	-21.8449503360666\\
45	-21.962547192139\\
47	-22.0832215586085\\
49	-22.2143183093924\\
51	-22.3215569502543\\
53	-22.4514364435285\\
55	-22.5709401236675\\
57	-22.7092024264745\\
59	-22.8352656406105\\
61	-22.9503142754827\\
63	-23.0749079619856\\
65	-23.2070213236576\\
67	-23.3463542333278\\
69	-23.4696955093389\\
71	-23.6074098307469\\
73	-23.743383394608\\
75	-23.8671326608657\\
77	-23.9946162964724\\
79	-24.1152866888946\\
81	-24.2376896818312\\
83	-24.3523905648951\\
85	-24.4776630438982\\
87	-24.5956487792977\\
89	-24.7152495883205\\
91	-24.8467152270977\\
93	-24.9524195461243\\
95	-25.0777944929268\\
97	-25.2039671840594\\
99	-25.3296916715794\\
101	-25.465323619217\\
103	-25.5900005681586\\
};
\addlegendentry{\rectwin{A}}

\addplot [line width=\plotlinewidth, color=mycolor2, mark=+, mark options={solid, mycolor2}]
  table[row sep=crcr]{%
1	-18.9656918816061\\
3	-28.4854718200486\\
5	-28.6647258014531\\
7	-28.8295807014382\\
9	-28.986721915529\\
11	-29.1472849085448\\
13	-29.3159133684264\\
15	-29.4781865336636\\
17	-29.6356580142755\\
19	-29.7989094229376\\
21	-29.9627847036679\\
23	-30.1286901646725\\
25	-30.2885808323716\\
27	-30.4588504479489\\
29	-30.6245202853033\\
31	-30.7926033108444\\
33	-30.9673031573267\\
35	-31.1292682247113\\
37	-31.2992693988777\\
39	-31.4663994139542\\
41	-31.6349299434256\\
43	-31.8087219804692\\
45	-31.9745965965068\\
47	-32.1613112904033\\
49	-32.330582289692\\
51	-32.5106744558917\\
53	-32.6865821862853\\
55	-32.8596884679792\\
57	-33.0341147058673\\
59	-33.2142219822189\\
61	-33.3923838120499\\
63	-33.5728744596989\\
65	-33.7532659974692\\
67	-33.9357648586547\\
69	-34.1221939417959\\
71	-34.3019286587757\\
73	-34.4840732083807\\
75	-34.6673222634845\\
77	-34.8517117295179\\
79	-35.0403770568852\\
81	-35.2259071191344\\
83	-35.4175622102763\\
85	-35.6144322051123\\
87	-35.807192934725\\
89	-35.9999835206199\\
91	-36.196577188769\\
93	-36.4026055688914\\
95	-36.6008072065242\\
97	-36.7954280215444\\
99	-36.9912379227306\\
101	-37.2016779616057\\
103	-37.4186572882108\\
};
\addlegendentry{\cosinewin{A}}

\addplot [line width=\plotlinewidth, color=mycolor3, mark=asterisk, mark options={solid, mycolor3}]
  table[row sep=crcr]{%
1	-20.5873844102935\\
3	-30.7767170486522\\
5	-30.9350143791559\\
7	-31.0982576567043\\
9	-31.2622895981837\\
11	-31.4343642365596\\
13	-31.6093754023123\\
15	-31.7735647425291\\
17	-31.9436594177221\\
19	-32.1076516065724\\
21	-32.27676421729\\
23	-32.4471968501498\\
25	-32.6252599714863\\
27	-32.8018041235641\\
29	-32.9831918847913\\
31	-33.155074195549\\
33	-33.3412168097529\\
35	-33.5127175840141\\
37	-33.6884174539359\\
39	-33.8636727638502\\
41	-34.0427949616877\\
43	-34.2214488489156\\
45	-34.4082125452434\\
47	-34.6040381070202\\
49	-34.7813111313772\\
51	-34.973141501704\\
53	-35.161115350016\\
55	-35.3412713154899\\
57	-35.5253957284362\\
59	-35.723441885986\\
61	-35.9084162622483\\
63	-36.0966382220227\\
65	-36.2909598328905\\
67	-36.4790994094284\\
69	-36.6687617251237\\
71	-36.847065991342\\
73	-37.0378650246919\\
75	-37.2365159046258\\
77	-37.4456336587406\\
79	-37.6341599428056\\
81	-37.8323363355223\\
83	-38.0164080296888\\
85	-38.2259921989622\\
87	-38.4317433628469\\
89	-38.6251569726588\\
91	-38.8414860360039\\
93	-39.0507750480336\\
95	-39.2717940262565\\
97	-39.4846965966713\\
99	-39.6938803025636\\
101	-39.8963494070308\\
103	-40.1123808273917\\
};
\addlegendentry{\roothannwin{A}}
\end{axis}
\end{tikzpicture}%

%% file: figures/PTWOLA_Results_L512_SWDFT.tex
\definecolor{mycolor1}{rgb}{0.34667,0.53600,0.69067}%
\definecolor{mycolor2}{rgb}{0.91529,0.28157,0.28784}%
\definecolor{mycolor3}{rgb}{0.44157,0.74902,0.43216}%
\begin{tikzpicture}[font=\scriptsize]

\begin{axis}[%
  width=\fwidth,
  height=\fheight,
  at={(0,0)},
scale only axis,
xmin=1,
xmax=103,
ymin=10, 
ymax=40,
xlabel={Subband filter length ($T$)},
ylabel = {\gls{ERLE} (dB)},
axis background/.style={fill=white},
axis x line*=bottom,
axis y line*=left,
legend columns = -1,
legend style={at={(0,1)}, anchor=north west, legend cell align=left, align=left, draw=white!15!black},
xmajorgrids,
ymajorgrids,
mark repeat=10,
mark size=2.0pt
]
\addplot [line width=\plotlinewidth, color=mycolor1, dashed,  mark=o, mark options={solid, mycolor1},forget plot]
  table[row sep=crcr]{%
1	10.6051665014754\\
3	10.737366376289\\
5	10.8836767550876\\
7	11.0245772927217\\
9	11.1688308913023\\
11	11.3074455625998\\
13	11.4385947470857\\
15	11.5623757478223\\
17	11.6992730397971\\
19	11.8519783242884\\
21	11.9796886090673\\
23	12.1004561251141\\
25	12.2353132318337\\
27	12.3631096127255\\
29	12.503200691617\\
31	12.6310302806971\\
33	12.7778279656162\\
35	12.9113475788527\\
37	13.0499600876434\\
39	13.1980311531821\\
41	13.3449025583358\\
43	13.4887499895067\\
45	13.625111860067\\
47	13.7582962199608\\
49	13.9050074079282\\
51	14.0267832288214\\
53	14.1680610181002\\
55	14.2959601881329\\
57	14.448533336062\\
59	14.590320796862\\
61	14.7274745826712\\
63	14.8661774212107\\
65	15.0140131052044\\
67	15.1707925773489\\
69	15.3063263643638\\
71	15.4508488812707\\
73	15.5909475334296\\
75	15.7301926084473\\
77	15.8722057620257\\
79	16.0134608721892\\
81	16.1429138608378\\
83	16.2724043012512\\
85	16.4146702163478\\
87	16.5534988846484\\
89	16.6944635165828\\
91	16.8521867447493\\
93	16.9858221785857\\
95	17.1280931614133\\
97	17.2629624995459\\
99	17.4000791079419\\
101	17.5367082323482\\
103	17.6730591132138\\
};

\addplot [line width=\plotlinewidth, color=mycolor2, dashed,  mark=+, mark options={solid, mycolor2},forget plot]
  table[row sep=crcr]{%
1	12.9072514701428\\
3	23.0919152132787\\
5	23.2481853951259\\
7	23.3991427062118\\
9	23.5423037576916\\
11	23.6933770000245\\
13	23.8472501130863\\
15	23.9926915553671\\
17	24.1285466470945\\
19	24.2815681052229\\
21	24.4392690763481\\
23	24.5897619939971\\
25	24.7232610595573\\
27	24.8680454355318\\
29	25.0229131404828\\
31	25.1839557719281\\
33	25.3453262175854\\
35	25.4944359742565\\
37	25.6458014997592\\
39	25.7954195742453\\
41	25.950670813687\\
43	26.1135039865679\\
45	26.2579652580027\\
47	26.4361028265085\\
49	26.5837918255469\\
51	26.7337196704013\\
53	26.8857637734786\\
55	27.0302530823632\\
57	27.1864046399162\\
59	27.3528797759464\\
61	27.5106616298746\\
63	27.6673245646228\\
65	27.8225751120819\\
67	27.9851554871959\\
69	28.1397207882227\\
71	28.3025036632937\\
73	28.4603188686779\\
75	28.6237064784014\\
77	28.7992883297948\\
79	28.9689895312065\\
81	29.1378210474839\\
83	29.3092850620923\\
85	29.4695681397774\\
87	29.6420240528918\\
89	29.8257516819543\\
91	29.9963489019019\\
93	30.1892613349499\\
95	30.3631924842617\\
97	30.5491320620117\\
99	30.7217120272064\\
101	30.9124595173663\\
103	31.116091866831\\
};

\addplot [line width=\plotlinewidth, color=mycolor3, dashed,  mark=asterisk, mark options={solid, mycolor3},forget plot]
  table[row sep=crcr]{%
1	15.8117047839966\\
3	21.5500504911652\\
5	21.6881256478488\\
7	21.8252655710722\\
9	21.964508658942\\
11	22.0961179540856\\
13	22.2440516811496\\
15	22.3855579904707\\
17	22.5300932099616\\
19	22.6626718916476\\
21	22.8030007480231\\
23	22.9476257221521\\
25	23.0807326507954\\
27	23.2240726759966\\
29	23.3619266688981\\
31	23.507514725765\\
33	23.6544143062418\\
35	23.8106776350604\\
37	23.9561924585678\\
39	24.0896828211828\\
41	24.2369175116656\\
43	24.3759333642718\\
45	24.525845674439\\
47	24.6805824053473\\
49	24.8255316434436\\
51	24.9748992242741\\
53	25.1272481653752\\
55	25.2609905206531\\
57	25.4067702534537\\
59	25.5762266322476\\
61	25.7316351559715\\
63	25.8676528107467\\
65	26.024895204783\\
67	26.1781865558848\\
69	26.3206407566591\\
71	26.4652852564223\\
73	26.6184971742207\\
75	26.7711442422752\\
77	26.9243689916545\\
79	27.0769003150561\\
81	27.2201275253212\\
83	27.3837456761446\\
85	27.5287246115918\\
87	27.6905780322835\\
89	27.8441501973748\\
91	27.9989042897457\\
93	28.1558878375954\\
95	28.3220043736376\\
97	28.4911817848348\\
99	28.6532000880129\\
101	28.8187881403168\\
103	28.993453153207\\
};

\addplot [line width=\plotlinewidth, color=mycolor1,  mark=o, mark options={solid, mycolor1}]
  table[row sep=crcr]{%
1	19.2232712591939\\
3	19.3594485090614\\
5	19.5033792215405\\
7	19.6404651446565\\
9	19.7795075539169\\
11	19.9156313855862\\
13	20.0542070509497\\
15	20.1857843854399\\
17	20.3251055881632\\
19	20.4820365057255\\
21	20.6139053312615\\
23	20.743476062025\\
25	20.8806044711466\\
27	21.0076407514589\\
29	21.1538530399666\\
31	21.2922288680045\\
33	21.4365979262754\\
35	21.5705615594137\\
37	21.7062423830764\\
39	21.8510343515222\\
41	21.9999971529931\\
43	22.1350985289182\\
45	22.2676746415736\\
47	22.4010447990272\\
49	22.5479045590797\\
51	22.6689770479161\\
53	22.8125236113258\\
55	22.9465735836608\\
57	23.1001617144404\\
59	23.2413217909049\\
61	23.3736027466725\\
63	23.5129328386292\\
65	23.6595207371192\\
67	23.8152411419694\\
69	23.9511679256937\\
71	24.1052035053928\\
73	24.2581922634782\\
75	24.3977171277031\\
77	24.5402559871682\\
79	24.6748854457105\\
81	24.8124238244453\\
83	24.9424797188106\\
85	25.0836192146245\\
87	25.2194985829335\\
89	25.3540854968023\\
91	25.5003918488064\\
93	25.6237237078614\\
95	25.7659521390377\\
97	25.9078744684129\\
99	26.049698029893\\
101	26.2015738085919\\
103	26.3425663999993\\
};
\addlegendentry{\rectwin{A}}

\addplot [line width=\plotlinewidth, color=mycolor2,  mark=+, mark options={solid, mycolor2}]
  table[row sep=crcr]{%
1	17.4916133226291\\
3	24.8099025245419\\
5	24.9617557022263\\
7	25.1107701789079\\
9	25.2518712359028\\
11	25.3996778642891\\
13	25.5512616008483\\
15	25.6945984276777\\
17	25.8292488743426\\
19	25.9798255757871\\
21	26.134689807617\\
23	26.2829816840084\\
25	26.4145323522288\\
27	26.5569684780695\\
29	26.7087723275262\\
31	26.8665305423697\\
33	27.0223094749424\\
35	27.1695452589062\\
37	27.3161423513392\\
39	27.4625540067236\\
41	27.6136928042775\\
43	27.7708968770877\\
45	27.9126330824556\\
47	28.0857975635852\\
49	28.228356348238\\
51	28.3750721927588\\
53	28.5191332815397\\
55	28.6587336104724\\
57	28.8097608066987\\
59	28.9712614887411\\
61	29.1245241840709\\
63	29.2767925888527\\
65	29.4285328936786\\
67	29.5878942813092\\
69	29.736003088572\\
71	29.8948248512544\\
73	30.0480304416245\\
75	30.2073716981549\\
77	30.3760645098337\\
79	30.5390062636031\\
81	30.7028507896987\\
83	30.8683273556086\\
85	31.0228136283456\\
87	31.1898766344011\\
89	31.365903556081\\
91	31.5299988117364\\
93	31.7155627691092\\
95	31.8830839050493\\
97	32.0595206937465\\
99	32.2259880176265\\
101	32.409919585721\\
103	32.6071887699892\\
};
\addlegendentry{\cosinewin{A}}

\addplot [line width=\plotlinewidth, color=mycolor3,  mark=asterisk, mark options={solid, mycolor3}]
  table[row sep=crcr]{%
1	19.1537717891022\\
3	28.7967787562436\\
5	28.9237611320653\\
7	29.0488210284506\\
9	29.1812584988038\\
11	29.3132045014571\\
13	29.4523314884629\\
15	29.5794748783114\\
17	29.7119208258797\\
19	29.8395437600722\\
21	29.9702714405252\\
23	30.1034571053051\\
25	30.2398261892061\\
27	30.3702636794611\\
29	30.5072123948215\\
31	30.6339810464562\\
33	30.7781538540724\\
35	30.915049661458\\
37	31.0494705404022\\
39	31.1794905231924\\
41	31.3085227327393\\
43	31.4459576000619\\
45	31.5871735617069\\
47	31.7345796193814\\
49	31.8682758740336\\
51	32.0112762474479\\
53	32.1508603076187\\
55	32.2793788965166\\
57	32.4180456205764\\
59	32.5758967404551\\
61	32.7172837348006\\
63	32.8500489923327\\
65	33.0040560189365\\
67	33.1440067722632\\
69	33.2808065939704\\
71	33.4154751499869\\
73	33.5576149078263\\
75	33.699009193121\\
77	33.8467304716527\\
79	33.9778215488633\\
81	34.1248459872212\\
83	34.2657350949398\\
85	34.4102649281962\\
87	34.5658169584091\\
89	34.7123046070182\\
91	34.8612147903672\\
93	35.0068090686698\\
95	35.1737158360301\\
97	35.3323033409811\\
99	35.4893106555701\\
101	35.6406959045061\\
103	35.7936812623895\\
};
\addlegendentry{\roothannwin{A}}
\end{axis}
\end{tikzpicture}%

%% file: figures/PTWOLA_L_512_ERLE_MSE_analysis_norm_windows.tex
\definecolor{mycolor1}{rgb}{0.34667,0.53600,0.69067}%
\definecolor{mycolor2}{rgb}{0.91529,0.28157,0.28784}%
\definecolor{mycolor3}{rgb}{0.44157,0.74902,0.43216}%
\begin{tikzpicture} [font=\scriptsize]

\begin{axis}[%
  width=\fwidth,
  height=\fheight,
  at={(0,0)},
scale only axis,
xmin=1,
xmax=101,
ymin=10, 
ymax=40,
xlabel={Subband filter length ($T$)},
ylabel = {\gls{ERLE} (dB)},
axis background/.style={fill=white},
axis x line*=bottom,
axis y line*=left,
legend columns = -1,
legend style={at={(0,1)}, anchor=north west, legend cell align=left, align=left, draw=white!15!black},
xmajorgrids,
ymajorgrids,
mark repeat=10,
mark size=2.0pt
]
\addplot [line width=\plotlinewidth, color=mycolor1, dashed, mark=o, mark options={solid, mycolor1}]
  table[row sep=crcr]{%
1	13.4087715019493\\
3	13.5120186933971\\
5	13.6151713526349\\
7	13.7182021791997\\
9	13.8211896223991\\
11	13.9241627045476\\
13	14.027142336731\\
15	14.1302275643546\\
17	14.233386591603\\
19	14.3366246680249\\
21	14.4399518180484\\
23	14.5435585656251\\
25	14.6474756536602\\
27	14.7516462603797\\
29	14.856023515071\\
31	14.9605708977064\\
33	15.0652193609097\\
35	15.1698900723887\\
37	15.2745263477804\\
39	15.3791256936137\\
41	15.4836388788452\\
43	15.5880894928678\\
45	15.6924840231478\\
47	15.7968389716701\\
49	15.9011549633525\\
51	16.0053523659905\\
53	16.1094166582716\\
55	16.213375806959\\
57	16.3172782830831\\
59	16.4210984365958\\
61	16.5247934355717\\
63	16.6283323519578\\
65	16.7317154894432\\
67	16.8348797117001\\
69	16.9379192312845\\
71	17.040923579437\\
73	17.1440436890425\\
75	17.2472520939373\\
77	17.350500196028\\
79	17.4537302188638\\
81	17.556931776209\\
83	17.6601853436787\\
85	17.7634697988339\\
87	17.8667262241328\\
89	17.9699449574373\\
91	18.0730936909\\
93	18.1761317685921\\
95	18.2790876782847\\
97	18.3820314051368\\
99	18.4850033146507\\
101	18.5879909348755\\
};
\addlegendentry{\rectwin{A}}

\addplot [line width=\plotlinewidth, color=mycolor2, dashed,  mark=+, mark options={solid, mycolor2}]
  table[row sep=crcr]{%
1	8.30697835288396\\
5	26.314498741709\\
9	26.5684632909276\\
13	26.8323736347787\\
17	27.0941976520178\\
21	27.3560230713328\\
25	27.6115602330712\\
29	27.8857526833868\\
33	28.1692796936582\\
37	28.4431097939517\\
41	28.7124615527665\\
45	28.9884246010687\\
49	29.2757837230834\\
53	29.5550443287519\\
57	29.8228963794714\\
61	30.09602244901\\
65	30.3556132866852\\
69	30.6173527465331\\
73	30.8766616761806\\
77	31.1509202958731\\
81	31.412155001386\\
85	31.6547674547108\\
89	31.9211780057939\\
93	32.1969768822969\\
97	32.4685075342434\\
101	32.7472809308254\\
};
\addlegendentry{\cosinewin{A}}

\addplot [line width=\plotlinewidth, color=mycolor3, dashed,  mark=asterisk, mark options={solid, mycolor3}]
  table[row sep=crcr]{%
1	15.8470079930966\\
3	21.560467618905\\
5	21.6648403035798\\
7	21.7652211021288\\
9	21.8687710941633\\
11	21.9726010922585\\
13	22.0718615265961\\
15	22.1778276218093\\
17	22.2776376530003\\
19	22.380222041601\\
21	22.4825299900974\\
23	22.5839580691717\\
25	22.6850969418397\\
27	22.7878837017931\\
29	22.89116657564\\
31	22.9943937326418\\
33	23.0947439107585\\
35	23.1981856809725\\
37	23.3042246374495\\
39	23.4112778862021\\
41	23.5184928055631\\
43	23.6180842855536\\
45	23.7241245834912\\
47	23.8352646610941\\
49	23.9316891993943\\
51	24.0353020088547\\
53	24.1475197636454\\
55	24.2558560934514\\
57	24.356187570368\\
59	24.4661590316063\\
61	24.5785398384403\\
63	24.6869223651807\\
65	24.7929617260623\\
67	24.8981976181153\\
69	25.0060491764546\\
71	25.1118767824901\\
73	25.2176038127285\\
75	25.3288571588799\\
77	25.4368547302662\\
79	25.5477616258898\\
81	25.6605188319657\\
83	25.7739290533565\\
85	25.8798470907544\\
87	25.9930095003305\\
89	26.1066556525131\\
91	26.2185214916938\\
93	26.3371575474434\\
95	26.4479182842952\\
97	26.5649145000958\\
99	26.686894337426\\
101	26.8047136226292\\
};
\addlegendentry{\roothannwin{A}}
\end{axis}
\end{tikzpicture}%

%% file: figures/PTWOLA_Results_L512_SWDFTvsconstR_norm.tex
\definecolor{mycolor1}{rgb}{0.34667,0.53600,0.69067}%
\definecolor{mycolor2}{rgb}{0.91529,0.28157,0.28784}%
\definecolor{mycolor3}{rgb}{0.44157,0.74902,0.43216}%
\definecolor{mycolor4}{rgb}{1.00000,0.54902,0.10196}%
\definecolor{mycolor5}{rgb}{0.49020,0.18039,0.56078}%
\definecolor{mycolor6}{rgb}{0.69020,0.40000,0.23922}%
\begin{tikzpicture}[font=\scriptsize]

\begin{axis}[%
  width=\fwidth,
  height=\fheight,
  at={(0,0)},
scale only axis,
xmin=1,
xmax=103,
ymin=10,
ymax=33,
xlabel={Total subband filter length ($T$)},
ylabel = {\gls{ERLE} (dB)},
axis background/.style={fill=white},
axis x line*=bottom,
axis y line*=left,
legend columns = -1,
legend style={at={(0,1)}, anchor=north west, legend cell align=left, align=left, draw=white!15!black},
xmajorgrids,
ymajorgrids,
mark repeat=10,
mark size=2.0pt
]
\addplot [line width=\plotlinewidth, color=mycolor1,  mark=o, mark options={solid, mycolor1}]
  table[row sep=crcr]{%
1	10.6051665014754\\
3	10.737366376289\\
5	10.8836767550876\\
7	11.0245772927217\\
9	11.1688308913023\\
11	11.3074455625998\\
13	11.4385947470857\\
15	11.5623757478223\\
17	11.6992730397971\\
19	11.8519783242884\\
21	11.9796886090673\\
23	12.1004561251141\\
25	12.2353132318337\\
27	12.3631096127255\\
29	12.503200691617\\
31	12.6310302806971\\
33	12.7778279656162\\
35	12.9113475788527\\
37	13.0499600876434\\
39	13.1980311531821\\
41	13.3449025583358\\
43	13.4887499895067\\
45	13.625111860067\\
47	13.7582962199608\\
49	13.9050074079282\\
51	14.0267832288214\\
53	14.1680610181002\\
55	14.2959601881329\\
57	14.448533336062\\
59	14.590320796862\\
61	14.7274745826712\\
63	14.8661774212107\\
65	15.0140131052044\\
67	15.1707925773489\\
69	15.3063263643638\\
71	15.4508488812707\\
73	15.5909475334296\\
75	15.7301926084473\\
77	15.8722057620257\\
79	16.0134608721892\\
81	16.1429138608378\\
83	16.2724043012512\\
85	16.4146702163478\\
87	16.5534988846484\\
89	16.6944635165828\\
91	16.8521867447493\\
93	16.9858221785857\\
95	17.1280931614133\\
97	17.2629624995459\\
99	17.4000791079419\\
101	17.5367082323482\\
103	17.6730591132138\\
};
\addlegendentry{\rectwin{A}}

\addplot [line width=\plotlinewidth, color=mycolor2, mark=+, mark options={solid, mycolor2}]
  table[row sep=crcr]{%
1	12.9072514701428\\
3	23.0919152132787\\
5	23.2481853951259\\
7	23.3991427062118\\
9	23.5423037576916\\
11	23.6933770000245\\
13	23.8472501130863\\
15	23.9926915553671\\
17	24.1285466470945\\
19	24.2815681052229\\
21	24.4392690763481\\
23	24.5897619939971\\
25	24.7232610595573\\
27	24.8680454355318\\
29	25.0229131404828\\
31	25.1839557719281\\
33	25.3453262175854\\
35	25.4944359742565\\
37	25.6458014997592\\
39	25.7954195742453\\
41	25.950670813687\\
43	26.1135039865679\\
45	26.2579652580027\\
47	26.4361028265085\\
49	26.5837918255469\\
51	26.7337196704013\\
53	26.8857637734786\\
55	27.0302530823632\\
57	27.1864046399162\\
59	27.3528797759464\\
61	27.5106616298746\\
63	27.6673245646228\\
65	27.8225751120819\\
67	27.9851554871959\\
69	28.1397207882227\\
71	28.3025036632937\\
73	28.4603188686779\\
75	28.6237064784014\\
77	28.7992883297948\\
79	28.9689895312065\\
81	29.1378210474839\\
83	29.3092850620923\\
85	29.4695681397774\\
87	29.6420240528918\\
89	29.8257516819543\\
91	29.9963489019019\\
93	30.1892613349499\\
95	30.3631924842617\\
97	30.5491320620117\\
99	30.7217120272064\\
101	30.9124595173663\\
103	31.116091866831\\
};
\addlegendentry{\cosinewin{A}}

\addplot [line width=\plotlinewidth, color=mycolor3, mark=asterisk, mark options={solid, mycolor3}]
  table[row sep=crcr]{%
1	15.8117047839966\\
3	21.5500504911652\\
5	21.6881256478488\\
7	21.8252655710722\\
9	21.964508658942\\
11	22.0961179540856\\
13	22.2440516811496\\
15	22.3855579904707\\
17	22.5300932099616\\
19	22.6626718916476\\
21	22.8030007480231\\
23	22.9476257221521\\
25	23.0807326507954\\
27	23.2240726759966\\
29	23.3619266688981\\
31	23.507514725765\\
33	23.6544143062418\\
35	23.8106776350604\\
37	23.9561924585678\\
39	24.0896828211828\\
41	24.2369175116656\\
43	24.3759333642718\\
45	24.525845674439\\
47	24.6805824053473\\
49	24.8255316434436\\
51	24.9748992242741\\
53	25.1272481653752\\
55	25.2609905206531\\
57	25.4067702534537\\
59	25.5762266322476\\
61	25.7316351559715\\
63	25.8676528107467\\
65	26.024895204783\\
67	26.1781865558848\\
69	26.3206407566591\\
71	26.4652852564223\\
73	26.6184971742207\\
75	26.7711442422752\\
77	26.9243689916545\\
79	27.0769003150561\\
81	27.2201275253212\\
83	27.3837456761446\\
85	27.5287246115918\\
87	27.6905780322835\\
89	27.8441501973748\\
91	27.9989042897457\\
93	28.1558878375954\\
95	28.3220043736376\\
97	28.4911817848348\\
99	28.6532000880129\\
101	28.8187881403168\\
103	28.993453153207\\
};
\addlegendentry{\roothannwin{A}}

\addplot [line width=\plotlinewidth, color=mycolor4, dashed,  mark phase = 5,mark=square, mark options={solid, mycolor4},forget plot]
  table[row sep=crcr]{%
  1	10.6051665014754\\
  3	10.737366376289\\
  5	10.8836767550876\\
  7	11.0245772927217\\
  9	11.1688308913022\\
  11	11.3074455625998\\
  13	11.4385947470857\\
  15	11.5623757478223\\
  17	11.6992730397971\\
  19	11.8519783242884\\
  21	11.9796886090673\\
  23	12.1004561251141\\
  25	12.2353132318337\\
  27	12.3631096127255\\
  29	12.503200691617\\
  31	12.6310302806971\\
  33	12.7778279656162\\
  35	12.9113475788527\\
  37	13.0499600876434\\
  39	13.1980311531821\\
  41	13.3449025583358\\
  43	13.4887499895067\\
  45	13.625111860067\\
  47	13.7582962199608\\
  49	13.9050074079282\\
  51	14.0267832288214\\
  53	14.1680610181002\\
  55	14.2959601881329\\
  57	14.448533336062\\
  59	14.590320796862\\
  61	14.7274745826712\\
  63	14.8661774212107\\
  65	15.0140131052044\\
  67	15.1707925773489\\
  69	15.3063263643638\\
  71	15.4508488812707\\
  73	15.5909475334296\\
  75	15.7301926084473\\
  77	15.8722057620257\\
  79	16.0134608721892\\
  81	16.1429138608378\\
  83	16.2724043012512\\
  85	16.4146702163478\\
  87	16.5534988846484\\
  89	16.6944635165828\\
  91	16.8521867447493\\
  93	16.9858221785857\\
  95	17.1280931614133\\
  97	17.2629624995459\\
  99	17.4000791079419\\
  101	17.5367082323482\\
  103	17.6730591132138\\
  };
\addlegendentry{\rectwin{A}}

\addplot [line width=\plotlinewidth, color=mycolor5, dashed, mark=diamond,  mark phase = 5, mark options={solid, mycolor5},forget plot]
  table[row sep=crcr]{%
3	23.121634657737\\
5	23.2629277232501\\
7	23.4109405031293\\
9	23.5521927907795\\
11	23.7013443481698\\
13	23.8570665259304\\
15	24.0016874754588\\
17	24.1384535595899\\
19	24.2919080945033\\
21	24.4476929797764\\
23	24.5974223601694\\
25	24.7308370767876\\
27	24.8749929391248\\
29	25.0303378846377\\
31	25.191695607349\\
33	25.3527163827974\\
35	25.5010007532778\\
37	25.6525274809832\\
39	25.8019518542093\\
41	25.9574174088059\\
43	26.1204608680307\\
45	26.2652601809713\\
47	26.4438635313465\\
49	26.5914671368247\\
51	26.7411530006353\\
53	26.895094832009\\
55	27.0402420194502\\
57	27.1957085255252\\
59	27.3619009846585\\
61	27.5203832698345\\
63	27.6769853846378\\
65	27.832484161144\\
67	27.9954408264768\\
69	28.1502677282649\\
71	28.3134569652857\\
73	28.4719597677464\\
75	28.6352157490443\\
77	28.8103644105072\\
79	28.9806400214056\\
81	29.1500277732501\\
83	29.3217046761074\\
85	29.4817634399341\\
87	29.653327965593\\
89	29.8365653375022\\
91	30.0070201258366\\
93	30.1994282308876\\
95	30.3740299198234\\
97	30.5589664893732\\
99	30.7320114349849\\
101	30.92340239786\\
103	31.1272319499183\\
};
\addlegendentry{\cosinewin{A}}

\addplot [line width=\plotlinewidth, color=mycolor6, dashed, mark=star,  mark phase = 5, mark options={solid, mycolor6},forget plot]
  table[row sep=crcr]{%
  5	21.8359629421523\\
  7	21.9807339086225\\
  9	22.1239336664734\\
  11	22.2637409763782\\
  13	22.4035348707723\\
  15	22.5533321631812\\
  17	22.6943453035\\
  19	22.8322773633442\\
  21	22.969934167546\\
  23	23.1249269560919\\
  25	23.2691779399012\\
  27	23.4067267729629\\
  29	23.5478727757068\\
  31	23.6882105140602\\
  33	23.8408095988418\\
  35	23.9892355644455\\
  37	24.1407773881592\\
  39	24.2794551024271\\
  41	24.4235497978886\\
  43	24.5696868810673\\
  45	24.7181432234112\\
  47	24.8677961722528\\
  49	25.0129803654438\\
  51	25.1601957714104\\
  53	25.3014168884815\\
  55	25.4369480127072\\
  57	25.5786692318818\\
  59	25.7212061025987\\
  61	25.8716025580748\\
  63	26.0212941850395\\
  65	26.1599397139401\\
  67	26.2977945641564\\
  69	26.4414987111224\\
  71	26.5877988403232\\
  73	26.7243325588465\\
  75	26.868508291998\\
  77	27.0014773043565\\
  79	27.1529581626974\\
  81	27.3031057778096\\
  83	27.4507518124681\\
  85	27.5958314755084\\
  87	27.733317564334\\
  89	27.8788931264395\\
  91	28.020291580821\\
  93	28.1700079619176\\
  95	28.330639856485\\
  97	28.4686409449906\\
  99	28.6182029235629\\
  101	28.7537913822496\\
  103	28.9051999305449\\
};
\addlegendentry{\roothannwin{A}}
\end{axis}
\end{tikzpicture}%

%% file: figures/PTWOLA_Results_L512_SWDFTvsconstR_des.tex
\definecolor{mycolor1}{rgb}{0.34667,0.53600,0.69067}%
\definecolor{mycolor2}{rgb}{0.91529,0.28157,0.28784}%
\definecolor{mycolor3}{rgb}{0.44157,0.74902,0.43216}%
\definecolor{mycolor4}{rgb}{1.00000,0.54902,0.10196}%
\definecolor{mycolor5}{rgb}{0.49020,0.18039,0.56078}%
\definecolor{mycolor6}{rgb}{0.69020,0.40000,0.23922}%
\definecolor{mycolor7}{rgb}{1.00000,0.07451,0.65098}%
\begin{tikzpicture}[font=\scriptsize]

\begin{axis}[%
  width=\fwidth,
  height=\fheight,
  at={(0,0)},
scale only axis,
xmin=1,
xmax=103,
ymin=19,
ymax=37,
xlabel={Total subband filter length ($T$)},
ylabel = {\gls{ERLE} (dB)},
axis background/.style={fill=white},
axis x line*=bottom,
axis y line*=left,
legend columns = -1,
legend style={at={(0,1)}, anchor=north west, legend cell align=left, align=left, draw=white!15!black},
xmajorgrids,
ymajorgrids,
mark repeat=10,
mark size=2.0pt
]
\addplot [line width=\plotlinewidth, color=mycolor1, mark=o, mark options={solid, mycolor1}]
  table[row sep=crcr]{%
1	19.2232712591939\\
3	19.3594485090614\\
5	19.5033792215405\\
7	19.6404651446565\\
9	19.7795075539169\\
11	19.9156313855862\\
13	20.0542070509497\\
15	20.1857843854399\\
17	20.3251055881632\\
19	20.4820365057255\\
21	20.6139053312615\\
23	20.743476062025\\
25	20.8806044711466\\
27	21.0076407514589\\
29	21.1538530399666\\
31	21.2922288680045\\
33	21.4365979262754\\
35	21.5705615594137\\
37	21.7062423830764\\
39	21.8510343515222\\
41	21.9999971529931\\
43	22.1350985289182\\
45	22.2676746415736\\
47	22.4010447990272\\
49	22.5479045590797\\
51	22.6689770479161\\
53	22.8125236113258\\
55	22.9465735836608\\
57	23.1001617144404\\
59	23.2413217909049\\
61	23.3736027466725\\
63	23.5129328386292\\
65	23.6595207371192\\
67	23.8152411419694\\
69	23.9511679256937\\
71	24.1052035053928\\
73	24.2581922634782\\
75	24.3977171277031\\
77	24.5402559871682\\
79	24.6748854457105\\
81	24.8124238244453\\
83	24.9424797188106\\
85	25.0836192146245\\
87	25.2194985829335\\
89	25.3540854968023\\
91	25.5003918488064\\
93	25.6237237078614\\
95	25.7659521390377\\
97	25.9078744684129\\
99	26.049698029893\\
101	26.2015738085919\\
103	26.3425663999993\\
};
\addlegendentry{\rectwin{A}}

\addplot [line width=\plotlinewidth, color=mycolor2, mark=+, mark options={solid, mycolor2}]
  table[row sep=crcr]{%
1	17.4916133226291\\
3	24.8099025245419\\
5	24.9617557022263\\
7	25.1107701789079\\
9	25.2518712359028\\
11	25.3996778642891\\
13	25.5512616008483\\
15	25.6945984276777\\
17	25.8292488743426\\
19	25.9798255757871\\
21	26.134689807617\\
23	26.2829816840084\\
25	26.4145323522288\\
27	26.5569684780695\\
29	26.7087723275262\\
31	26.8665305423697\\
33	27.0223094749424\\
35	27.1695452589062\\
37	27.3161423513392\\
39	27.4625540067236\\
41	27.6136928042775\\
43	27.7708968770877\\
45	27.9126330824556\\
47	28.0857975635852\\
49	28.228356348238\\
51	28.3750721927588\\
53	28.5191332815397\\
55	28.6587336104724\\
57	28.8097608066987\\
59	28.9712614887411\\
61	29.1245241840709\\
63	29.2767925888527\\
65	29.4285328936786\\
67	29.5878942813092\\
69	29.736003088572\\
71	29.8948248512544\\
73	30.0480304416245\\
75	30.2073716981549\\
77	30.3760645098337\\
79	30.5390062636031\\
81	30.7028507896987\\
83	30.8683273556086\\
85	31.0228136283456\\
87	31.1898766344011\\
89	31.365903556081\\
91	31.5299988117364\\
93	31.7155627691092\\
95	31.8830839050493\\
97	32.0595206937465\\
99	32.2259880176265\\
101	32.409919585721\\
103	32.6071887699892\\
};
\addlegendentry{\cosinewin{A}}

\addplot [line width=\plotlinewidth, color=mycolor3, mark=asterisk, mark options={solid, mycolor3}]
  table[row sep=crcr]{%
1	19.1537717891022\\
3	28.7967787562436\\
5	28.9237611320653\\
7	29.0488210284506\\
9	29.1812584988038\\
11	29.3132045014571\\
13	29.4523314884629\\
15	29.5794748783114\\
17	29.7119208258797\\
19	29.8395437600722\\
21	29.9702714405252\\
23	30.1034571053051\\
25	30.2398261892061\\
27	30.3702636794611\\
29	30.5072123948215\\
31	30.6339810464562\\
33	30.7781538540724\\
35	30.915049661458\\
37	31.0494705404022\\
39	31.1794905231924\\
41	31.3085227327393\\
43	31.4459576000619\\
45	31.5871735617069\\
47	31.7345796193814\\
49	31.8682758740336\\
51	32.0112762474479\\
53	32.1508603076187\\
55	32.2793788965166\\
57	32.4180456205764\\
59	32.5758967404551\\
61	32.7172837348006\\
63	32.8500489923327\\
65	33.0040560189365\\
67	33.1440067722632\\
69	33.2808065939704\\
71	33.4154751499869\\
73	33.5576149078263\\
75	33.699009193121\\
77	33.8467304716527\\
79	33.9778215488633\\
81	34.1248459872212\\
83	34.2657350949398\\
85	34.4102649281962\\
87	34.5658169584091\\
89	34.7123046070182\\
91	34.8612147903672\\
93	35.0068090686698\\
95	35.1737158360301\\
97	35.3323033409811\\
99	35.4893106555701\\
101	35.6406959045061\\
103	35.7936812623895\\
};
\addlegendentry{\roothannwin{A}}

\addplot [line width=\plotlinewidth, color=mycolor4, dashed,  mark phase = 5, mark=square, mark options={solid, mycolor4},forget plot]
  table[row sep=crcr]{%
1	19.2921086093296\\
3	19.4223863336769\\
5	19.5688442743102\\
7	19.7123926910353\\
9	19.8415628781666\\
11	19.9877085446229\\
13	20.1190349669744\\
15	20.2580733290635\\
17	20.4049912734098\\
19	20.5483809574622\\
21	20.6800023572052\\
23	20.810161247823\\
25	20.9444872244662\\
27	21.0772597137063\\
29	21.2301810610623\\
31	21.3621798882414\\
33	21.5095446247562\\
35	21.6317475058831\\
37	21.7817450943493\\
39	21.9207833815845\\
41	22.0630931030659\\
43	22.1962526037622\\
45	22.3375730842829\\
47	22.4702278333943\\
49	22.6161280152925\\
51	22.7435800687543\\
53	22.8742536592596\\
55	23.0255374073526\\
57	23.1708402510801\\
59	23.3031207752007\\
61	23.4420151533051\\
63	23.5810458495264\\
65	23.7413899301556\\
67	23.8775054583049\\
69	24.0357956737366\\
71	24.1922038657342\\
73	24.3351982094272\\
75	24.4700496148782\\
77	24.6036103484072\\
79	24.7457786463858\\
81	24.8755708324098\\
83	25.0047038890042\\
85	25.1548021600008\\
87	25.293942965541\\
89	25.4175298311581\\
91	25.5651858266617\\
93	25.7052200742509\\
95	25.825213086244\\
97	25.9721005955205\\
99	26.1270010759096\\
101	26.2741469060581\\
103	26.4185216035683\\
};
\addlegendentry{rectdes R = 0}

\addplot [line width=\plotlinewidth, color=mycolor5, dashed,  mark phase = 5, mark=diamond, mark options={solid, mycolor5},forget plot]
  table[row sep=crcr]{%
3	24.837474596404\\
5	24.9758745931098\\
7	25.1219074336743\\
9	25.2614293157354\\
11	25.4070729653791\\
13	25.5605048167071\\
15	25.7030886031159\\
17	25.8385338516661\\
19	25.9897593938816\\
21	26.1428592089543\\
23	26.2903286287822\\
25	26.4215572010206\\
27	26.56339369992\\
29	26.715970808104\\
31	26.8742784910405\\
33	27.0298032981793\\
35	27.1763593516493\\
37	27.3231706146567\\
39	27.4692112702141\\
41	27.6205625129547\\
43	27.7778186133194\\
45	27.9197776016194\\
47	28.0934939328334\\
49	28.2359549652983\\
51	28.3824400555832\\
53	28.528508156912\\
55	28.6687362010865\\
57	28.8192525325442\\
59	28.9802696531597\\
61	29.1341460736938\\
63	29.2864997133802\\
65	29.4385619910412\\
67	29.5982434801495\\
69	29.7467635512424\\
71	29.906035749324\\
73	30.0599865285568\\
75	30.2190096202425\\
77	30.3873652546785\\
79	30.5507716645925\\
81	30.7151037446884\\
83	30.8806049220311\\
85	31.0346554469524\\
87	31.2009535208055\\
89	31.3765397706333\\
91	31.5406683037932\\
93	31.7257275926652\\
95	31.8939377327646\\
97	32.0694786915396\\
99	32.2363719217889\\
101	32.4207132345264\\
103	32.6181545648384\\
};
\addlegendentry{Cosdes R = 1}

\addplot [line width=\plotlinewidth, color=mycolor6, dashdotted,  mark phase = 5, mark=star, mark options={solid, mycolor6},forget plot]
  table[row sep=crcr]{%
5	26.5844773496044\\
7	26.6916352765471\\
9	26.8107052581207\\
11	26.9196674099964\\
13	27.0346657556473\\
15	27.1552335082038\\
17	27.2645369465976\\
19	27.3700269553518\\
21	27.4868057185402\\
23	27.6048858817361\\
25	27.7157283309965\\
27	27.8195614455569\\
29	27.9316338369251\\
31	28.0453325782924\\
33	28.159949463759\\
35	28.2766493569031\\
37	28.3824841212309\\
39	28.4907315388868\\
41	28.6002171467583\\
43	28.7015059503022\\
45	28.8112966637169\\
47	28.9067976492858\\
49	29.0231237776681\\
51	29.1168931319359\\
53	29.211335097047\\
55	29.3141038145636\\
57	29.4020568078738\\
59	29.4981210142193\\
61	29.5991228016034\\
63	29.6899960108261\\
65	29.778837490768\\
67	29.8686613786297\\
69	29.9554150850351\\
71	30.0444995929217\\
73	30.1300208686253\\
75	30.2210407409177\\
77	30.304374336055\\
79	30.3970662863567\\
81	30.4839329592959\\
83	30.5653913880721\\
85	30.6482736092486\\
87	30.7191015420395\\
89	30.797877233142\\
91	30.8810886298593\\
93	30.9537158002955\\
95	31.0351323299932\\
97	31.1078585416317\\
99	31.1806288476631\\
101	31.2473187653293\\
103	31.3160402933677\\
};
\addlegendentry{SqrtHanndes R = 2}

\addplot [line width=\plotlinewidth, color=mycolor7, dashed,  mark phase = 5, mark=x, mark options={solid, mycolor7},forget plot]
  table[row sep=crcr]{%
9	28.6242751542542\\
11	28.7745860774732\\
13	28.9344695469268\\
15	29.092067300217\\
17	29.2500888278547\\
19	29.4180789090492\\
21	29.5723074642881\\
23	29.7261134107752\\
25	29.8949587650145\\
27	30.0607213233106\\
29	30.2270159118911\\
31	30.3752208947624\\
33	30.5333822691641\\
35	30.7037422482731\\
37	30.8764243229677\\
39	31.047012574108\\
41	31.2077080603886\\
43	31.3685640076257\\
45	31.5286738721754\\
47	31.6913243028474\\
49	31.8624618241722\\
51	32.0135562551539\\
53	32.1957938479979\\
55	32.3437893174563\\
57	32.4931367739767\\
59	32.6528933645959\\
61	32.7997767964996\\
63	32.9563718141962\\
65	33.1168227333923\\
67	33.2713513625759\\
69	33.420601945136\\
71	33.569351820589\\
73	33.7182123883738\\
75	33.8653672581287\\
77	34.0118497642108\\
79	34.15487269403\\
81	34.3043979387815\\
83	34.4602658865539\\
85	34.6077820642072\\
87	34.7530005265819\\
89	34.9027141628546\\
91	35.0367599023645\\
93	35.1790764890868\\
95	35.3312598673537\\
97	35.4682522059124\\
99	35.6163329591593\\
101	35.7510955225594\\
103	35.8821196307576\\
};
\addlegendentry{SqrtHanndes R = 4,forget plot}

\end{axis}
\end{tikzpicture}%

%% file: figures/PTWOLA_Results_L512_SWDFTvsconstnf_norm.tex
\definecolor{mycolor1}{rgb}{0.34667,0.53600,0.69067}%
\definecolor{mycolor2}{rgb}{0.91529,0.28157,0.28784}%
\definecolor{mycolor3}{rgb}{0.44157,0.74902,0.43216}%
\begin{tikzpicture}[font=\scriptsize]

\begin{axis}[%
  width=\fwidth,
  height=\fheight,
  at={(0,0)},
scale only axis,
xmin=1,
xmax=103,
ymin=8,
ymax=32,
xlabel={Total subband filter length ($T$)},
ylabel = {\gls{ERLE} (dB)},
axis background/.style={fill=white},
axis x line*=bottom,
axis y line*=left,
legend columns = -1,
legend style={at={(0,1)}, anchor=north west, legend cell align=left, align=left, draw=white!15!black},
xmajorgrids,
ymajorgrids,
mark repeat=10,
mark size=2.0pt
]
\addplot [line width=\plotlinewidth, color=mycolor1, mark=o, mark options={solid, mycolor1}]
  table[row sep=crcr]{%
1	10.6051665014754\\
3	10.737366376289\\
5	10.8836767550876\\
7	11.0245772927217\\
9	11.1688308913023\\
11	11.3074455625998\\
13	11.4385947470857\\
15	11.5623757478223\\
17	11.6992730397971\\
19	11.8519783242884\\
21	11.9796886090673\\
23	12.1004561251141\\
25	12.2353132318337\\
27	12.3631096127255\\
29	12.503200691617\\
31	12.6310302806971\\
33	12.7778279656162\\
35	12.9113475788527\\
37	13.0499600876434\\
39	13.1980311531821\\
41	13.3449025583358\\
43	13.4887499895067\\
45	13.625111860067\\
47	13.7582962199608\\
49	13.9050074079282\\
51	14.0267832288214\\
53	14.1680610181002\\
55	14.2959601881329\\
57	14.448533336062\\
59	14.590320796862\\
61	14.7274745826712\\
63	14.8661774212107\\
65	15.0140131052044\\
67	15.1707925773489\\
69	15.3063263643638\\
71	15.4508488812707\\
73	15.5909475334296\\
75	15.7301926084473\\
77	15.8722057620257\\
79	16.0134608721892\\
81	16.1429138608378\\
83	16.2724043012512\\
85	16.4146702163478\\
87	16.5534988846484\\
89	16.6944635165828\\
91	16.8521867447493\\
93	16.9858221785857\\
95	17.1280931614133\\
97	17.2629624995459\\
99	17.4000791079419\\
101	17.5367082323482\\
103	17.6730591132138\\
};
\addlegendentry{\rectwin{A}}

\addplot [line width=\plotlinewidth, color=mycolor2,  mark=+, mark options={solid, mycolor2}]
  table[row sep=crcr]{%
1	12.9072514701428\\
3	23.0919152132787\\
5	23.2481853951259\\
7	23.3991427062118\\
9	23.5423037576916\\
11	23.6933770000245\\
13	23.8472501130863\\
15	23.9926915553671\\
17	24.1285466470945\\
19	24.2815681052229\\
21	24.4392690763481\\
23	24.5897619939971\\
25	24.7232610595573\\
27	24.8680454355318\\
29	25.0229131404828\\
31	25.1839557719281\\
33	25.3453262175854\\
35	25.4944359742565\\
37	25.6458014997592\\
39	25.7954195742453\\
41	25.950670813687\\
43	26.1135039865679\\
45	26.2579652580027\\
47	26.4361028265085\\
49	26.5837918255469\\
51	26.7337196704013\\
53	26.8857637734786\\
55	27.0302530823632\\
57	27.1864046399162\\
59	27.3528797759464\\
61	27.5106616298746\\
63	27.6673245646228\\
65	27.8225751120819\\
67	27.9851554871959\\
69	28.1397207882227\\
71	28.3025036632937\\
73	28.4603188686779\\
75	28.6237064784014\\
77	28.7992883297948\\
79	28.9689895312065\\
81	29.1378210474839\\
83	29.3092850620923\\
85	29.4695681397774\\
87	29.6420240528918\\
89	29.8257516819543\\
91	29.9963489019019\\
93	30.1892613349499\\
95	30.3631924842617\\
97	30.5491320620117\\
99	30.7217120272064\\
101	30.9124595173663\\
103	31.116091866831\\
};
\addlegendentry{\cosinewin{A}}

\addplot [line width=\plotlinewidth, color=mycolor3,  mark=asterisk, mark options={solid, mycolor3}]
  table[row sep=crcr]{%
1	15.8117047839966\\
3	21.5500504911652\\
5	21.6881256478488\\
7	21.8252655710722\\
9	21.964508658942\\
11	22.0961179540856\\
13	22.2440516811496\\
15	22.3855579904707\\
17	22.5300932099616\\
19	22.6626718916476\\
21	22.8030007480231\\
23	22.9476257221521\\
25	23.0807326507954\\
27	23.2240726759966\\
29	23.3619266688981\\
31	23.507514725765\\
33	23.6544143062418\\
35	23.8106776350604\\
37	23.9561924585678\\
39	24.0896828211828\\
41	24.2369175116656\\
43	24.3759333642718\\
45	24.525845674439\\
47	24.6805824053473\\
49	24.8255316434436\\
51	24.9748992242741\\
53	25.1272481653752\\
55	25.2609905206531\\
57	25.4067702534537\\
59	25.5762266322476\\
61	25.7316351559715\\
63	25.8676528107467\\
65	26.024895204783\\
67	26.1781865558848\\
69	26.3206407566591\\
71	26.4652852564223\\
73	26.6184971742207\\
75	26.7711442422752\\
77	26.9243689916545\\
79	27.0769003150561\\
81	27.2201275253212\\
83	27.3837456761446\\
85	27.5287246115918\\
87	27.6905780322835\\
89	27.8441501973748\\
91	27.9989042897457\\
93	28.1558878375954\\
95	28.3220043736376\\
97	28.4911817848348\\
99	28.6532000880129\\
101	28.8187881403168\\
103	28.993453153207\\
};
\addlegendentry{\roothannwin{A}}

\addplot [line width=\plotlinewidth, color=mycolor1, dashed ,forget plot,  mark=o,  mark options={solid, mycolor1}]
  table[row sep=crcr]{%
5	10.8836767550876\\
7	12.9454853612196\\
9	12.5060049364275\\
11	13.3798001747273\\
13	13.2479569499792\\
15	13.6718288549766\\
17	13.6177793148418\\
19	13.8597918935269\\
21	13.8336871104321\\
23	13.9974530077481\\
25	13.9827193388521\\
27	14.105963271284\\
29	14.1012296597335\\
31	14.1932331631532\\
33	14.1917169298057\\
35	14.2625627468004\\
37	14.2628912507479\\
39	14.3248187689322\\
41	14.3265653484623\\
43	14.3837221669545\\
45	14.3862015246691\\
47	14.437447123882\\
49	14.4419069132486\\
51	14.4855890507734\\
53	14.4907386198545\\
55	14.5314508834468\\
57	14.5363716109555\\
59	14.5759800024132\\
61	14.5811760044722\\
63	14.6196254294499\\
65	14.6262167351937\\
67	14.6615830147216\\
69	14.6695162745748\\
71	14.70371301652\\
73	14.7111685715812\\
75	14.7451503443146\\
77	14.7527806900594\\
79	14.7844011404793\\
81	14.7918772440588\\
83	14.8240121066746\\
85	14.8322274731882\\
87	14.8627466919457\\
89	14.871707493425\\
91	14.9022630067105\\
93	14.9110795992447\\
95	14.940032762465\\
97	14.9487917039243\\
99	14.9770497663804\\
101	14.9853096245264\\
103	15.0113089049299\\
};
\addlegendentry{rectn nf = 5}

\addplot [line width=\plotlinewidth, color=mycolor2, dashed ,forget plot,  mark=+, mark options={solid, mycolor2}]
  table[row sep=crcr]{%
5	8.86670658304267\\
7	23.3432645998745\\
9	22.9590400226843\\
11	23.6917855637852\\
13	23.6657886616288\\
15	23.9130478888827\\
17	23.9181670574682\\
19	24.0553451867189\\
21	24.0630510498011\\
23	24.1580807890561\\
25	24.168929910232\\
27	24.2410501420511\\
29	24.2507988101705\\
31	24.3079745008229\\
33	24.3196436301098\\
35	24.3682250601046\\
37	24.3788843024492\\
39	24.4193126358902\\
41	24.430612554116\\
43	24.4687180266131\\
45	24.4802178008268\\
47	24.5142340777735\\
49	24.5241418333814\\
51	24.5541546510634\\
53	24.5629307252661\\
55	24.589867402063\\
57	24.5997318240512\\
59	24.6263441522453\\
61	24.6373817307374\\
63	24.6628728596124\\
65	24.6742727358871\\
67	24.700545244559\\
69	24.7129240479511\\
71	24.7386543260029\\
73	24.7496405957018\\
75	24.7748013270524\\
77	24.7860178768143\\
79	24.811428295776\\
81	24.8221003971883\\
83	24.8451743450248\\
85	24.8565362306575\\
87	24.879590985099\\
89	24.8916630466206\\
91	24.9156057270587\\
93	24.9261190157336\\
95	24.9486867009045\\
97	24.9605612331736\\
99	24.9836863875799\\
101	24.9949979967204\\
103	25.0170081614846\\
};
\addlegendentry{Cosn nf = 5}

\addplot [line width=\plotlinewidth, color=mycolor3, dashed ,forget plot,  mark=asterisk, mark options={solid, mycolor3}]
  table[row sep=crcr]{%
5	11.5849041316751\\
7	20.0558426428392\\
9	22.1239336664734\\
11	22.8488526662499\\
13	22.8735353366254\\
15	23.199932220965\\
17	23.1751233068064\\
19	23.3517120264071\\
21	23.3339221635284\\
23	23.454221052879\\
25	23.4456642566949\\
27	23.538781410125\\
29	23.5335764356555\\
31	23.6060663565981\\
33	23.6044076079859\\
35	23.664851600577\\
37	23.6649501347836\\
39	23.7147858241126\\
41	23.7161745138483\\
43	23.7628171270301\\
45	23.7662139338841\\
47	23.8077527631387\\
49	23.8118047260238\\
51	23.8481614110532\\
53	23.8515467279849\\
55	23.8847466764129\\
57	23.8888824920354\\
59	23.9207734045656\\
61	23.9263541780372\\
63	23.9568843708943\\
65	23.9627353403004\\
67	23.9939934124804\\
69	24.0016558024298\\
71	24.0323842232256\\
73	24.039176927451\\
75	24.0685266491938\\
77	24.0752656561151\\
79	24.1049756893084\\
81	24.1117262669342\\
83	24.1390477582206\\
85	24.1463729098063\\
87	24.1733292950396\\
89	24.1814075550283\\
91	24.2101160182761\\
93	24.2168490769413\\
95	24.2436966072988\\
97	24.2519402847126\\
99	24.2794977944287\\
101	24.2872629577367\\
103	24.3126504836974\\
};
\addlegendentry{SqrtHanndes nf = 5}

\addplot [color=mycolor1, dotted ,forget plot, mark phase = 5, mark=o,line width=1pt, mark options={solid, mycolor1}]
  table[row sep=crcr]{%
51	14.0267832288214\\
53	15.9082007332818\\
55	15.5822951600149\\
57	16.3825775235985\\
59	16.2875679080038\\
61	16.6573597455974\\
63	16.6163215198485\\
65	16.8327748484333\\
67	16.8127530574333\\
69	16.9544440569142\\
71	16.94557936411\\
73	17.0462590027397\\
75	17.0438149093826\\
77	17.1277806637871\\
79	17.1281137621439\\
81	17.198551256028\\
83	17.2015526178175\\
85	17.2595879921761\\
87	17.2649259920046\\
89	17.3187637924045\\
91	17.3263677666298\\
93	17.3735604479998\\
95	17.3819193577523\\
97	17.4222765341159\\
99	17.430987839389\\
101	17.4686734828074\\
103	17.4782527173789\\
};
\addlegendentry{rectn nf = 51}

\addplot [color=mycolor2,  dotted ,forget plot, mark=+, mark phase = 5,line width=1pt, mark options={solid, mycolor2}]
  table[row sep=crcr]{%
51	9.60417165709812\\
53	26.8146940235434\\
55	26.460252542158\\
57	27.228075404273\\
59	27.1769632640975\\
61	27.4061935971881\\
63	27.4098791834903\\
65	27.5304935270412\\
67	27.5339706404416\\
69	27.6164873143744\\
71	27.6231170033905\\
73	27.6877251359029\\
75	27.6970476728665\\
77	27.7493857665238\\
79	27.7596828902441\\
81	27.8041633183912\\
83	27.8127778236805\\
85	27.8500917086157\\
87	27.8603440280417\\
89	27.8969254587405\\
91	27.9085570767731\\
93	27.9420045897277\\
95	27.9533672338135\\
97	27.981346518163\\
99	27.9921522744809\\
101	28.019011523558\\
103	28.030792573323\\
};
\addlegendentry{Cosn nf = 51}

\addplot [color=mycolor3, dotted ,forget plot, mark phase = 5,line width=1pt, mark=asterisk, mark options={solid, mycolor3}]
  table[row sep=crcr]{%
51	12.9782020166166\\
53	22.2144669416117\\
55	25.4369480127072\\
57	26.1657237978228\\
59	26.2985558050564\\
61	26.5825628319402\\
63	26.5687832941724\\
65	26.7264228625886\\
67	26.7096274563431\\
69	26.8164139217109\\
71	26.8059147601021\\
73	26.887176044179\\
75	26.8831512024184\\
77	26.9487826066147\\
79	26.947657145411\\
81	27.004902036232\\
83	27.0051062126098\\
85	27.0489261355277\\
87	27.0510636584847\\
89	27.0951414501411\\
91	27.1001152222806\\
93	27.1412015255674\\
95	27.1471502593513\\
97	27.1807756477693\\
99	27.1862701580605\\
101	27.21827219124\\
103	27.2248501521039\\
};
\addlegendentry{SqrtHanndes nf = 51}

\end{axis}

\end{tikzpicture}%

%% file: figures/PTWOLA_Results_L512_SWDFTvsconstnf_des.tex
\definecolor{mycolor1}{rgb}{0.34667,0.53600,0.69067}%
\definecolor{mycolor2}{rgb}{0.91529,0.28157,0.28784}%
\definecolor{mycolor3}{rgb}{0.44157,0.74902,0.43216}%
\begin{tikzpicture}[font=\scriptsize]

\begin{axis}[%
  width=\fwidth,
  height=\fheight,
  at={(0,0)},
scale only axis,
xmin=1,
xmax=103,
ymin=8,
ymax=36,
xlabel={Total subband filter length ($T$)},
ylabel = {\gls{ERLE} (dB)},
axis background/.style={fill=white},
axis x line*=bottom,
axis y line*=left,
legend columns = -1,
legend style={at={(1,0)}, anchor=south east, legend cell align=left, align=left, draw=white!15!black},
xmajorgrids,
ymajorgrids,
mark repeat=10,
mark size=2.0pt
]
\addplot [line width=\plotlinewidth, color=mycolor1, mark=o, mark options={solid, mycolor1}]
  table[row sep=crcr]{%
1	19.2232712591939\\
3	19.3594485090614\\
5	19.5033792215405\\
7	19.6404651446565\\
9	19.7795075539169\\
11	19.9156313855862\\
13	20.0542070509497\\
15	20.1857843854399\\
17	20.3251055881632\\
19	20.4820365057255\\
21	20.6139053312615\\
23	20.743476062025\\
25	20.8806044711466\\
27	21.0076407514589\\
29	21.1538530399666\\
31	21.2922288680045\\
33	21.4365979262754\\
35	21.5705615594137\\
37	21.7062423830764\\
39	21.8510343515222\\
41	21.9999971529931\\
43	22.1350985289182\\
45	22.2676746415736\\
47	22.4010447990272\\
49	22.5479045590797\\
51	22.6689770479161\\
53	22.8125236113258\\
55	22.9465735836608\\
57	23.1001617144404\\
59	23.2413217909049\\
61	23.3736027466725\\
63	23.5129328386292\\
65	23.6595207371192\\
67	23.8152411419694\\
69	23.9511679256937\\
71	24.1052035053928\\
73	24.2581922634782\\
75	24.3977171277031\\
77	24.5402559871682\\
79	24.6748854457105\\
81	24.8124238244453\\
83	24.9424797188106\\
85	25.0836192146245\\
87	25.2194985829335\\
89	25.3540854968023\\
91	25.5003918488064\\
93	25.6237237078614\\
95	25.7659521390377\\
97	25.9078744684129\\
99	26.049698029893\\
101	26.2015738085919\\
103	26.3425663999993\\
};
\addlegendentry{\rectwin{A}}

\addplot [line width=\plotlinewidth, color=mycolor2,  mark=+, mark options={solid, mycolor2}]
  table[row sep=crcr]{%
1	17.4916133226291\\
3	24.8099025245419\\
5	24.9617557022263\\
7	25.1107701789079\\
9	25.2518712359028\\
11	25.3996778642891\\
13	25.5512616008483\\
15	25.6945984276777\\
17	25.8292488743426\\
19	25.9798255757871\\
21	26.134689807617\\
23	26.2829816840084\\
25	26.4145323522288\\
27	26.5569684780695\\
29	26.7087723275262\\
31	26.8665305423697\\
33	27.0223094749424\\
35	27.1695452589062\\
37	27.3161423513392\\
39	27.4625540067236\\
41	27.6136928042775\\
43	27.7708968770877\\
45	27.9126330824556\\
47	28.0857975635852\\
49	28.228356348238\\
51	28.3750721927588\\
53	28.5191332815397\\
55	28.6587336104724\\
57	28.8097608066987\\
59	28.9712614887411\\
61	29.1245241840709\\
63	29.2767925888527\\
65	29.4285328936786\\
67	29.5878942813092\\
69	29.736003088572\\
71	29.8948248512544\\
73	30.0480304416245\\
75	30.2073716981549\\
77	30.3760645098337\\
79	30.5390062636031\\
81	30.7028507896987\\
83	30.8683273556086\\
85	31.0228136283456\\
87	31.1898766344011\\
89	31.365903556081\\
91	31.5299988117364\\
93	31.7155627691092\\
95	31.8830839050493\\
97	32.0595206937465\\
99	32.2259880176265\\
101	32.409919585721\\
103	32.6071887699892\\
};
\addlegendentry{\cosinewin{A}}

\addplot [line width=\plotlinewidth, color=mycolor3,  mark=asterisk, mark options={solid, mycolor3}]
  table[row sep=crcr]{%
1	19.1537717891022\\
3	28.7967787562436\\
5	28.9237611320653\\
7	29.0488210284506\\
9	29.1812584988038\\
11	29.3132045014571\\
13	29.4523314884629\\
15	29.5794748783114\\
17	29.7119208258797\\
19	29.8395437600722\\
21	29.9702714405252\\
23	30.1034571053051\\
25	30.2398261892061\\
27	30.3702636794611\\
29	30.5072123948215\\
31	30.6339810464562\\
33	30.7781538540724\\
35	30.915049661458\\
37	31.0494705404022\\
39	31.1794905231924\\
41	31.3085227327393\\
43	31.4459576000619\\
45	31.5871735617069\\
47	31.7345796193814\\
49	31.8682758740336\\
51	32.0112762474479\\
53	32.1508603076187\\
55	32.2793788965166\\
57	32.4180456205764\\
59	32.5758967404551\\
61	32.7172837348006\\
63	32.8500489923327\\
65	33.0040560189365\\
67	33.1440067722632\\
69	33.2808065939704\\
71	33.4154751499869\\
73	33.5576149078263\\
75	33.699009193121\\
77	33.8467304716527\\
79	33.9778215488633\\
81	34.1248459872212\\
83	34.2657350949398\\
85	34.4102649281962\\
87	34.5658169584091\\
89	34.7123046070182\\
91	34.8612147903672\\
93	35.0068090686698\\
95	35.1737158360301\\
97	35.3323033409811\\
99	35.4893106555701\\
101	35.6406959045061\\
103	35.7936812623895\\
};
\addlegendentry{\roothannwin{A}}

\addplot [line width=\plotlinewidth, color=mycolor1, dashed,  mark=o, mark options={solid, mycolor1},forget plot]
  table[row sep=crcr]{%
5	19.5033792215405\\
7	18.1558416629442\\
9	18.6094389507451\\
11	19.3174661434739\\
13	20.0390132079262\\
15	20.6793437375546\\
17	21.2770793014595\\
19	21.8133955553252\\
21	22.2669397712122\\
23	22.649761384353\\
25	23.0050131610499\\
27	23.3653205358846\\
29	23.7106348234676\\
31	24.0295962907577\\
33	24.3242302291439\\
35	24.5743191778054\\
37	24.764689400009\\
39	24.9099978969747\\
41	25.0515394727551\\
43	25.1835277781376\\
45	25.322503038223\\
47	25.4817561572289\\
49	25.6325560082452\\
51	25.7788993020141\\
53	25.9152311460226\\
55	26.0316516659701\\
57	26.1322181852863\\
59	26.2225046719773\\
61	26.3180099869163\\
63	26.4249564975799\\
65	26.5352542761515\\
67	26.6412989069168\\
69	26.7198693758276\\
71	26.7853877082276\\
73	26.8543874233187\\
75	26.9427051062076\\
77	27.0392613145838\\
79	27.112855051701\\
81	27.1629114380034\\
83	27.2249598350932\\
85	27.3110640823063\\
87	27.3734076356061\\
89	27.4362850953692\\
91	27.5270346824687\\
93	27.6143759207737\\
95	27.6819285957338\\
97	27.7639865858258\\
99	27.8525186236591\\
101	27.928753366606\\
103	27.9688852343804\\
};
\addlegendentry{rectdes nf = 5}

\addplot [line width=\plotlinewidth, color=mycolor2, dashed,  mark=+, mark options={solid, mycolor2},forget plot]
  table[row sep=crcr]{%
5	9.4267801846633\\
7	25.0557066607492\\
9	24.431984648947\\
11	25.3989978875946\\
13	25.2923091419898\\
15	25.6533352601323\\
17	25.6019132045722\\
19	25.8039683907228\\
21	25.7716893593052\\
23	25.9124024503909\\
25	25.8894401103738\\
27	25.9939449326979\\
29	25.9822811733498\\
31	26.0626552026832\\
33	26.0555618884754\\
35	26.1244295567317\\
37	26.120058401758\\
39	26.1763598005629\\
41	26.1741331274397\\
43	26.2265959503553\\
45	26.226121327427\\
47	26.2721463731419\\
49	26.2727540306846\\
51	26.3122332864681\\
53	26.3131924656012\\
55	26.3484227010632\\
57	26.35079986981\\
59	26.3851873716506\\
61	26.3884091825996\\
63	26.4216412742454\\
65	26.4261450806681\\
67	26.4588632776952\\
69	26.4647411751429\\
71	26.4971469972839\\
73	26.5026166138112\\
75	26.5342531432654\\
77	26.5385092734399\\
79	26.5707233729187\\
81	26.5756492028139\\
83	26.6040456087894\\
85	26.6103872231972\\
87	26.6381139817568\\
89	26.6452732421379\\
91	26.6743706784234\\
93	26.6798371431601\\
95	26.7077979361158\\
97	26.7145595025895\\
99	26.7426284542336\\
101	26.7498263326934\\
103	26.7765645613932\\
};
\addlegendentry{Cosdes nf = 5}

\addplot [line width=\plotlinewidth, color=mycolor3, dashed,  mark=asterisk, mark options={solid, mycolor3},forget plot]
  table[row sep=crcr]{%
5	12.0638028371614\\
7	20.2877020938472\\
9	26.7572399761643\\
11	28.0015744754175\\
13	28.8622885253132\\
15	29.4468509094382\\
17	29.5920395632393\\
19	29.8947943190563\\
21	29.8962178013879\\
23	30.1025214595299\\
25	30.0758803443207\\
27	30.2404281862907\\
29	30.2244021662719\\
31	30.3371493721355\\
33	30.3201044122785\\
35	30.422854219826\\
37	30.4079091663942\\
39	30.4888774560436\\
41	30.4730853577424\\
43	30.5477302842332\\
45	30.5372882040408\\
47	30.6046690046809\\
49	30.5979444003884\\
51	30.653989722208\\
53	30.6447842363314\\
55	30.6949248402\\
57	30.687466489625\\
59	30.7327508387438\\
61	30.7256610057499\\
63	30.7716793820546\\
65	30.7640280799714\\
67	30.8052009177177\\
69	30.8002565736654\\
71	30.8460118652287\\
73	30.8465271923304\\
75	30.8867673096825\\
77	30.8822007000276\\
79	30.9264461102993\\
81	30.9226042891913\\
83	30.9589550127837\\
85	30.9580468674634\\
87	30.9913450755847\\
89	30.9898188884646\\
91	31.0298087661247\\
93	31.0287564969256\\
95	31.0648890690192\\
97	31.0632596916667\\
99	31.1024533861133\\
101	31.1050238585475\\
103	31.1405210972436\\
};
\addlegendentry{SqrtHanndes nf = 5}

\addplot [color=mycolor1, dotted,  mark phase = 5, line width=1pt, mark=o, mark options={solid, mycolor1},forget plot]
  table[row sep=crcr]{%
51	22.6689770479161\\
53	21.1919110545914\\
55	21.672124427196\\
57	22.4155886223262\\
59	23.1956503757176\\
61	23.8788826711749\\
63	24.458585415219\\
65	24.9958080866338\\
67	25.4682310932565\\
69	25.882331166491\\
71	26.2463779348488\\
73	26.5391070154749\\
75	26.7840652296737\\
77	27.015801034651\\
79	27.2498617243126\\
81	27.4769231078737\\
83	27.6939917780451\\
85	27.8804433686964\\
87	28.0474434605881\\
89	28.2237787440299\\
91	28.3970827627893\\
93	28.5661695435112\\
95	28.7379708944032\\
97	28.8922950976942\\
99	29.0245276499015\\
101	29.1343531099863\\
103	29.2352007975311\\
};
\addlegendentry{rectdes nf = 51}

\addplot [color=mycolor2,  dotted, mark=+, mark phase = 5, line width=1pt,mark options={solid, mycolor2},forget plot]
  table[row sep=crcr]{%
51	9.33721253936773\\
53	28.4520064312896\\
55	28.0208977391614\\
57	29.0134614491552\\
59	28.9054677736009\\
61	29.2296410421876\\
63	29.1922941375661\\
65	29.3673934625534\\
67	29.3376867720269\\
69	29.452912417954\\
71	29.4363851969857\\
73	29.5267029733436\\
75	29.5167193270467\\
77	29.5881832689937\\
79	29.583481399405\\
81	29.6432999484003\\
83	29.641477476326\\
85	29.6903165334329\\
87	29.6889280024259\\
89	29.7364446115417\\
91	29.7390004854831\\
93	29.7809168038893\\
95	29.7869052872265\\
97	29.8211788566061\\
99	29.8264650028059\\
101	29.8591486180596\\
103	29.8651836952984\\
};
\addlegendentry{Cosdes nf = 51}

\addplot [color=mycolor3, dotted, mark phase = 5, mark=asterisk, line width=1pt, mark options={solid, mycolor3},forget plot]
  table[row sep=crcr]{%
51	12.2522096242564\\
53	21.6446420863358\\
55	29.259895947469\\
57	30.9660685923204\\
59	32.5663166747131\\
61	33.216996826501\\
63	33.5120466714626\\
65	33.8532881461446\\
67	33.8991071824643\\
69	34.1058729935072\\
71	34.1045753789037\\
73	34.2590231983767\\
75	34.247201474182\\
77	34.3634590366587\\
79	34.3453806558293\\
81	34.4527665255902\\
83	34.4514548635702\\
85	34.5213750715275\\
87	34.4951130854343\\
89	34.5666050172048\\
91	34.5581647533521\\
93	34.6232984326314\\
95	34.6319583905298\\
97	34.6804745871901\\
99	34.6729895678011\\
101	34.7205725530296\\
103	34.716350517669\\
};
\addlegendentry{SqrtHanndes nf = 51}

\end{axis}

\end{tikzpicture}%

%% file: WOLA_subband_conclusion.tex
\section{Conclusion}\label{WOLA_subband_filtering_Sec:conclusions}
\noindent This paper has addressed the constraints inherent in conventional WOLA filter banks, and introduced a generalized \gls{WOLA} filter bank for subband adaptive filtering and investigated its performance for subband system identification. 

\noindent The \gls{MSE} performance of the generalized \gls{WOLA} filter bank has been analyzed, highlighting the critical factors for accurate full-band system identification, including the order of the subband filters, the full-band system impulse response length, decimation factor, and the design of the analysis and synthesis filter bank. 

\noindent Both analytical and empirical evidence has confirmed that the proposed generalized \gls{WOLA} filter bank substantially enhances performance in subband system identification compared to the conventional \gls{WOLA} filter bank.  To mitigate the increased computational complexity associated with the generalized \gls{WOLA} filter bank, the \gls{PTWOLA} was proposed, which maintains computational complexity on par with conventional \gls{WOLA}, while demonstrating performance that is comparable to, or slightly exceeds, that of the generalized \gls{WOLA} filter bank for the same subband filter order.

For future research, a promising direction is to extend the application of the generalized \gls{WOLA} and \gls{PTWOLA} frameworks from single-channel system identification to multi-channel spatial filtering. Conventional \gls{WOLA} filter banks are widely used for tasks such as adaptive beamforming for noise reduction and Direction of Arrival (DOA) estimation. The improved modeling capabilities demonstrated in this paper could offer significant performance enhancements in these multi-channel scenarios.